\documentclass[longauth]{aa}  

\usepackage{graphicx}
\usepackage{txfonts}
\usepackage{lscape}             % to rotate a single page table, example in appendix.
\usepackage{placeins}           % useful with \FloatBarrier, to keep 
\usepackage{tabularx}
\usepackage{multirow}
\usepackage{booktabs}
\usepackage{float}
\usepackage{xcolor}
\usepackage{natbib}
\usepackage{comment}
\usepackage[breaklinks, colorlinks, citecolor=blue, linkcolor=blue]{hyperref}
\begin{document}

   \title{Searching for hot water world candidates with CHEOPS}
   \subtitle{II. A low-density super-Earth orbiting the K~dwarf TOI-2211}
   \titlerunning{TOI-2211}

%%%%%%%%%%%%%%%%%%%%%%%%%%%%%%%%%%%%%%%%
% Please do not include ORCIDs next to author names.
% Only ORCIDs authenticated by individual authors in EDP Sciences editorial system will be taken into account.
% ORCIDs included here will be removed.
%%%%%%%%%%%%%%%%%%%%%%%%%%%%%%%%%%%%%%%%

   \author{
        J.~A.~Egger\inst{\ref{inst:1}}\,$^{\href{https://orcid.org/0000-0003-1628-4231}{\protect\includegraphics[height=0.19cm]{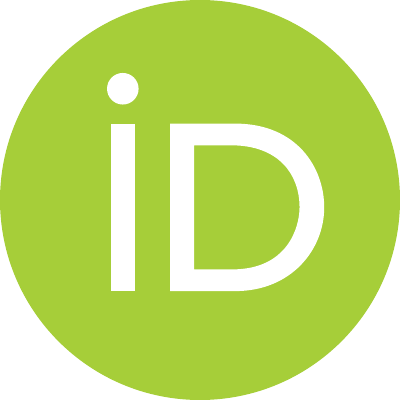}}}$, 
        P.~Cortés-Zuleta\inst{\ref{inst:2}}\,$^{\href{https://orcid.org/0000-0002-6174-4666}{\protect\includegraphics[height=0.19cm]{figures/orcid.pdf}}}$, 
        D.~Kubyshkina\inst{\ref{inst:3}}, 
        H.~P.~Osborn\inst{\ref{inst:4},\ref{inst:5}}\,$^{\href{https://orcid.org/0000-0002-4047-4724}{\protect\includegraphics[height=0.19cm]{figures/orcid.pdf}}}$, 
        S.~Marques\inst{\ref{inst:6}}\,$^{\href{https://orcid.org/0009-0007-3825-7791}{\protect\includegraphics[height=0.19cm]{figures/orcid.pdf}}}$, 
        I.~Crossfield\inst{\ref{inst:7}}, 
        A.~Deline\inst{\ref{inst:8}}, 
        S.~Ulmer-Moll\inst{\ref{inst:9},\ref{inst:10}}\,$^{\href{https://orcid.org/0000-0003-2417-7006}{\protect\includegraphics[height=0.19cm]{figures/orcid.pdf}}}$, 
        A.~E.~Simon\inst{\ref{inst:6},\ref{inst:4}}\,$^{\href{https://orcid.org/0000-0001-9773-2600}{\protect\includegraphics[height=0.19cm]{figures/orcid.pdf}}}$, 
        G.~Scandariato\inst{\ref{inst:11}}\,$^{\href{https://orcid.org/0000-0003-2029-0626}{\protect\includegraphics[height=0.19cm]{figures/orcid.pdf}}}$, 
        A.~C.~M.~Correia\inst{\ref{inst:12}}\,$^{\href{https://orcid.org/0000-0002-8946-8579}{\protect\includegraphics[height=0.19cm]{figures/orcid.pdf}}}$, 
        Y.~N.~E.~Eschen\inst{\ref{inst:13}}\,$^{\href{https://orcid.org/0009-0006-6397-2503}{\protect\includegraphics[height=0.19cm]{figures/orcid.pdf}}}$, 
        D.~Gandolfi\inst{\ref{inst:14}}\,$^{\href{https://orcid.org/0000-0001-8627-9628}{\protect\includegraphics[height=0.19cm]{figures/orcid.pdf}}}$, 
        V.~Adibekyan\inst{\ref{inst:15}}\,$^{\href{https://orcid.org/0000-0002-0601-6199}{\protect\includegraphics[height=0.19cm]{figures/orcid.pdf}}}$, 
        A.~Bonfanti\inst{\ref{inst:3}}\,$^{\href{https://orcid.org/0000-0002-1916-5935}{\protect\includegraphics[height=0.19cm]{figures/orcid.pdf}}}$, 
        C.~Pezzotti\inst{\ref{inst:10}}, 
        P.~H.~Premnath\inst{\ref{inst:16}}, 
        S.~G.~Sousa\inst{\ref{inst:17},\ref{inst:18}}\,$^{\href{https://orcid.org/0000-0001-9047-2965}{\protect\includegraphics[height=0.19cm]{figures/orcid.pdf}}}$, 
        T.~G.~Wilson\inst{\ref{inst:13}}\,$^{\href{https://orcid.org/0000-0001-8749-1962}{\protect\includegraphics[height=0.19cm]{figures/orcid.pdf}}}$, 
        L.~Fossati\inst{\ref{inst:3}}\,$^{\href{https://orcid.org/0000-0003-4426-9530}{\protect\includegraphics[height=0.19cm]{figures/orcid.pdf}}}$, 
        R.~Eltschinger\inst{\ref{inst:6},\ref{inst:4}}, 
        Y.~Alibert\inst{\ref{inst:4},\ref{inst:6}}\,$^{\href{https://orcid.org/0000-0002-4644-8818}{\protect\includegraphics[height=0.19cm]{figures/orcid.pdf}}}$,
        D.~Degen\inst{\ref{inst:5}}\,$^{\href{https://orcid.org/0009-0008-1068-481X}{\protect\includegraphics[height=0.19cm]{figures/orcid.pdf}}}$, 
        G.~Lacedelli\inst{\ref{inst:19}}\,$^{\href{https://orcid.org/0000-0002-4197-7374}{\protect\includegraphics[height=0.19cm]{figures/orcid.pdf}}}$, 
        T.~Zingales\inst{\ref{inst:20},\ref{inst:21}}\,$^{\href{https://orcid.org/0000-0001-6880-5356}{\protect\includegraphics[height=0.19cm]{figures/orcid.pdf}}}$, 
        B.~Akinsanmi\inst{\ref{inst:8}}\,$^{\href{https://orcid.org/0000-0001-6519-1598}{\protect\includegraphics[height=0.19cm]{figures/orcid.pdf}}}$, 
        R.~Alonso\inst{\ref{inst:19},\ref{inst:22}}\,$^{\href{https://orcid.org/0000-0001-8462-8126}{\protect\includegraphics[height=0.19cm]{figures/orcid.pdf}}}$, 
        T.~Bárczy\inst{\ref{inst:23}}\,$^{\href{https://orcid.org/0000-0002-7822-4413}{\protect\includegraphics[height=0.19cm]{figures/orcid.pdf}}}$, 
        D.~Barrado\inst{\ref{inst:24}}\,$^{\href{https://orcid.org/0000-0002-5971-9242}{\protect\includegraphics[height=0.19cm]{figures/orcid.pdf}}}$, 
        S.~C.~C.~Barros\inst{\ref{inst:17},\ref{inst:18}}\,$^{\href{https://orcid.org/0000-0003-2434-3625}{\protect\includegraphics[height=0.19cm]{figures/orcid.pdf}}}$, 
        W.~Baumjohann\inst{\ref{inst:3}}\,$^{\href{https://orcid.org/0000-0001-6271-0110}{\protect\includegraphics[height=0.19cm]{figures/orcid.pdf}}}$, 
        W.~Benz\inst{\ref{inst:6},\ref{inst:4}}\,$^{\href{https://orcid.org/0000-0001-7896-6479}{\protect\includegraphics[height=0.19cm]{figures/orcid.pdf}}}$, 
        N.~Billot\inst{\ref{inst:8}}\,$^{\href{https://orcid.org/0000-0003-3429-3836}{\protect\includegraphics[height=0.19cm]{figures/orcid.pdf}}}$, 
        L.~Borsato\inst{\ref{inst:21}}\,$^{\href{https://orcid.org/0000-0003-0066-9268}{\protect\includegraphics[height=0.19cm]{figures/orcid.pdf}}}$, 
        A.~Brandeker\inst{\ref{inst:25}}\,$^{\href{https://orcid.org/0000-0002-7201-7536}{\protect\includegraphics[height=0.19cm]{figures/orcid.pdf}}}$, 
        C.~Broeg\inst{\ref{inst:6},\ref{inst:4}}\,$^{\href{https://orcid.org/0000-0001-5132-2614}{\protect\includegraphics[height=0.19cm]{figures/orcid.pdf}}}$, 
        M.~Buder\inst{\ref{inst:26}}, 
        A.~Castro-González\inst{\ref{inst:8}}, 
        A.~Collier~Cameron\inst{\ref{inst:2}}\,$^{\href{https://orcid.org/0000-0002-8863-7828}{\protect\includegraphics[height=0.19cm]{figures/orcid.pdf}}}$, 
        C.~Corral~Van~Damme\inst{\ref{inst:1}}, 
        Sz.~Csizmadia\inst{\ref{inst:26}}\,$^{\href{https://orcid.org/0000-0001-6803-9698}{\protect\includegraphics[height=0.19cm]{figures/orcid.pdf}}}$, 
        P.~E.~Cubillos\inst{\ref{inst:3},\ref{inst:28}}, 
        M.~B.~Davies\inst{\ref{inst:29}}\,$^{\href{https://orcid.org/0000-0001-6080-1190}{\protect\includegraphics[height=0.19cm]{figures/orcid.pdf}}}$, 
        M.~Deleuil\inst{\ref{inst:30}}\,$^{\href{https://orcid.org/0000-0001-6036-0225}{\protect\includegraphics[height=0.19cm]{figures/orcid.pdf}}}$, 
        O.~D.~S.~Demangeon\inst{\ref{inst:17},\ref{inst:18}}\,$^{\href{https://orcid.org/0000-0001-7918-0355}{\protect\includegraphics[height=0.19cm]{figures/orcid.pdf}}}$, 
        B.-O.~Demory\inst{\ref{inst:4},\ref{inst:31},\ref{inst:6}}\,$^{\href{https://orcid.org/0000-0002-9355-5165}{\protect\includegraphics[height=0.19cm]{figures/orcid.pdf}}}$, 
        A.~Derekas\inst{\ref{inst:32}}, 
        B.~Edwards\inst{\ref{inst:33}}, 
        D.~Ehrenreich\inst{\ref{inst:8},\ref{inst:34}}\,$^{\href{https://orcid.org/0000-0001-9704-5405}{\protect\includegraphics[height=0.19cm]{figures/orcid.pdf}}}$, 
        A.~Erikson\inst{\ref{inst:26}}, 
        A.~Fortier\inst{\ref{inst:6},\ref{inst:4}}\,$^{\href{https://orcid.org/0000-0001-8450-3374}{\protect\includegraphics[height=0.19cm]{figures/orcid.pdf}}}$, 
        M.~Fridlund\inst{\ref{inst:35},\ref{inst:36}}\,$^{\href{https://orcid.org/0000-0002-0855-8426}{\protect\includegraphics[height=0.19cm]{figures/orcid.pdf}}}$, 
        K.~Gazeas\inst{\ref{inst:37}}\,$^{\href{https://orcid.org/0000-0002-8855-3923}{\protect\includegraphics[height=0.19cm]{figures/orcid.pdf}}}$, 
        M.~Gillon\inst{\ref{inst:38}}\,$^{\href{https://orcid.org/0000-0003-1462-7739}{\protect\includegraphics[height=0.19cm]{figures/orcid.pdf}}}$, 
        M.~Güdel\inst{\ref{inst:39}}, 
        M.~N.~Günther\inst{\ref{inst:1}}\,$^{\href{https://orcid.org/0000-0002-3164-9086}{\protect\includegraphics[height=0.19cm]{figures/orcid.pdf}}}$, 
        J.~Hasiba\inst{\ref{inst:3}}, 
        A.~Heitzmann\inst{\ref{inst:8}}\,$^{\href{https://orcid.org/0000-0002-8091-7526}{\protect\includegraphics[height=0.19cm]{figures/orcid.pdf}}}$, 
        Ch.~Helling\inst{\ref{inst:3},\ref{inst:40}}, 
        K.~G.~Isaak\inst{\ref{inst:1}}\,$^{\href{https://orcid.org/0000-0001-8585-1717}{\protect\includegraphics[height=0.19cm]{figures/orcid.pdf}}}$, 
        L.~L.~Kiss\inst{\ref{inst:41},\ref{inst:42}}, 
        D.~Kitzmann\inst{\ref{inst:6},\ref{inst:4}}, 
        J.~Korth\inst{\ref{inst:43},\ref{inst:44}}\,$^{\href{https://orcid.org/0000-0002-0076-6239}{\protect\includegraphics[height=0.19cm]{figures/orcid.pdf}}}$, 
        K.~W.~F.~Lam\inst{\ref{inst:26}}\,$^{\href{https://orcid.org/0000-0002-9910-6088}{\protect\includegraphics[height=0.19cm]{figures/orcid.pdf}}}$, 
        J.~Laskar\inst{\ref{inst:45}}\,$^{\href{https://orcid.org/0000-0003-2634-789X}{\protect\includegraphics[height=0.19cm]{figures/orcid.pdf}}}$, 
        A.~Lecavelier~des~Etangs\inst{\ref{inst:46}}\,$^{\href{https://orcid.org/0000-0002-5637-5253}{\protect\includegraphics[height=0.19cm]{figures/orcid.pdf}}}$, 
        A.~Leleu\inst{\ref{inst:8},\ref{inst:6}}\,$^{\href{https://orcid.org/0000-0003-2051-7974}{\protect\includegraphics[height=0.19cm]{figures/orcid.pdf}}}$, 
        M.~Lendl\inst{\ref{inst:8}}\,$^{\href{https://orcid.org/0000-0001-9699-1459}{\protect\includegraphics[height=0.19cm]{figures/orcid.pdf}}}$, 
        M.~Lund\inst{\ref{inst:47}}, 
        D.~Magrin\inst{\ref{inst:21}}\,$^{\href{https://orcid.org/0000-0003-0312-313X}{\protect\includegraphics[height=0.19cm]{figures/orcid.pdf}}}$, 
        P.~F.~L.~Maxted\inst{\ref{inst:48}}\,$^{\href{https://orcid.org/0000-0003-3794-1317}{\protect\includegraphics[height=0.19cm]{figures/orcid.pdf}}}$, 
        B.~Merín\inst{\ref{inst:49}}\,$^{\href{https://orcid.org/0000-0002-8555-3012}{\protect\includegraphics[height=0.19cm]{figures/orcid.pdf}}}$, 
        C.~Mordasini\inst{\ref{inst:6},\ref{inst:4}}, 
        M.~Munari\inst{\ref{inst:11}}\,$^{\href{https://orcid.org/0000-0003-0990-050X}{\protect\includegraphics[height=0.19cm]{figures/orcid.pdf}}}$, 
        V.~Nascimbeni\inst{\ref{inst:21}}\,$^{\href{https://orcid.org/0000-0001-9770-1214}{\protect\includegraphics[height=0.19cm]{figures/orcid.pdf}}}$, 
        G.~Olofsson\inst{\ref{inst:25}}\,$^{\href{https://orcid.org/0000-0003-3747-7120}{\protect\includegraphics[height=0.19cm]{figures/orcid.pdf}}}$, 
        R.~Ottensamer\inst{\ref{inst:39}}, 
        I.~Pagano\inst{\ref{inst:11}}\,$^{\href{https://orcid.org/0000-0001-9573-4928}{\protect\includegraphics[height=0.19cm]{figures/orcid.pdf}}}$, 
        E.~Pallé\inst{\ref{inst:19},\ref{inst:22}}\,$^{\href{https://orcid.org/0000-0003-0987-1593}{\protect\includegraphics[height=0.19cm]{figures/orcid.pdf}}}$, 
        A.~Peel\inst{\ref{inst:6},\ref{inst:4}}, 
        G.~Peter\inst{\ref{inst:26}}\,$^{\href{https://orcid.org/0000-0001-6101-2513}{\protect\includegraphics[height=0.19cm]{figures/orcid.pdf}}}$, 
        D.~Piazza\inst{\ref{inst:50}}, 
        G.~Piotto\inst{\ref{inst:21},\ref{inst:20}}\,$^{\href{https://orcid.org/0000-0002-9937-6387}{\protect\includegraphics[height=0.19cm]{figures/orcid.pdf}}}$, 
        D.~Pollacco\inst{\ref{inst:13}}, 
        D.~Queloz\inst{\ref{inst:5},\ref{inst:51}}\,$^{\href{https://orcid.org/0000-0002-3012-0316}{\protect\includegraphics[height=0.19cm]{figures/orcid.pdf}}}$, 
        R.~Ragazzoni\inst{\ref{inst:21},\ref{inst:20}}\,$^{\href{https://orcid.org/0000-0002-7697-5555}{\protect\includegraphics[height=0.19cm]{figures/orcid.pdf}}}$, 
        N.~Rando\inst{\ref{inst:1}}, 
        H.~Rauer\inst{\ref{inst:52},\ref{inst:53}}\,$^{\href{https://orcid.org/0000-0002-6510-1828}{\protect\includegraphics[height=0.19cm]{figures/orcid.pdf}}}$, 
        I.~Ribas\inst{\ref{inst:54},\ref{inst:55}}\,$^{\href{https://orcid.org/0000-0002-6689-0312}{\protect\includegraphics[height=0.19cm]{figures/orcid.pdf}}}$, 
        N.~C.~Santos\inst{\ref{inst:17},\ref{inst:18}}\,$^{\href{https://orcid.org/0000-0003-4422-2919}{\protect\includegraphics[height=0.19cm]{figures/orcid.pdf}}}$, 
        D.~Ségransan\inst{\ref{inst:8}}\,$^{\href{https://orcid.org/0000-0003-2355-8034}{\protect\includegraphics[height=0.19cm]{figures/orcid.pdf}}}$, 
        A.~M.~S.~Smith\inst{\ref{inst:26}}\,$^{\href{https://orcid.org/0000-0002-2386-4341}{\protect\includegraphics[height=0.19cm]{figures/orcid.pdf}}}$, 
        M.~Stalport\inst{\ref{inst:10},\ref{inst:38}}, 
        S.~Sulis\inst{\ref{inst:30}}\,$^{\href{https://orcid.org/0000-0001-8783-526X}{\protect\includegraphics[height=0.19cm]{figures/orcid.pdf}}}$, 
        Gy.~M.~Szabó\inst{\ref{inst:32}}\,$^{\href{https://orcid.org/0000-0002-0606-7930}{\protect\includegraphics[height=0.19cm]{figures/orcid.pdf}}}$, 
        S.~Udry\inst{\ref{inst:8}}\,$^{\href{https://orcid.org/0000-0001-7576-6236}{\protect\includegraphics[height=0.19cm]{figures/orcid.pdf}}}$, 
        V.~Van~Grootel\inst{\ref{inst:10}}\,$^{\href{https://orcid.org/0000-0003-2144-4316}{\protect\includegraphics[height=0.19cm]{figures/orcid.pdf}}}$, 
        J.~Venturini\inst{\ref{inst:8}}\,$^{\href{https://orcid.org/0000-0001-9527-2903}{\protect\includegraphics[height=0.19cm]{figures/orcid.pdf}}}$, 
        E.~Villaver\inst{\ref{inst:19},\ref{inst:22}}, 
        N.~A.~Walton\inst{\ref{inst:56}}\,$^{\href{https://orcid.org/0000-0003-3983-8778}{\protect\includegraphics[height=0.19cm]{figures/orcid.pdf}}}$, and
        S.~Wolf\inst{\ref{inst:50}}
    }
    
   \authorrunning{J. A. Egger et al.}

   \institute{
        \label{inst:1} European Space Agency (ESA), European Space Research and Technology Centre (ESTEC), Keplerlaan 1, 2201 AZ Noordwijk, The Netherlands \and
        \label{inst:2} Centre for Exoplanet Science, SUPA, School of Physics and Astronomy, University of St Andrews, North Haugh, St Andrews KY16 9SS, UK \and
        \label{inst:3} Space Research Institute, Austrian Academy of Sciences, Schmiedlstrasse 6, A-8042 Graz, Austria \and
        \label{inst:4} Center for Space and Habitability, University of Bern, Gesellschaftsstrasse 6, 3012 Bern, Switzerland \and
        \label{inst:5} ETH Zurich, Department of Physics, Wolfgang-Pauli-Strasse 2, CH-8093 Zurich, Switzerland \and
        \label{inst:6} Space Research and Planetary Sciences, Physics Institute, University of Bern, Gesellschaftsstrasse 6, 3012 Bern, Switzerland \and
        \label{inst:7} Department of Physics and Astronomy, University of Kansas, Lawrence, KS, USA \and
        \label{inst:8} Observatoire astronomique de l'Université de Genève, Chemin Pegasi 51, 1290 Versoix, Switzerland \and
        \label{inst:9} Leiden Observatory, University of Leiden, Einsteinweg 55, 2333 CA Leiden, The Netherlands \and
        \label{inst:10} Space sciences, Technologies and Astrophysics Research (STAR) Institute, Université de Liège, Allée du 6 Août 19C, 4000 Liège, Belgium \and
        \label{inst:11} INAF, Osservatorio Astrofisico di Catania, Via S. Sofia 78, 95123 Catania, Italy \and
        \label{inst:12} CFisUC, Departamento de Física, Universidade de Coimbra, 3004-516 Coimbra, Portugal \and
        \label{inst:13} Department of Physics, University of Warwick, Gibbet Hill Road, Coventry CV4 7AL, United Kingdom \and
        \label{inst:14} Dipartimento di Fisica, Università degli Studi di Torino, via Pietro Giuria 1, I-10125, Torino, Italy \and
        \label{inst:15} Instituto de Astrofisica e Ciencias do Espaco, Universidade do Porto, CAUP, Rua das Estrelas, 4150-762 Porto, Portugal
        Departamento de Fisica e Astronomia, Faculdade de Ciencias, Universidade do Porto, Rua do Campo Alegre, 4169-007 Porto, Portugal \and
        \label{inst:16} Department of Physics \& Astronomy, University of California, Irvine, CA 92697, USA \and
        \label{inst:17} Instituto de Astrofisica e Ciencias do Espaco, Universidade do Porto, CAUP, Rua das Estrelas, 4150-762 Porto, Portugal \and
        \label{inst:18} Departamento de Fisica e Astronomia, Faculdade de Ciencias, Universidade do Porto, Rua do Campo Alegre, 4169-007 Porto, Portugal \and
        \label{inst:19} Instituto de Astrofísica de Canarias, Vía Láctea s/n, 38200 La Laguna, Tenerife, Spain \and
        \label{inst:20} Dipartimento di Fisica e Astronomia "Galileo Galilei", Università degli Studi di Padova, Vicolo dell'Osservatorio 3, 35122 Padova, Italy \and
        \label{inst:21} INAF, Osservatorio Astronomico di Padova, Vicolo dell'Osservatorio 5, 35122 Padova, Italy \and
        \label{inst:22} Departamento de Astrofísica, Universidad de La Laguna, Astrofísico Francisco Sanchez s/n, 38206 La Laguna, Tenerife, Spain \and
        \label{inst:23} Admatis, 5. Kandó Kálmán Street, 3534 Miskolc, Hungary \and
        \label{inst:24} Depto. de Astrofísica, Centro de Astrobiología (CSIC-INTA), ESAC campus, 28692 Villanueva de la Cañada (Madrid), Spain \and
        \label{inst:25} Department of Astronomy, Stockholm University, AlbaNova University Center, 10691 Stockholm, Sweden \and
        \label{inst:26} Centre for Exoplanet Science, SUPA School of Physics and Astronomy, University of St Andrews, North Haugh, St Andrews KY16 9SS, UK \and
        \label{inst:27} Institute of Space Research, German Aerospace Center (DLR), Rutherfordstrasse 2, 12489 Berlin, Germany \and
        \label{inst:28} INAF, Osservatorio Astrofisico di Torino, Via Osservatorio, 20, I-10025 Pino Torinese To, Italy \and
        \label{inst:29} Centre for Mathematical Sciences, Lund University, Box 118, 221 00 Lund, Sweden \and
        \label{inst:30} Aix Marseille Univ, CNRS, CNES, LAM, 38 rue Frédéric Joliot-Curie, 13388 Marseille, France \and
        \label{inst:31} ARTORG Center for Biomedical Engineering Research, University of Bern, Bern, Switzerland \and
        \label{inst:32} ELTE Gothard Astrophysical Observatory, 9700 Szombathely, Szent Imre h. u. 112, Hungary \and
        \label{inst:33} SRON Netherlands Institute for Space Research, Niels Bohrweg 4, 2333 CA Leiden, Netherlands \and
        \label{inst:34} Centre Vie dans l’Univers, Faculté des sciences, Université de Genève, Quai Ernest-Ansermet 30, 1211 Genève 4, Switzerland \and
        \label{inst:35} Leiden Observatory, University of Leiden, PO Box 9513, 2300 RA Leiden, The Netherlands \and
        \label{inst:36} Department of Space, Earth and Environment, Chalmers University of Technology, Onsala Space Observatory, 439 92 Onsala, Sweden \and
        \label{inst:37} National and Kapodistrian University of Athens, Department of Physics, University Campus, Zografos GR-157 84, Athens, Greece \and
        \label{inst:38} Astrobiology Research Unit, Université de Liège, Allée du 6 Août 19C, B-4000 Liège, Belgium \and
        \label{inst:39} Department of Astrophysics, University of Vienna, Türkenschanzstrasse 17, 1180 Vienna, Austria \and
        \label{inst:40} Institute for Theoretical Physics and Computational Physics, Graz University of Technology, Petersgasse 16, 8010 Graz, Austria \and
        \label{inst:41} Konkoly Observatory, Research Centre for Astronomy and Earth Sciences, 1121 Budapest, Konkoly Thege Miklós út 15-17, Hungary \and
        \label{inst:42} ELTE E\"otv\"os Lor\'and University, Institute of Physics, P\'azm\'any P\'eter s\'et\'any 1/A, 1117 Budapest, Hungary \and
        \label{inst:43} Instituto de Astrofísica de Canarias (IAC), 38205 La Laguna, Tenerife, Spain \and
        \label{inst:44} Departamento de Astrofísica, Universidad de la Laguna (ULL), E-38206 La Laguna, Tenerife Spain \and
        \label{inst:45} IMCCE, UMR8028 CNRS, Observatoire de Paris, PSL Univ., Sorbonne Univ., 77 av. Denfert-Rochereau, 75014 Paris, France \and
        \label{inst:46} Institut d'astrophysique de Paris, UMR7095 CNRS, Université Pierre \& Marie Curie, 98bis blvd. Arago, 75014 Paris, France \and
        \label{inst:47} NASA Exoplanet Science Institute, IPAC, California Institute of Technology, Pasadena, CA 91125, USA \and
        \label{inst:48} Astrophysics Group, Lennard Jones Building, Keele University, Staffordshire, ST5 5BG, United Kingdom \and
        \label{inst:49} European Space Agency, ESA - European Space Astronomy Centre, Camino Bajo del Castillo s/n, 28692 Villanueva de la Cañada, Madrid, Spain \and
        \label{inst:50} Weltraumforschung und Planetologie, Physikalisches Institut, University of Bern, Gesellschaftsstrasse 6, 3012 Bern, Switzerland \and
        \label{inst:51} Cavendish Laboratory, JJ Thomson Avenue, Cambridge CB3 0HE, UK \and
        \label{inst:52} German Aerospace Center (DLR), Markgrafenstrasse 37, 10117 Berlin, Germany \and
        \label{inst:53} Institut fuer Geologische Wissenschaften, Freie Universitaet Berlin, Malteserstrasse 74-100,12249 Berlin, Germany \and
        \label{inst:54} Institut de Ciencies de l'Espai (ICE, CSIC), Campus UAB, Can Magrans s/n, 08193 Bellaterra, Spain \and
        \label{inst:55} Institut d'Estudis Espacials de Catalunya (IEEC), 08860 Castelldefels (Barcelona), Spain \and
        \label{inst:56} Institute of Astronomy, University of Cambridge, Madingley Road, Cambridge, CB3 0HA, United Kingdom
   }

   \date{Received 24 June 2026; accepted 5 October 2026}
 
   \abstract{Small exoplanets on close-in orbits are exposed to high levels of stellar irradiation, which can effectively remove low-mass primordial H/He envelopes through atmospheric escape. Under these conditions, planets whose masses and radii are incompatible with both bare rocky interiors and extended H-dominated atmospheres are therefore expected to host high mean molecular weight atmospheres, making them promising candidates for volatile-rich hot water worlds. We present the first detailed characterisation of the low-density super-Earth TOI-2211\,b (P$\sim$3.09~d), a recently validated transiting planet orbiting a K~type star (G~mag: 9.1). Combining TESS and CHEOPS photometry with radial velocity measurements from HARPS, ESPRESSO, and KPF, we precisely constrain the planet's radius and mass to $1.457\pm 0.047$~R$_\oplus$ and $2.69 \pm 0.56$~M$_\oplus$. We also identify an outer, non-transiting planetary candidate (P$\sim$15.6~d, Msin(i) $ = 8.53 \pm 1.04$~M$_\oplus$) in the radial velocity data. We investigate the internal structure of the transiting planet and assess the stability of the inferred volatile envelopes through a detailed evaporation analysis, thereby identifying TOI-2211\,b as a promising hot water world candidate. TOI-2211\,b therefore joins the small but growing sample of these close-in low-density super-Earths, which represent particularly interesting targets for testing current planet formation and evolution theories. Additionally, we investigate the full architecture of the planetary system using a generative model of planet formation and evolution, with the resulting posterior distributions showing a clear over-density consistent with the observed orbital period and minimal mass of the non-transiting planetary candidate identified in the radial velocity data.}


   \maketitle
   \nolinenumbers

%%%%%%%%%%%%%%%%%%%%%%%%%%%%%%%%%%%%%%%%%%%%%%%%%%%%%%%%%%%%%%
\section{Introduction}
\label{sec:introduction}

Planets with radii between the ones of Earth and Neptune do not have an analogue in our Solar system, but are among the most commonly detected in our galaxy \citep{Mayor+2011,Fressin+2013,Hsu+2019,Kunimoto+2020}. Indeed, roughly 60\% of all confirmed planets listed by the NASA exoplanet archive fall into this category.\footnote{\url{https://exoplanetarchive.ipac.caltech.edu/}, query from 12~May 2026.} The distribution of these planets' radii shows a distinct bimodality, with the smaller super-Earths separated from the larger sub-Neptunes by the so-called radius valley \citep[][for orbital periods smaller than $\sim$100~days]{Fulton+2017}. Such an evaporation valley was previously predicted based on theoretical models by several independent groups \citep{Owen+Wu2013,Jin+2014,Lopez+Fortney2014}, all assuming sub-Neptunes to host H/He envelopes. Since then, models based on this assumption have been shown to accurately reproduce the observational properties of the radius valley well \citep[e.g.][]{Owen+Wu2017,Fulton+Petigura2018,VanEylen+2018,Mordasini2020,Ho+VanEylen2023,Bonfanti+2024,Ho+2024}.

In more recent years, this traditional picture of sub-Neptunes with H/He envelopes has, however, been challenged, as many interior compositions are compatible with their observed mass and radius values \citep{Valencia+2007,Seager+2007,Rogers+Seager2010a,Zeng+2019}. For instance, planet formation models predict the existence of a large number of water-rich planets on short orbital periods, having formed outside the water ice line and then migrated inwards \citep[e.g.][]{Ward1997,Tanaka+2002,Emsenhuber+2021a,Emsenhuber+2021b, Emsenhuber+2023}. Multiple studies have shown that such water-rich sub-Neptunes can explain the observed location of the radius valley \citep{Venturini+2020b,Venturini+2020a,Izidoro+2022,Burn+2024,Venturini+2024,Chakrabarty+2026,Shibata+Izidoro2026}. Other works also proposed that, from a formation perspective, some of these planets should be carbon-rich soot planets \citep{Bergin+2023,Lin+Seager2025,Li+2026}. 

In addition, interior-atmosphere interactions have been proposed as a mechanism to alter the primordial atmospheric composition of these planets. For example, volatiles can be partitioned into the interior and later on outgassed by the cooling magma ocean, a process that may also lead to the endogenic production of water \citep[e.g.][]{Kite+2020,Dorn+Lichtenberg2021,Schlichting+Young2022,Seo+2024,Rigby+2024,Luo+2024,Bower+2025,Ito+2025,Werlen+2025,Steinmeyer+2026}. Multiple studies have also investigated how such interactions could shape the compositions of planetary atmospheres across the radius valley, resulting in a broad range of atmospheric compositions. More specifically, \cite{Heng+2025} reported a gradient of mean molecular weight across the radius valley, while \cite{Cherubim+2025} found an oxidation gradient across the valley, with prominent populations of He- and O$_2$-dominated atmospheres on close-in planets. \cite{Woitke+2021} derive a chemical classification for the composition of planetary atmospheres below 600~K, with a particular focus on the co-existence of CH$_4$, CO$_2$, and H$_2$O.

With the advent of JWST \citep{Gardner+2006}, it has now become possible to probe the atmospheric properties of some super-Earths and sub-Neptunes observationally, with the small planets observed so far spanning a broad range of bulk properties and equilibrium temperatures. While many results remain tentative and further observations are needed, the first studies indicate a large diversity of atmospheric compositions, as summarised for example in \cite{Madhusudhan+2025} or \cite{Roy+2026}. For several planets, the observed spectra show features consistent with molecules such as H$_2$O, CO$_2$, and CH$_4$. This was interpreted as evidence for a range of atmospheric compositions, from H$_2$-dominated atmospheres with different enrichment levels (e.g. K2-18~b, \citealp{Madhusudhan+2023}; TOI-270~d, \citealp{Benneke+2024,Holmberg+Madhusudhan2024}; TOI-421~b, \citealp{Davenport+2025}) to steam-dominated atmospheres \citep[GJ~9827~d,][]{Piaulet+2024}. However, other planets showcase flat, featureless spectra, which can either be explained by the presence of clouds and hazes or by a high mean molecular weight atmosphere (e.g. TOI-836~c, \citealp{Wallack+2024}; HD~15337~c, \citealp{Wallack+2026}). 

For a specific subset of planets, some of the intrinsic degeneracy of their internal structure and composition can already be lifted based solely on their measured bulk density. These planets are on orbits close to their host star and receive high levels of stellar irradiation, which can cause part of their atmospheres to evaporate \citep{Owen+Wu2017,Lopez2017,Jin+Mordasini2018,kubyshkina_2021mesa}. H/He atmospheres are more susceptible to this atmospheric escape due to their low mean molecular weight than heavier, volatile-enriched atmospheres \citep{Johnstone+2015,Yoshida+2022,Yoshida+2024,GarciaMunoz+2024,Yoshida+Gaidos2025,Kubyshkina+2026}. As a result, planets that are less dense than a bare rocky core but denser than those with a Mercury-like core and a $\sim$1\% H/He envelope need to contain volatiles heavier than H/He in their envelopes. Similar arguments have previously been made in other works, for example for $\pi$~Men\,c \citep{GarciaMunoz+2020, GarciaMunoz+2021}, GJ~9827\,d \citep{Piaulet+2024}, Kepler-138\,d \citep{Piaulet+2023}, and HD~219134\,b \citep{Dorn+Heng2018,Kubyshkina+2018,Ligi+2019,Gupta+Schlichting2021}, as well as in \cite{Castro-Gonzalez+2023} for a population of low-density super-Earths.

This motivates our CHEOPS \citep[CHaracterising ExOPlanet Satellite;][]{Benz+2021,Fortier+2024} guaranteed time observation (GTO) programme (Hot Water Worlds, CH\_PR140068), which aims to identify new planets located in these sparsely populated regions of the mass-radius diagram. We refer to these regions as the Hot Water World triangles. In the first paper of this series, \cite{Egger+2025}, we introduced the concept behind our programme in more detail and presented the observational results for its first two targets, TOI-238~b and TOI-1685~b. In this second paper, we now present the first detailed characterisation of the recently validated transiting low-density super-Earth TOI-2211\,b using data from TESS, CHEOPS, HARPS, ESPRESSO, and KPF, along with an additional radial velocity planetary candidate orbiting TOI-2211.

We present a summary of the observations used in this paper in Section~\ref{sec:observations}, perform a characterisation of the host star TOI-2211 in Section~\ref{sec:star}, and model the planetary signals in the photometric and RV data in Section~\ref{sec:data+analysis}. In Section~\ref{sec:discussion}, we then use the derived planetary properties to model the internal structure and potential atmospheric evolution of TOI-2211\,b. In the same section, we also investigate additional constraints on the new planet candidate identified in the radial velocity data, studying transit constraints from TESS and modelling the dynamical stability of a two-planet system. We also attempt to infer the broader architecture of the planetary system using a generative model based on the Bern model of planet formation and evolution. Finally, we provide a summary of our work and draw conclusions in Section~\ref{sec:conclusion}.

%%%%%%%%%%%%%%%%%%%%%%%%%%%%%%%%%%%%%%%%%%%%%%%%%%%%%%%%%%%%%%
\section{Observations}
\label{sec:observations}

\subsection{TESS}
The Transiting Exoplanet Survey Satellite \citep[TESS;][]{Ricker+2015} first observed TOI-2211 in July~2020 as part of sector~27. Later on, TOI-2211 was reobserved by TESS in sectors~67, 92, and 95. A summary of these TESS observations is given in Table~\ref{tab:tess_dat}, including the cadence of the observations, the camera and CCD TOI-2211 was observed in, and how many transits of TOI-2211\,b are covered by each sector. In the following, we used the Pre-search Data Conditioning Simple Aperture Photometry (PDCSAP) photometry of all four available TESS sectors \citep{Smith2012,Stumpe2012,Stumpe2014}, obtained through the Mikulski Archive for Space Telescopes (MAST). %We checked for contamination by nearby stars, but do not find 

\subsection{CHEOPS}
In addition to the available TESS photometry, we obtained eight visits of TOI-2211 with the CHaracterising ExOplanet Satellite \citep[CHEOPS;][]{Benz+2021,Fortier+2024} as part of the Guaranteed Time Observations (GTO) programme `Hot Water Worlds' (CH\_PR140068), spanning a total of 68.5~hours between 2024-06-17 and 2025-08-24. The goal of these additional observations was to refine the planetary radius of TOI-2211\,b, a strategy which has previously been shown to be successful due to the higher photometric precision of CHEOPS compared to TESS \citep[e.g.][]{Luque+2023, Bonfanti+2024, Rosario+2024, Krenn+2024, Egger+2024, Egger+2025, Zingales+2025, Leonardi+2025}. A complete observation log of these CHEOPS observations is provided in Table~\ref{tab:cheops_dat}. We used PIPE\footnote{\url{https://github.com/alphapsa/PIPE}} \citep{Brandeker+2024} to extract Point Spread Function (PSF) photometry of these CHEOPS light curves. This helps to reduce contamination by hot pixels, background stars, or cosmic ray hits.

\subsection{HARPS}\label{obs:harps}
TOI-2211 was additionally observed with the High Accuracy Radial Velocity Planet Searcher \citep[HARPS;][]{Pepe+2000}, which is mounted at the 3.6 m telescope at the ESO La Silla Observatory in Chile. HARPS is a fibre-fed, high-resolution spectrograph with a resolving power of $R=115\,000$. The observations were carried out between 2024-07-07 and 2024-09-27 as part of the ESO program with ID 113.26G5 (PI: Egger). The exposure time was 1200 s for all the observations. We gathered 19 spectra with an average signal-to-noise at 550 nm of 41, and airmass at the beginning of the observations ranging between 1.0 and 1.2. The spectra were reduced with the standard HARPS data reduction software v3.3.6 \citep[DRS;][]{Pepe+2002,Lovis+Pepe2007} and the RVs were computed applying the cross-correlation technique using a G9 mask. The HARPS RV time series, shown in the bottom left panel of Figure~\ref{fig:joint_fit}, has a mean RV precision of 2.4 m/s and a RV scatter of 3.4 m/s.

\subsection{ESPRESSO}
The Echelle SPectrograph for Rocky Exoplanets and Stable Spectroscopic Observations \citep[ESPRESSO;][]{Pepe+2014} is a ultra-stable, high-resolution, fibre-fed spectrograph mounted on the 8.2 m Very Large Telescope (VLT) at Paranal Observatory in Chile. We gathered 30 additional spectra of TOI-2211 with ESPRESSO between 2025-05-05 and 2025-07-26, as part of the ESO program with ID 115.2842 (PI: Egger). All the observations had an exposure time of 600 s. The average signal-to-noise at 550 nm is 78 and the airmass at the beginning of the observations ranged between 1.0 and 1.5. The spectra were reduced with the standard ESPRESSO pipeline v3.3.10 \citep{Pepe+2021} and the RVs were computed with the cross-correlation technique using a G9 mask. The ESPRESSO RV time series, shown in the bottom left panel of Figure~\ref{fig:joint_fit}, has a mean RV precision of 0.6 m/s and a RV scatter of 2.5 m/s.

\subsection{KPF}
The Keck Planet Finder \citep[KPF;][]{Gibson+2016} is a high-resolution, fibre-fed spectrograph mounted on the 10 m Keck I telescope at W.M. Keck Observatory on Maunakea, Hawai'i. We gathered 27 spectra of TOI-2211 with KPF between 2024-05-24 and 2025-12-03, as part of the Keck Strategic Mission Support (KSMS) program of \cite{Crossfield+2025}. All observations had an exposure time of 346 s. The average signal-to-noise at 548 nm is 155 and the airmass at the beginning of the observations ranged between 1.4 and 2.1. The spectra were reduced with the standard KPF Data Reduction Pipeline (DRP Version 2.9.1) and the RVs were extracted via the cross-correlation technique. We note that a thermal offset occurred around BJD 2460790, and we account for this as an RV offset in our analysis. After quality control cuts, we were left with 19 spectra. The KPF RV time series, shown in the bottom left panel of Figure~\ref{fig:joint_fit}, has a mean RV precision of 0.97 m/s and a RV scatter of 2.8 m/s.

%%%%%%%%%%%%%%%%%%%%%%%%%%%%%%%%%%%%%%%%%%%%%%%%%%%%%%%%%%%%%%
\section{Host star characterisation}
\label{sec:star}
%spectroscopic parameters
The stellar spectroscopic parameters ($T_{\mathrm{eff}}$, $\log g$, [Fe/H]) were estimated using the \textsc{ARES+MOOG} methodology, as described in more detail in \cite{Santos-13}, \cite{Sousa-14}, and \cite{Sousa-21}. For this purpose, we used the latest version of \textsc{ARES}\footnote{The latest version, \textsc{ARES v2}, can be downloaded at \url{https://github.com/sousasag/ARES}.}
 \citep{Sousa-07, Sousa-15} to measure the equivalent widths (EW) of iron lines in a combined HARPS spectrum of TOI-2211. We adopted the list of FeI and FeII lines presented in \citet[][]{Tsantaki-2013}, which is more suitable for stars with temperatures about 5200 K and below.

A minimization procedure was applied to impose ionization and excitation equilibrium, allowing us to converge to the best set of spectroscopic parameters. For this step, we used a grid of Kurucz model atmospheres \citep{Kurucz-93} together with the radiative transfer code \textsc{MOOG} \citep{Sneden-73}.

In this analysis, we also derived a more accurate trigonometric surface gravity using \textit{Gaia} DR3 data, following the procedure described in \citet[][]{Sousa-21}. Stellar abundances were determined using the classical curve of growth method under the assumption of local thermodynamic equilibrium, employing the same codes and model atmospheres as for the stellar parameter determination. For the derivation of chemical abundances of Mg and Si, we closely followed the methods described in \citet{Adibekyan-12, Adibekyan-15}.

%radius
We used a Markov chain Monte Carlo (MCMC) modified infrared flux method \citep[IRFM --][]{blackwell_1977,schanche_2020} to determine the stellar radius of TOI-2211. This was done by extracting synthetic photometry from a spectral energy distribution we constructed using stellar atmosphere models \citep{castelli_2003} and our spectroscopically derived stellar parameters, that we compared to observed broadband fluxes in the following bandpasses:  2MASS $J$, $H$, and $K$, WISE $W1$ and $W2$, and \textit{Gaia} $G$, $G_\mathrm{BP}$, and $G_\mathrm{RP}$ \citep{skrutskie_2006,wright_2010,gaiacollaboration_2022}. From this approach we retrieved the stellar bolometric flux that we converted into the effective temperature and stellar radius by using the offset-corrected \textit{Gaia} parallax \citep{lindegren_2021}.

%mass and age
The effective temperature $T_{\mathrm{eff}}$, metallicity [Fe/H], and radius $R_{\star}$ along with their uncertainties constituted the input set to compute the isochronal mass $M_{\star}$ and age $t_{\star}$ using two different stellar evolutionary models. In detail, we applied the isochrone placement routine \citep{bonfanti2015,bonfanti2016} and its capability of interpolating the input set within pre-computed grids of PARSEC\footnote{\textsl{PA}dova and T\textsl{R}ieste \textsl{S}tellar \textsl{E}volutionary \textsl{C}ode: \url{https://stev.oapd.inaf.it/cgi-bin/cmd}} v1.2S \citep{marigo2017} isochrones and tracks to infer a first pair of estimates for $M_{\star}$ and $t_{\star}$, $0.812\pm0.040\,M_{\odot}$ and $7.7\pm4.8$\,Gyr. A second pair of mass and age values was computed using the CLES \citep[Code Liègeois d'Évolution Stellaire;][]{scuflaire2008} code, that builds the best-fit evolutionary track accounting for the input observables following the Levenberg-Marquadt optimisation scheme \citep{salmon2021}, $0.797\pm0.055\,M_{\odot}$ and $13.7\pm8.0$\,Gyr.
As described in \citet{bonfanti2021}, we then checked the mutual consistency of the two respective pairs of outcomes via a $\chi^2$-based criterion and finally merged the distributions obtaining $M_{\star}=0.806_{-0.049}^{+0.047}\,M_{\odot}$ and $t_{\star}=9.9_{-6.2}^{+3.9}$\,Gyr.

\begin{figure}
    \centering
    \includegraphics[width=0.9\linewidth]{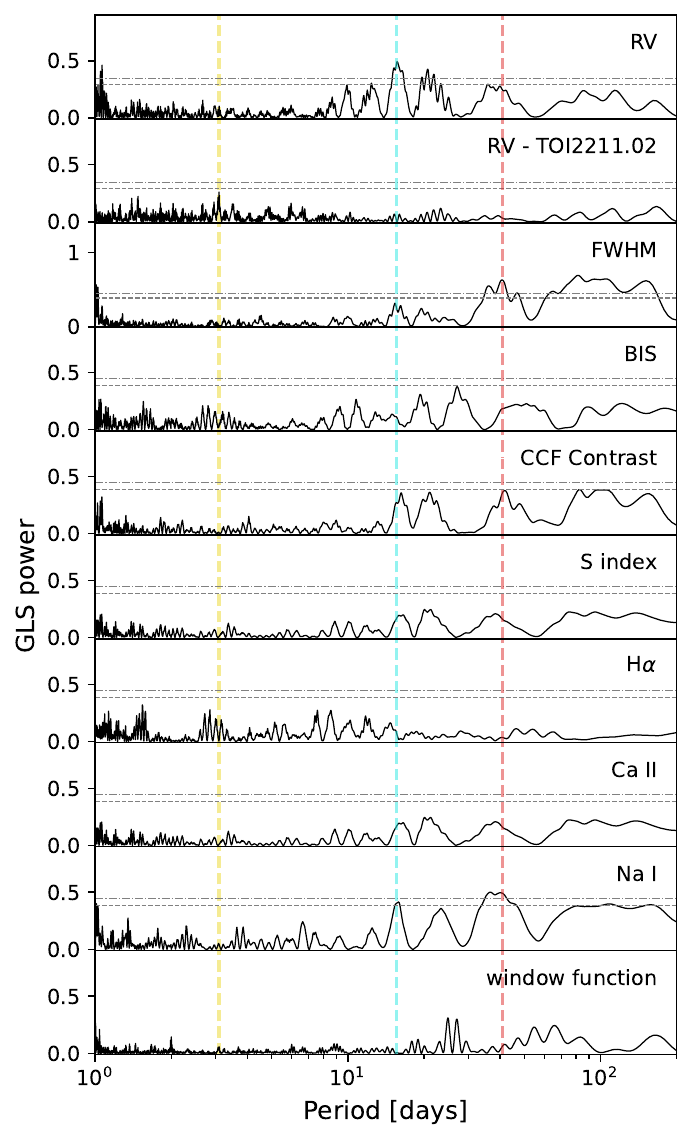}
    \caption{Generalised Lomb-Scargle periodograms of the whole RV dataset and the activity indicators from HARPS and ESPRESSO data. The yellow and cyan dashed vertical lines depict the transiting planet signal at 3.09 days and the non-transiting planet candidate at 15.6 days, respectively. The red dashed vertical line indicates the potential stellar rotation period at 41 days. The dashed and dash-dotted horizontal lines mark the 0.1 and 1$\%$ FAP, respectively. }
    \label{fig:periodograms}
\end{figure}

%stellar rotation period
Given the old nature of the host star, gyrochronology predicts a stellar rotation period of $36^{+7}_{-14}$ days \citep{Barnes+2007}. In support of this result, the $\log R'_{\rm HK}$ value measured from the ESPRESSO spectra is $-5.09$, suggesting low activity levels and a stellar rotation period of $\sim$66 days by following \citet{Noyes+1984}. We analysed the periodicities of the activity indicators from the HARPS and ESPRESSO spectra to determine the stellar rotation period using the Generalised Lomb-Scargle (GLS) periodogram \citep{GLS}. The periodograms of the available activity indicators are shown in Figure \ref{fig:periodograms}. We do not find significant periodicities with false alarm probability (FAP; \citealt{Baluev2008}) below 10$\%$ in the BIS, CCF Contrast, S index, H$\alpha$ or Ca II. The FWHM shows two periodicities with low FAP of $\log_{10}(\rm{FAP})=-9.02$ and $-7.28$ at 82 and 41 days, respectively. We note that the significant periodicity at 82 days is a broad, not well constrained peak. The periodogram of the Na I line also shows a significant periodicity around 41 days.

The low frequency periodicities observed in many of the activity indicators' periodograms, although of low significance, are likely to be an artefact driven by the large gap of $\sim$250 days between the HARPS and ESPRESSO observing seasons. Our current dataset does not allow us to resolve the stellar rotation period between 41 or 82 days because of their harmonic relationship. Moreover, there is a sparse coverage of the 82 days signal, making it challenging to robustly confirm it as the stellar rotation period. We therefore adopt 41 days as a plausible stellar rotation period, which is supported by gyrochronology, for the purposes of interpreting the activity diagnostics. However, we note that the current data do not uniquely determine the rotation period; values near $66-82$ days remain possible and additional RVs and activity monitoring would be required to distinguish between these alternatives.

\begin{table}
\renewcommand{\arraystretch}{1.2}
\caption{Stellar properties of TOI-2211.}
\centering
\begin{tabular}{lrr}
\hline\hline
\multicolumn{3}{l}{Star names} \\
\hline
\multicolumn{1}{l}{HIP} & \multicolumn{2}{l}{101503} \\
\multicolumn{1}{l}{HD} & \multicolumn{2}{l}{195755} \\
\multicolumn{1}{l}{TOI} & \multicolumn{2}{l}{2211} \\
\multicolumn{1}{l}{TIC} & \multicolumn{2}{l}{212253390} \\
\multicolumn{1}{l}{TYC} & \multicolumn{2}{l}{6920-00789-1} \\
\multicolumn{1}{l}{\textit{Gaia} DR3} & \multicolumn{2}{l}{6796762076358897408} \\
\multicolumn{1}{l}{2MASS} & \multicolumn{2}{l}{09252337-1555185} \\
\hline
Parameter & Value & Source \\
\hline
$\alpha$ (J2000) & 20h 34m 09.28s & [1] \\
$\delta$ (J2000) & $-$29$^\circ$ 40' 19.60'' & [1] \\
RV~[$\mathrm{km\,s^{-1}}$] & $25.2 \pm 0.1$ & [1] \\
Parallax~[$\mathrm{mas}$] & $10.20 \pm 0.02$ & [1a] \\
Distance~[$\mathrm{pc}$] & $98.1 \pm 0.2$ & [1a] \\
log g [cgs] & 4.44 $\pm$ 0.03 & [1] \\
\hline
G~[$\mathrm{mag}$] & $9.078 \pm 0.003$ & [1] \\
%$G\_\mathbb{BP}$~[$\mathrm{mag}$] & $9.373 \pm 0.003$ & [1] \\
%$G\_\mathbb{RP}$~[$\mathrm{mag}$] & $8.611 \pm 0.004$ & [1] \\
J~[$\mathrm{mag}$] & $8.11 \pm 0.02$ & [2] \\
H~[$\mathrm{mag}$] & $7.86 \pm 0.03$ & [2] \\
K~[$\mathrm{mag}$] & $7.80 \pm 0.02$ & [2] \\
V~[$\mathrm{mag}$] & $9.33 \pm 0.02$ & [3] \\
B~[$\mathrm{mag}$] & $9.94 \pm 0.03$ & [3] \\
TESS~[$\mathrm{mag}$] & $8.666 \pm 0.006$ & [4]\\
\hline
T$_\textrm{eff}$ [K] & 5216 $\pm$ 66 & [5] \\
$[\mathrm{Fe/H}]$ [dex] & -0.13 $\pm$ 0.04 & [5] \\
$[\mathrm{Mg/H}]$ [dex] &  0.00 $\pm$ 0.04 & [5] \\
$[\mathrm{Si/H}]$ [dex] & -0.10 $\pm$ 0.05 & [5] \\
%v$_\mathrm{mic}$ & 0.74 $\pm$ 0.06 & km~s$^{-1}$ & \\
R$_\star$ [R$_\odot$] & 0.839 $\pm$ 0.007 & [5] \\
M$_\star$ [M$_\odot$] & $0.806_{-0.049}^{+0.047}$ & [5] \\
t$_\star$ [Gyr] & $9.9_{-6.2}^{+3.9}$ & [5] \\
\hline
\end{tabular}
\label{tab:stellar_params}
%\tablefoot{XXX.}
\tablebib{
[1]~\cite{GaiaCollab2021}; 
[1a] parallax corrected according to \citet{lindegren_2021};
[2] 2MASS \citep{skrutskie_2006};
[3] Tycho-2 \citep{Hoeg2000};
[4] TESS \citep{Stassun2019}; 
[5] This work.
}
\end{table}
\renewcommand{\arraystretch}{1}

%%%%%%%%%%%%%%%%%%%%%%%%%%%%%%%%%%%%%%%%%%%%%%%%%%%%%%%%%%%%%%
\section{Modelling the planetary signals}
\label{sec:data+analysis}

TOI-2211.01, a planetary candidate with an orbital period of 3.09~days, was first identified as a transiting planet candidate in December~2020 in the TESS data. A statistical validation of TOI-2211.01 was then first performed by \cite{Crossfield+2025} using \texttt{TRICERATOPS} \citep{Giacalone+2021}, classifying it as a likely planet. Later on, \cite{Lafarga+2026} validated TOI-2211.01 as part of their automatic search for transiting planets in TESS-SPOC full frame images with the \texttt{RAVEN} pipeline \citep{Hadjigeorghiou+2025}. They also identified an additional transiting planet candidate at an orbital period of 6.88~days, which, however, is not validated by their pipeline.

\subsection{Radial velocities}

The RV follow-up of TOI-2211 started with HARPS, from which we obtained 19 spectra in a time span of almost three months. The GLS periodogram of the HARPS RVs shows no significant periodicities, and the signal of the transiting planet at 3.09~days is detected below the $~2\sigma$ level. This is probably due to the relatively low RV precision (see Section \ref{obs:harps}). Therefore, we extended the RV follow-up with ESPRESSO in order to improve the RV precision. We obtained ESPRESSO RVs with a precision four times smaller than those of HARPS.

We computed the GLS periodogram of the whole RV dataset to identify the periodicities present. The mean RV of each instrument time series was removed to deal with the instrumental offsets. The period with the lowest FAP is located at 15.6~days, as shown in the top panel of Figure \ref{fig:periodograms}. After removal of this signal by fitting a sinusoid with a period of 15.6~days, the signal of the transiting planet at 3.09~days gains significance with FAP<$10\%$. No other signals show high power in the periodogram, as seen in the second panel of Figure \ref{fig:periodograms}. In particular, we do not find significant signals at the period of 6.88 days. The signal at 15.6~days does not show low FAP in any of the activity indicators periodograms, and most likely does not correspond to the stellar rotation period (see Section \ref{sec:star}) or its harmonics, suggesting a planetary nature. Moreover, we found a low Pearson's correlation coefficient between the RVs and the activity indicators with values between $-0.07$ and $0.16$. The RV periodogram also shows a group of high-significance signals with periods between $\sim$$20-30$ days. These periodicities are not present in the HARPS or ESPRESSO datasets, and only appear when including the KPF data. As seen in the second panel of Figure \ref{fig:periodograms}, these signals disappear when fitting a sinusoid with a period of 15.6~days.

\begin{figure}
     \centering
    \includegraphics[width=0.99\linewidth]{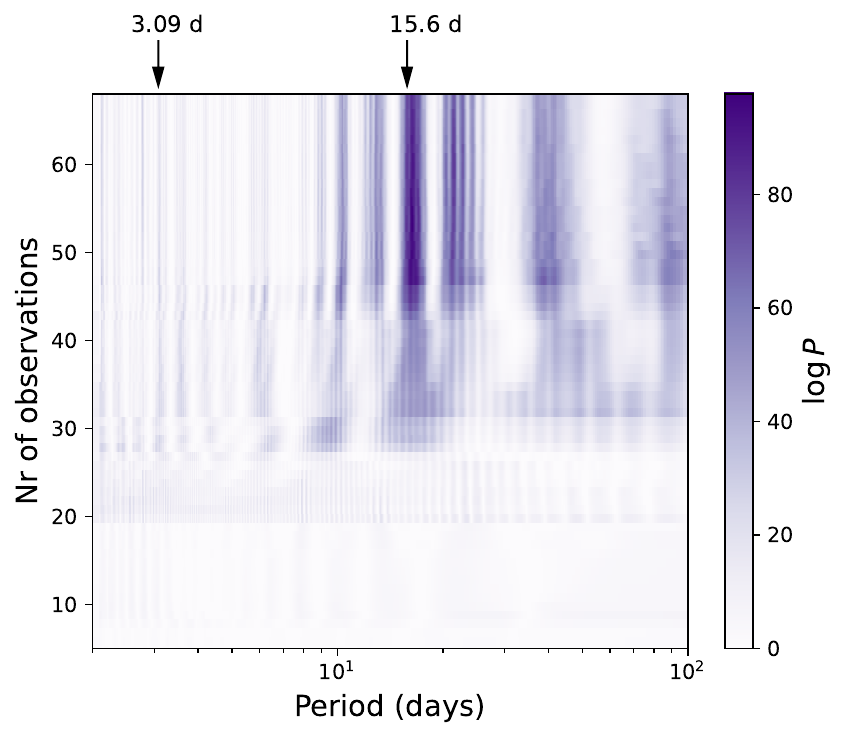}
    \caption{Stacked BGLS periodogram applied on the complete RV dataset. The colour bar indicates the log-likelihood (log$P$) level. The black arrows mark the transiting planet period at 3.08 days and non-transiting candidate at 15.6 days.}
    \label{fig:sbgls}
    \end{figure}

\begin{figure}
     \centering
    \includegraphics[width=0.9\linewidth]{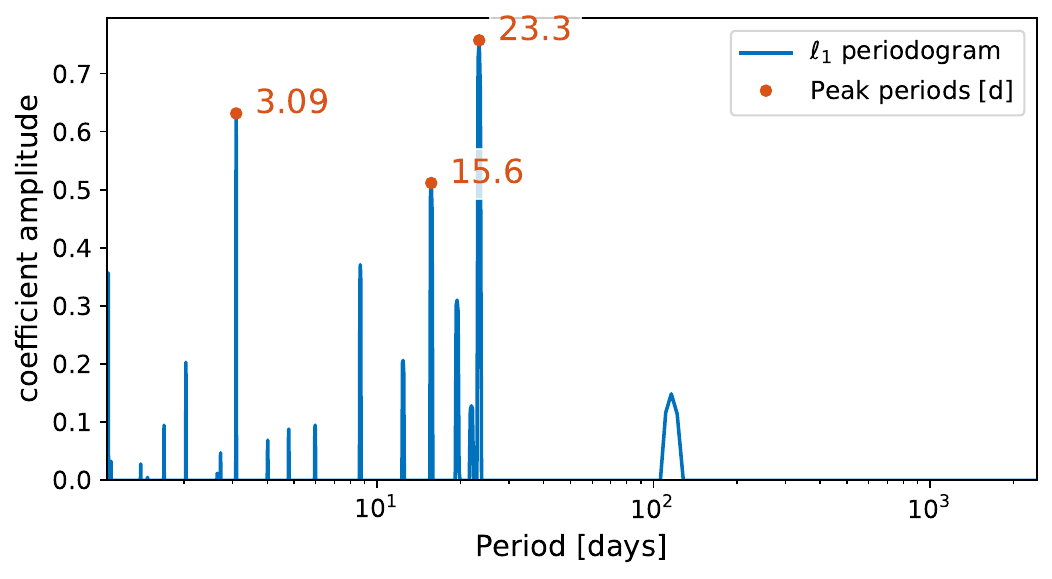}
    \caption{$\ell_1$-periodogram of the complete RV dataset. The three most significant periods are highlighted with a red circle. }
    \label{fig:l1}
    \end{figure}

To further asses the signals detected by the GLS periodogram, we applied the stacked Bayesian General Lomb-Scargle (BGLS) periodogram \citep{Mortier2017} to study their coherence and significance, in particular for the signal at 15.6~days since the transiting planet can be assumed as known. Figure \ref{fig:sbgls} shows the stacked BGLS periodogram, where the signal with the largest log-likelihood is located at $\sim$15.6 days. This signal shows a consistently increasing log-likelihood as we add more observations, as expected for planetary signals. The transiting planet signal at 3.09~days can be also seen in the BGLS periodogram although with less significance, consistent with the GLS periodogram. Additionally, we computed the $\ell_1$-periodogram \citep{Hara2017}, which is shown in Figure \ref{fig:l1}. We adopted a noise model composed of 1 m/s white noise. The three most significant peaks in the $\ell_1$-periodogram are located at the period of the transiting planet (3.08 d), and at 15.6 and 23.3 days, similarly to what it is seen in the GLS periodogram but with less aliasing. However, unlike the GLS periodogram, the 23.3 days signal in the $\ell_1$-periodogram is the most significant.

We used the \texttt{kima} framework \citep{kima} to asses the optimal number of Keplerian signals ($N_{p}$) present in the RV dataset. Each Keplerian model includes as free parameters the semi-amplitude $K$, the orbital period $P$, the eccentricity $e$, the argument of periastron $\omega$, and the mean anomaly, $\phi$. The transiting planet at 3.09~days is assumed as known by providing Gaussian priors in the orbital period and transit time. Therefore, \texttt{kima} explores the presence of additional planets in the system. The eccentricity was initially set free and the rest of the parameters have wide uniform priors (see Table \ref{tab:RVprior}). We also included a linear trend parameter in the RV model to account for possible long-term variations. 

We run models with free and fixed eccentricity and a maximum of three planets (i.e., $N_{p}=[1,2,3]$). The $\log(Z)$ of these models are listed in Table \ref{tab:RVcomparison}. We find that the $N_{p}=2$ is statistically preferred over $N_{p}=1$ with a Bayes factor of 73.7 and 81.5 for the cases of free and fixed eccentricity, respectively \citep[see e.g.][]{kass_raftery_95}. Despite the higher $\log(Z)$ for the three-planet model, the Bayes factor between $N_{p}=2$ and $N_{p}=3$ is not large enough to support this model, with only 3.7 and 2.5 for the cases of free and fixed eccentricity, respectively. We also note that there is not enough evidence to support eccentric orbits. Since the Bayes factor within \texttt{kima} to claim significant detection is set at 150, we treat the second signal as a planet candidate (TOI-2211.02). Extending the RV follow-up of TOI-2211 is required to prove the detection of the second planet in this system. Regarding the orbital period of TOI-2211.02, the maximum likelihood sample of the \texttt{kima} model with fixed $N_{p}=2$ indicates a value of 15.6~days, confirming that the group of signals between $\sim$$20-30$ days are rather an artefact possibly driven by KPF systematics.  

\begin{table}
\centering
\footnotesize
 \caption[]{Comparison of RV models.}\label{tab:RVcomparison}
\begin{tabular}{c| c c}
\hline \hline
& Free $e$ & Fixed $e=0$ \\
$Np$ & $\log(Z)$ & $\log(Z)$ \\
\hline
1 & -181.0 & -180.9 \\
2 & -176.7 & -176.5 \\
3 & -175.4 & -175.6\\

\hline
\hline

\end{tabular}
\end{table}

As an independent analysis, we used the \texttt{pyaneti} \citep{Barragan+2022} package to fit Keplerian signals considering one or two planets. For these models we also assumed the transiting planet as known with Gaussian priors in the orbital period and transit time. The rest of the Keplerian parameters were set with wide uniform priors, except for the eccentricity which was fixed at zero. A linear trend was also included. The two-planet model is strongly preferred over a one-planet model with a $\Delta\rm{BIC}=25$. Regarding the orbital period of the non-transiting candidate, we compared models with uniform priors in the orbital period of $\rm \mathcal{U}(15,16)$ and $\rm \mathcal{U}(20,30)$ days. We found that the two-planet model with a non-transiting candidate at 15.6 days is preferred with a $\Delta\rm{BIC}=7$.

We also tested a multidimensional Gaussian Process (GP) model (see \citealt{Rajpaul+2015}) with a quasi-periodic kernel included in \texttt{pyaneti} to account for stellar activity. We applied a uniform prior in Prot including the possible values from Section \ref{sec:star} ($\rm \mathcal{U}(30,70)$), wide uniform priors in the decay time ($\rm \mathcal{U}(30,200)$), and the harmonic complexity ($\rm \mathcal{U}(0.1,2.0)$). We used the FWHM as an ancillary time series since it seems to be rotationally modulated. However, the model does not yield well-constrained estimates of the stellar rotation period or the remaining GP hyperparameters. Therefore, we did not include these results and adopted the \texttt{pyaneti} RV model only with two Keplerians and a linear trend. It is possible that the low activity levels of the host star reduces the effectiveness of modelling the stellar activity signal simultaneously with the RVs. 

The RV analysis of TOI-2211 yields a two-planet system: one transiting at 3.09~days with an RV semi-amplitude of $K_{b}=1.3\pm0.3~\rm{m}\,\rm{s}^{-1}$, and an additional planet candidate at 15.6~days with a RV semi-amplitude of $K_{c}=2.5\pm0.3~\rm{m}.\rm{s}^{-1}$. The linear trend in the RV time series is found to be $0.011\pm0.007~\rm{m}\,\rm{s}^{-1}\,\rm{d}^{-1}$.

\subsection{Joint photometric and RV fit}

\begin{figure*}
    \centering
    \includegraphics[width=\linewidth]{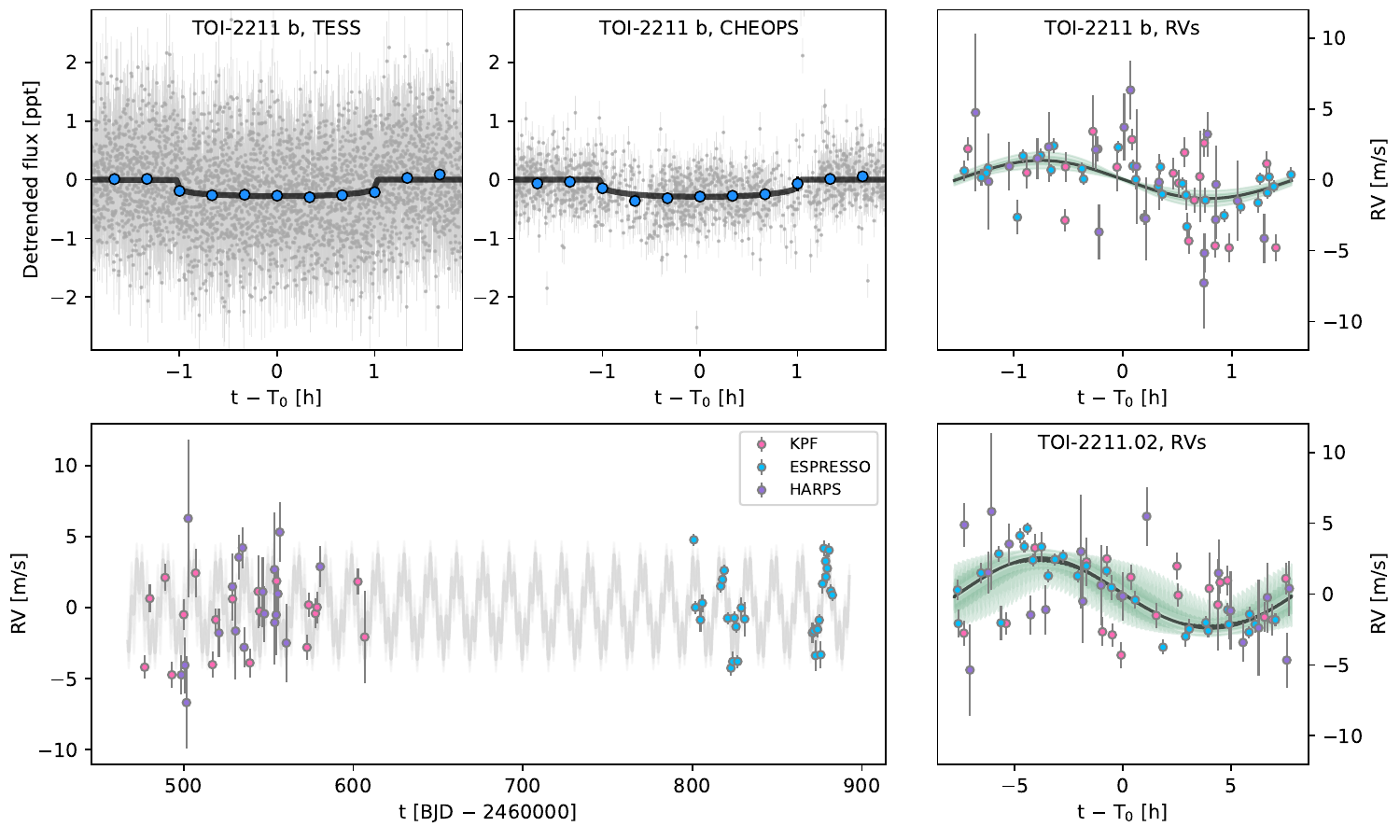}
    \caption{Collected photometric and RV data for TOI-2211, together with the best-fit models from the joint fit. Upper left and middle: Detrended TESS and CHEOPS light curves for TOI-2211\,b with the best-fit model in black, phase-folded to the orbital period and showing relative flux normalised to~0. Lower left: RV time series of HARPS in purple, ESPRESSO in blue, and KPF in pink, with the best-fit model in grey. Upper and lower right: Phase-folded RVs at the period of TOI-2211\,b (upper) and the planet candidate TOI-2211.02 (lower), with the best fit model in black. The mean RV as well as the RV signal from the respective other planet or planet candidate has been subtracted. 1- and 2-sigma error regions are shown in green.}
    \label{fig:joint_fit}
\end{figure*}

We use the open-source code \texttt{chexoplanet}\footnote{\url{https://github.com/hposborn/chexoplanet}} \citep[e.g.][]{Egger+2024} to jointly model the available TESS and CHEOPS data, as well as the RVs from multiple instruments. To reduce the impact of systematics, \texttt{chexoplanet} samples multiple systematic and transit models simultaneously. The first part is the linear and quadratic decorrelation of CHEOPS flux with respect to various hyperparameters, both on a per-visit and global basis. Initially each individual CHEOPS light curve is fitted including decorrelation with normalised parameters of the x and y centroid position, three aliases of the cosine and sine of the roll-angle, major residuals of the PSF fit, and time. Additionally on-board temperature and background flux are included, but split into high- and low-frequency signals using a spline with $\sim$10~min knot spacing to separate trends from jitter and noise. Bayesian model comparison then allows us to determine which of these hyperparameters improve the model (Bayes Factor >1). As a second step, we then determine the similarity of these decorrelation parameters across all observations using a leave-one-out comparison. Significant linear and quadratic parameters are then included in the global model. 

Furthermore, \texttt{chexoplanet} co-models shorter-frequency variations in the flux as a function of the spacecraft roll angle using a cubic spline (6.5~degree knot spacing). This removes modulation due to a combination of the rotating field of view of CHEOPS due to its nadir-locked orbit around the Earth and the strongly asymmetric PSF of each background star \citep{Benz+2021} as described for example in \citet{Lendl+2020} and \cite{bonfanti2021}. Due to slight differences in the year-to-year roll angle response, we fit two spline models for the 2024 and 2025 data respectively.

For the TESS data, we use the PDC\_SAP flux from the 120s observations, and remove long-timescale trends using a transit-masked bspline with knot distance of 0.9\,d. \texttt{exoplanet} \citep{Foreman-Mackey+2021} is used for the transit models, with normal priors centred on the theoretical limb-darkening coefficients derived by \cite{Claret2017} for TESS and \cite{Claret2021} for CHEOPS ($\sigma_u=0.1$). We used a broad log-normal prior for the planetary radius ratio, and broad log-normal priors for jitter terms accounting for additional white noise for each of the two instruments separately. Normal priors were used for timing and orbital periods using values and errors from the TOI catalogue. For the impact parameter, we used a uniform prior ranging from 0 to $1+R_p/R_S$.

For the RVs, we used normal priors for log jitter terms and zero-point offsets for each instrument informed by the median and standard deviation of each dataset. Due to the nature of Monte Carlo sampling, we cannot explore multiple local minima by setting broad uninformed priors on the ephemerides of outer planets, so we instead include a prior of $P=15.62\pm0.14$\,d and $T_{0}=2460821.12\pm0.64$ on the candidate identified by \texttt{kima} modelling ($\sigma$ here is set as twice the RV-only model uncertainties).
Eccentricity was parameterised using the Rayleigh distribution of small close-in planets \citet{VanEylen+2015}.

From this joint fit, we derive a planetary radius of $1.467 \pm 0.047$~R$_\oplus$ and a mass of $2.69 \pm 0.56$~M$_\oplus$ for the transiting planet TOI-2211\,b, as well as minimal mass of $8.53 \pm 1.04$~M$_\oplus$ for the outer planet candidate. The resulting best fit models are visualised in Figure~\ref{fig:joint_fit}, while the posterior distributions of the fitted and derived planetary parameters are summarised in Table~\ref{tab:planetary_params}. A more complete list of the priors and posteriors of all fit parameters can be found in Table~\ref{tab:joint_fit_priors_posteriors}.

In order to test reliability, we performed two analyses with two different priors on the semi-amplitude, one with a broad normal distribution $K=0\pm5\,\rm{m}.\rm{s}^{-1}$ and another with a broad log normal of $\log{K}=-1\pm5$. We use the former model results here, but both converged to a mass for planet b within $1\sigma$ ($2.69 \pm 0.56$ and $2.55 \pm 0.58$, respectively). To test the impact of the CHEOPS photometry on the result, we also ran an additional TESS-only model and found that including the CHEOPS data results in a $\sim$25\% improvement in radius precision.

To inspect the 6.88~d candidate proposed by \cite{Lafarga+2026}, we performed a 2-planet model using both the TESS and CHEOPS photometry and the priors from the TOI and \cite{Lafarga+2026}, respectively. This resulted in an as-expected transit for TOI-2211\,b (281~ppm in TESS), and no significant detection of any signal at the ephemeris of the \cite{Lafarga+2026} candidate ($\delta_c<2\sigma$), with an implied 3-sigma upper limit on the transit depth in TESS of 82~ppm. Re-assessing the light curves, we noted that the $T_0$ of the 6.88~d candidate is <1 minute from the time of a high-SNR transit of the 3.09~d planet. It is therefore very likely that the derived 6.88~d period is an alias of the true period, potentially in combination with instrumental noise at half the orbital period of TESS ($\sim$13.5~d). This likely also explains why \cite{Lafarga+2026} were unable to statistically confirm their derived signal.

On the RV side, the scatter observed in the RV residuals could suggest some remanent signals not being considered in the joint model. The scatter in the RV residuals is 2.3 m/s for the complete RV data set, with per-instrument values of 1.2 m/s (ESPRESSO), 2.6 m/s (KPF), and 3.2 m/s (HARPS). These differences between instruments could suggest an instrumental systematics origin. As an additional test, we computed the GLS periodogram of the RV residuals from the best-fit joint model to check if there is any periodicity left. We detect no periodicities with a FAP below 10\%. However, given the limited RV observing window, we cannot rule out the presence of additional planets in the system that could contribute to additional RV scatter. On the other hand, stellar activity could still contribute some level of RV scatter since we did not include the rotationally modulated stellar activity signal in the joint model.  Moreover, stochastic processes such as granulation could also contribute with sub-meter RV variability in K-type stars \citep{Dalal2023}. Extending the RV follow-up will potentially allow to properly determine the stellar rotation period and identify additional planets in this system.

\begin{table*}
\renewcommand{\arraystretch}{1.3}
\caption{Posterior distributions of the fitted and derived planetary parameters from the joint photometry and RV analysis for the transiting planet TOI-2211\,b and the outer planetary candidate TOI-2211.02 detected in the RVs.}
\centering
\begin{tabular}{lcc}
\hline\hline
Parameter & TOI-2211\,b & TOI-2211.02\\
\hline
Epoch, $t_0$ [BJD-2400000]  & $59038.1045 \pm 0.0020$ & $60820.91 \pm 0.33$ \\
Period, $P$ [d] & $3.0926838 \pm 0.0000040$ & $15.611^{+0.050}_{-0.056}$ \\
Eccentricity, $e$ & $0.056^{+0.035}_{-0.028}$ &  $0.058^{+0.035}_{-0.028}$ \\
Radius ratio, $R_p/R_s$ & $0.01592 \pm 0.00049$ & - \\
Impact parameter, $b$ & $0.52 \pm 0.09$  & - \\
RV semi-amplitude, $K$ [m\,s$^{-1}$] & $1.36 \pm 0.27$ & $2.52 \pm 0.28$ \\
\hline
Semi-major axis, $a$ [AU] & $0.03875 \pm 0.00078$ &  0.1138 $\pm$ 0.0023\\
Semi-major axis over stellar radius, $a/R_s$ & $9.94 \pm 0.22$ & 29.16 $\pm$ 0.63 \\
Incident flux, $S_p$ [S$_\oplus$] & 313 $\pm$ 21 & 36 $\pm$ 2 \\
Equilibrium temperature, $T_{eq}$ [K] & 1107 $\pm$ 18 & 683 $\pm$ 11 \\
Transit duration, $t_{14}$ [h] & 2.065 $\pm$ 0.041 & - \\
Radius, $R_p$ [R$_\oplus$] & 1.457 $\pm$ 0.047 & - \\
Mass, $M_p$ [M$_\oplus$] & 2.69 $\pm$ 0.56 & $8.52 \pm 1.04^\dagger$ \\
\hline
\end{tabular}
\label{tab:planetary_params}
\tablefoot{$\dagger$ denotes the minimal mass, $M_p\sin(i)$.}
\end{table*}
\renewcommand{\arraystretch}{1}

%%%%%%%%%%%%%%%%%%%%%%%%%%%%%%%%%%%%%%%%%%%%%%%%%%%%%%%%%%%%%%
\section{Discussion}
\label{sec:discussion}

\subsection{Analysing the internal structure of TOI-2211 b}\label{sec:internal_structure}

\begin{figure}
    \centering
    \includegraphics[width=\linewidth]{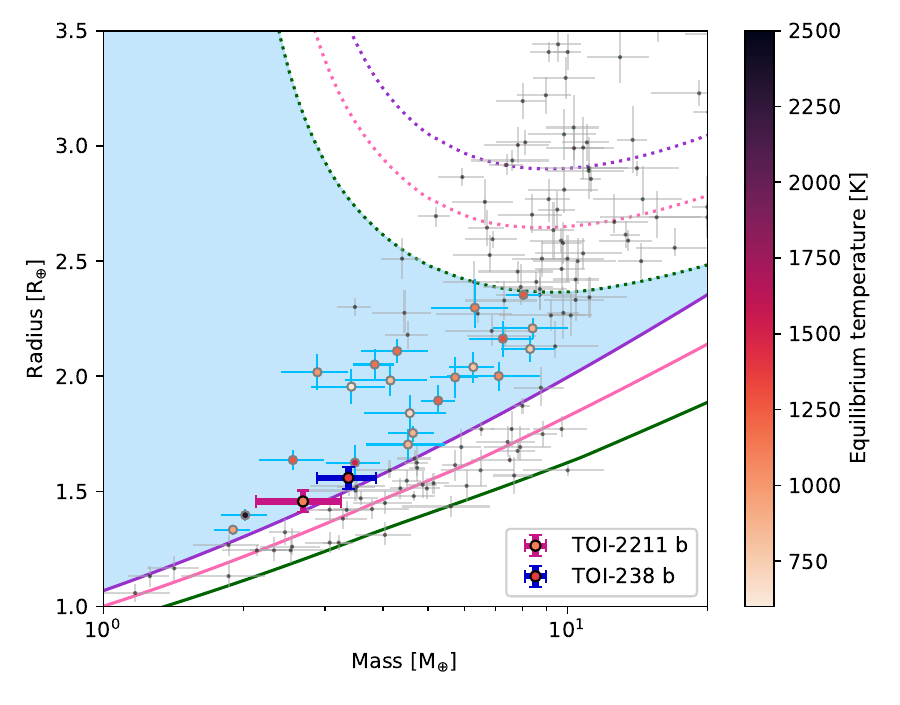}
    \caption{Mass-radius diagram showing the location of TOI-2211\,b in relation to the known exoplanet population \citep[data from the PlanetS catalogue;][]{Otegi+2020a, Parc+2024}. In addition to TOI-2211\,b, also TOI-238\,b as presented in Paper I \citep{Egger+2025} is labelled. Also shown are a number of iso-composition curves generated using the BICEPS forward model \citep{Haldemann+2024} for TOI-2211\,b's equilibrium temperature. Planets that fall into the Hot Water World triangles are depicted with blue error bars, the Hot Water World triangle corresponding to the equilibrium temperature of TOI-2211\,b is indicated in light blue. Markers of planets that fall into the triangle are colour-coded according to their equilibrium temperature.}
    \label{fig:mr_diagram}
\end{figure}

Based on the observational bulk properties of TOI-2211\,b derived in Section~\ref{sec:data+analysis}, we now analyse the implications this has for the composition and internal structure of this planet. As a first step, we study the location of TOI-2211\,b in a mass-radius diagram as depicted in Figure~\ref{fig:mr_diagram}, together with a number of mass-radius relations generated with the \texttt{BICEPS} planetary structure model \citep{Haldemann+2024}. The Hot Water World triangle as introduced in \cite{Egger+2025} that corresponds to the equilibrium temperature of TOI-2211\,b is marked in light blue.

Similarly to TOI-238\,b \citep[studied in Paper I;][]{Egger+2025}, we find that at the 1$\sigma$ level, TOI-2211\,b falls just above the iso-composition line for a bare rocky planetary core without any iron content. For a planet to lie above this line would imply that its observed mean density cannot be explained without at least a small volatile envelope. At the same time, the derived mass and radius values lie above the composition line for an Earth-like bare core by at least 2$\sigma$. We note that from a formation point of view, it is difficult to come up with a scenario where a planet would only accrete silicates and no iron at all, making a purely rocky, iron-free core composition unlikely. Simultaneously, the planet lies well below the iso-composition line for a Mercury-like core with an added H/He envelope of $\sim$1\%. As we show below, this means that only a very small low mean molecular weight envelope would be compatible with the planet's observed mass and radius values, which most likely would not have survived until the system's age given the short orbital period of TOI-2211\,b. If we compare TOI-2211\,b to other precisely characterised planets\footnote{Query of the PlanetS catalogue \citep{Otegi+2020a,Parc+2024} on 18~August 2026, requiring $T_{eq}>600$~K, $\Delta m<25\%$, $\Delta r<5\%$.} located in the Hot Water World triangles, we find that TOI-2211\,b is one of the smallest, both in terms of mass and radius, and also one of the most dense. Its equilibrium temperature lies well within the average range of $1100\pm350$~K of the planets falling into these regions.

As a next step, we will use the \texttt{plaNETic} framework\footnote{\url{https://github.com/joannegger/plaNETic}} \citep{Egger+2024} to investigate the internal structure of TOI-2211\,b in more detail. \texttt{plaNETic} is an open-source inference code based on the \texttt{BICEPS} planetary structure model \citep{Haldemann+2024} that uses a neural network as a fast surrogate model to speed up the inference of the internal structure of the observed planet. Here, we run a total of six models with different priors, allowing for a water-rich atmospheric composition on the one hand (prior~A) and fixing it to be water-poor on the other (prior~B). For each of these two options, we run three models based on different assumptions of the planetary Si/Mg/Fe ratios, assuming them to be stellar (option~1), iron-enriched compared to the host star \citep[option~2, using the fit from][]{Adibekyan+2021}, or uniformly sampling them independently from the host star values (option~3). A more detailed description of \texttt{plaNETic} and these chosen priors can be found in \cite{Egger+2024}.

\begin{figure*}
    \centering
    \includegraphics[width=\textwidth]{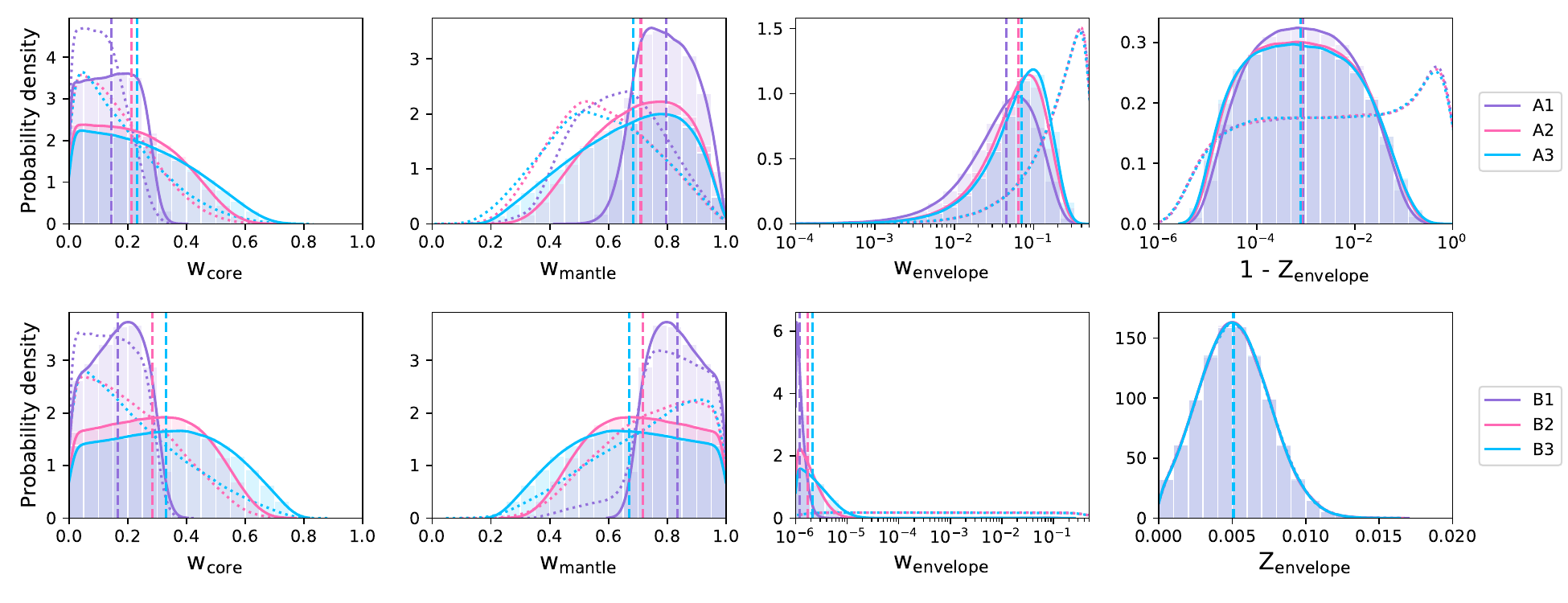}
    \caption{Results of the internal structure analysis for TOI-2211\,b using the \texttt{plaNETic} framework. Depicted are, from left to right, the posteriors of the mass fractions of the inner core, mantle, and volatile envelope layers, as well as the mass fraction of water in the volatile layer, denoted by Z$_\mathrm{envelope}$, for all six models. The top row shows results for the water-rich prior (A), the bottom row for the water-poor prior (B), with the different colours depicting the different Si/Mg/Fe prior choices (1: stellar, 2: iron-enriched, 3: uniform priors), as described in the main text. Solid lines represent posteriors, dotted lines priors, and dashed vertical lines show the median values of the distributions. }
    \label{fig:int_struct_2211}
\end{figure*}

The resulting posterior distributions of the most important internal structure parameters are visualised in Figure~\ref{fig:int_struct_2211}, while the inferred medians and one-sigma errors for the full set of parameters is provided in Table~\ref{tab:internal_structure_results_b}. We find that in both the water-rich and the water-poor case, we do not strongly constrain the core and mantle mass fractions of TOI-2211\,b, with posteriors that are very close to the chosen priors. In the case of the water-rich prior, we infer mainly water-rich envelopes with mass fractions of a few percent, with median values of the order of $\sim$5\%. For the water-poor prior, the inferred H/He-dominated envelopes are all very small, of the order of $10^{-6}$, and would not be stable against evaporation until the current age of the system, as we show in the next section.

We note that a water-dominated envelope is not the only explanation for a high mean molecular weight envelope; alternative compositions such as CO$_2$-rich or other volatile-dominated envelopes are also compatible with the observationally derived mass and radius values.

\subsection{Atmospheric evolution models for TOI-2211 b}
\label{sec:atmospheric_evolution}

\subsubsection{H/He-dominated atmospheres}
\begin{figure}
    \centering
    \includegraphics[width=1\linewidth]{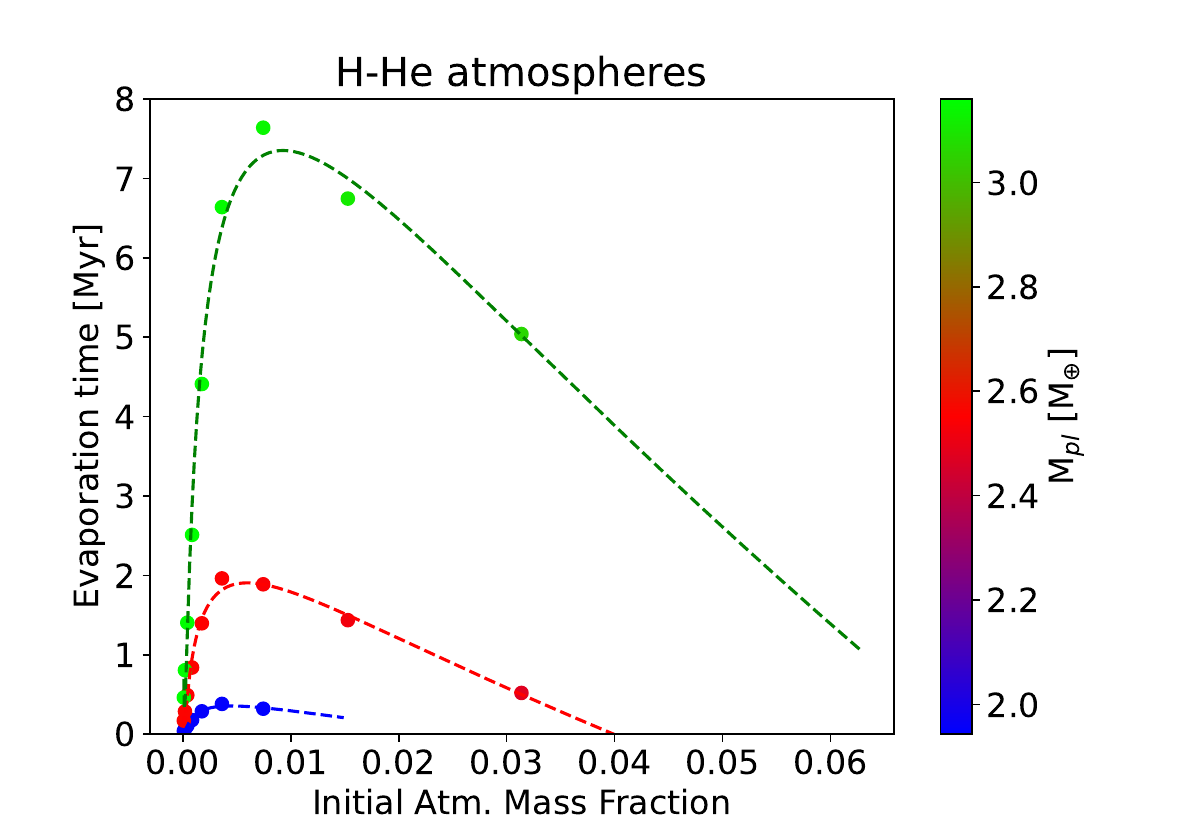}
    \caption{Evaporation time of H/He atmospheres with solar-like metallicity from TOI-2211\,b (calculated from the time of protoplanetary disk dispersal adopted at the age of 10\,Myr), assuming that the star evolved as a slow rotator with $P_{\rm rot}^{150} = 15$\,days. Dashed lines represent a cubic polynomial fit of the $\tau_{\rm esc}(f_{\rm atm}^0)$ for each of the three considered initial planetary masses.}
    \label{fig:H-He_evap_Tesc}
\end{figure}

To assess the stability of atmospheres with different compositions under the conditions of TOI-2211\,b, we followed the procedure first introduced in \citet{Egger+2025}. 
We started by estimating the evaporation timescale of H/He-dominated atmospheres with solar metallicities using the atmospheric evolution framework from \citet{kubyshkina_2021mesa,Kubyshkina+2022}, which relies on the Mors code \citep{johnstone2021,spada2013} for the stellar input. As we are interested in the maximum evaporation time, we only considered the slow rotator scenario for TOI-2211, assuming a stellar rotation period of $P_{\rm rot}^{150}=15$\,days at the age of 150\,Myr \citep[based on data from][]{Johnstone+2015}. For the initial planetary mass, we adopted three values within the 1$\sigma$ interval reported in Table\,\ref{tab:planetary_params} (1.96, 2.56, and 3.16\,$M_{\oplus}$); on top of that, we considered initial atmospheric mass fractions $f_{\rm atm}^0$ between 0.01-6.4\%. We note, however, that in our models the atmosphere becomes gravitationally unstable for mass fractions above $\sim$3.5\% for the given planetary parameters; therefore, these cases are not included in the results shown here. Figure\,\ref{fig:H-He_evap_Tesc} presents the dependence of the planetary evaporation timescale $\tau_{\rm esc}$ (starting from the time of the protoplanetary disk dispersal, adopted at 10\,Myr) on the initial atmospheric mass fraction and planetary mass. We find that the evaporation timescale for H/He-dominated atmospheres on TOI-2211\,b does not exceed $\sim$7.65\,Myr and maximizes at $f_{\rm atm}^0$ values between 0.5 and 1\%. The physical reasons behind this effect and the model's sensitivity to various free parameters were discussed in detail in \cite{kubyshkina_2021mesa}. These results rule out the existence of H/He-dominated atmosphere on TOI-2211\,b at present day with high confidence.

\subsubsection{Water-rich atmospheres}
As our next step, we model the evolution of water-rich atmospheres. 
We start from the present-day parameters outlined in Section\,\ref{sec:internal_structure} and Table\,\ref{tab:planetary_params} and step back in time, re-evaluating the atmospheric escape rate, atmospheric mass fraction, and planetary radius at every time step, until we reach the age of protoplanetary disk dispersal, as detailed in \cite{Egger+2025}. 
To this end, we use mass-radius relations derived for planets with atmospheric water mass fractions of $Z_{\rm envelope}=0.99$ using the same structure model as in Section~\ref{sec:internal_structure} along with two different atmospheric escape models. The first one was described in \citet{Egger+2025} and represents the energy-limited (EL) approximation \citep{Erkaev2007A&A...472..329E} with the heating efficiency parameter $\eta$ treated as a free parameter. 
The second atmospheric escape model, used here for the first time, employs atmospheric escape rate predictions of the hydrodynamic model CHAIN \citep[Cloudy e Hydro Ancora INsieme;][]{kubyshkina2024} for atmospheres with $Z_{\rm envelope}=0.9$\footnote{Highest $Z_{\rm envelope}$ within applicability limit}. 
This specific model setup was described in detail in 
\cite{Kubyshkina+2026}. Throughout the evolution simulations, we define the atmospheric escape rate for a given pair of $R_{\rm p}$ and $M_{\rm p}$ through interpolation
between tabulated escape rate values derived with CHAIN (Kubyshkina et al. in prep.) for planets orbiting a solar mass star with an equilibrium temperature of 1100\,K, which is fairly close to the irradiation parameters of TOI-2211\,b. Tabulated escape rates were calculated for six planetary masses between 1.64 and 4.39\,$M_{\oplus}$ and 
radii between 1.14 and 5.15\,$R_{\oplus}$ ($5-7$ values for each mass). 
The considered EUV fluxes, $(0.85-292)\times10^3$\,${\rm erg/s/cm^2}$, safely cover the range predicted for the evolution of TOI-2211\,b. 
With this evolution model, we examine the range of model parameters outlined in Table~\ref{tab:water_evo}. 
We note that the second escape model based on CHAIN does not depend on the heating efficiency parameter $\eta$, as the actual heating efficiency is calculated self-consistently within the model \citep[see a detailed discussion on the heating efficiency in][]{kubyshkina2024}. 

\begin{table}
    \centering
    \caption{Model parameters considered in the evolution simulations.}
    \begin{tabular}{lcc}
    \toprule
    \toprule
         parameter & values & points\\
         \midrule
         $R_{\rm p}$ [R$_{\oplus}$] & $1.42-1.53$ & 10 \\
         $M_{\rm p}$ [M$_{\oplus}$] & $1.97-3.15$ & 10 \\
         system age [Gyr] & 3.7, 9.9, 12 & 3 \\
         core silicate fraction [\%] & 30, 74, 99 & 3 \\
         core luminosity [erg/s] &  $10^{19}$, $10^{21}$, $10^{23}$ & 3 \\
         $P_{\rm rot}^{150}$ [d] & 1, 15$^*$ & 2 \\
         \midrule
         $\eta$ (EL only) & 0.01, 0.05, 0.10, 0.15 & 4 \\
         \bottomrule
    \end{tabular}
    \footnotesize{$^*$these values roughly bracket the most likely rotation scenarios for 0.8-1.2\,M$_{\odot}$ stars \citep{Johnstone+2015}.}
    \label{tab:water_evo}
\end{table}

\begin{figure}
    \centering
    \includegraphics[width=\linewidth, trim=0 1.5cm 0 3cm, clip]{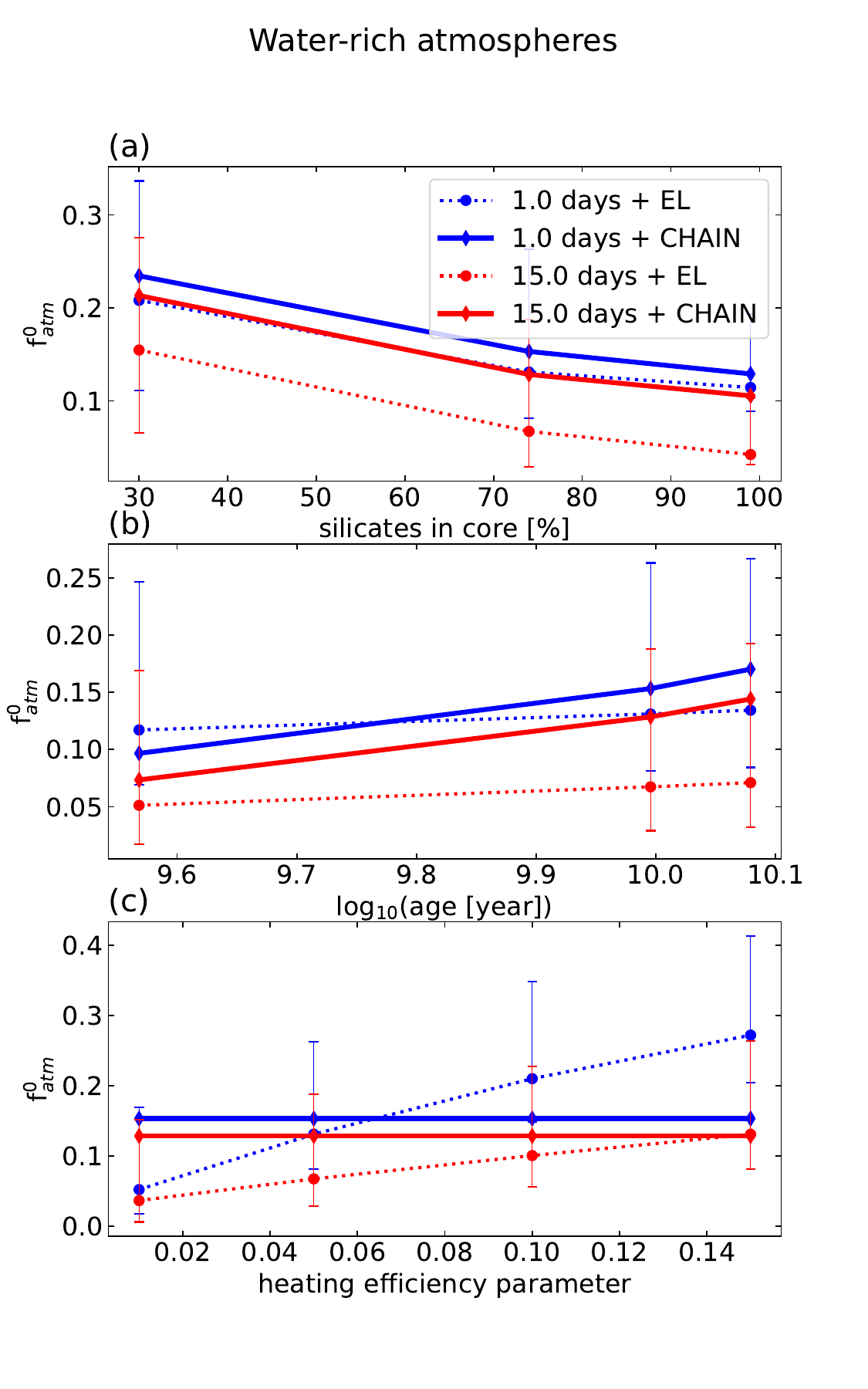}
    \caption{Primordial atmospheric mass fraction predicted by analysing the escape of water-rich atmospheres across the planetary lifetime, using the EL model from \citet{Egger+2025} (dotted lines) and interpolation between CHAIN models for water-rich atmospheres using the approach from \cite{Kubyshkina+2026} (solid lines). Blue and red lines correspond to the case of the host star evolving as a fast (1 day) and slow (15 days) rotator, respectively. Each panel summarises the dependency on a specific model parameter: (a) core composition (fraction of silicates), (b) present-day age of the system, and (c) heating efficiency parameter in the energy-limited model. }
    \label{fig:atm_evo_water}
\end{figure}

Figure~\ref{fig:atm_evo_water} summarises our results for both atmospheric escape models. In panels (a), (b), and (c), we show the primordial (at 10\,Myr) atmospheric mass fraction derived in our evolution simulation as a function of composition (silicate fraction in the core), adopted present-day age, and $\eta$, respectively. We found that the core luminosity has no considerable effect on our results, therefore it is not shown. For each panel, all parameters not listed explicitly are fixed at the ``average'' values, namely at 74\% for the  silicate fraction (panels b and c), 9.9\,Gyr for the age of the system (panels a and c), 5\% for $\eta$ (panels a and b), and $10^{21}$\,erg/s for the core luminosity (all panels). Thus, the uncertainties shown in the plots purely reflect the 
1$\sigma$ observational uncertainties on mass and radius. To keep the plots readable, we only show the uncertainties for the EL escape model; the uncertainties for the second model have approximately the same range and shift in the same way as the median values.

We find that the $f_{\rm atm}^0$ values are roughly consistent with what is expected from planetary formation models \citep[e.g.][]{Burn+2024} and lie between $\sim$$0.4$ and $41.3$\% if considering all the cases ($\sim$$3.6-33.5$\% for the CHAIN-based escape models). They show a weak dependence on the system's age. The difference between the slow and fast stellar rotators is more pronounced for the EL model, where the escape is fully set by stellar XUV, while the CHAIN-based escape model also accounts for blow-off like regime at early ages \citep[][]{Erkaev2015MNRAS.448.1916E}. For both escape models, the initial atmospheric mass fraction is higher for more metallic cores.
For the EL model, $f_{\rm atm}^0$ increases with increasing $\eta$, in particular in the case of the fast rotator. Comparing this dependency to the CHAIN-based escape model, we find that the latter predicts $f_{\rm atm}^0$ values similar to the ones predicted by the EL model with $\eta$ $\sim5-10$\%.
We note, however, that these $\eta$ values only represent an average over the whole evolution. If we define the scaling parameter $\eta_{\mathrm{fit}}$ for a given evolution step as a ratio of the escape rate defined with CHAIN ($\dot{M}_{\rm CHAIN}$) to the energy-limited escape rate assuming $\eta=1$, that is

\begin{equation}
    \eta_{\rm fit} = \frac{\dot{M}_{\rm CHAIN}}{\dot{M}_{\rm EL,\eta=1}},%\nonumber
\end{equation}

we find that $\eta_{\rm fit}$ varies between $\sim$$0.03-45\%$ across the evolution, with lower values at early ages.

These results suggest that water-rich atmospheres could be stable enough to survive under the harsh irradiation conditions of TOI-2211\,b on Gyr timescales and therefore are one possible explanation for the extended radius of the planet. 
Furthermore, this scenario does not necessarily require a massive amount of water accreted before the dissipation of the protoplanetary disk. The median $f_{\rm atm}^0$ values predicted by the models vary between $\sim$$5$ and $20$\% of the planetary mass, while the lower edge of the model uncertainties is consistent with present-day values predicted by internal structure modelling in Section\,\ref{sec:internal_structure}.

\subsection{Additional constraints on the 15.6~day planet candidate}

\subsubsection{Transit constraints from TESS}
\begin{figure}
    \centering
    \includegraphics[width=0.9\linewidth]{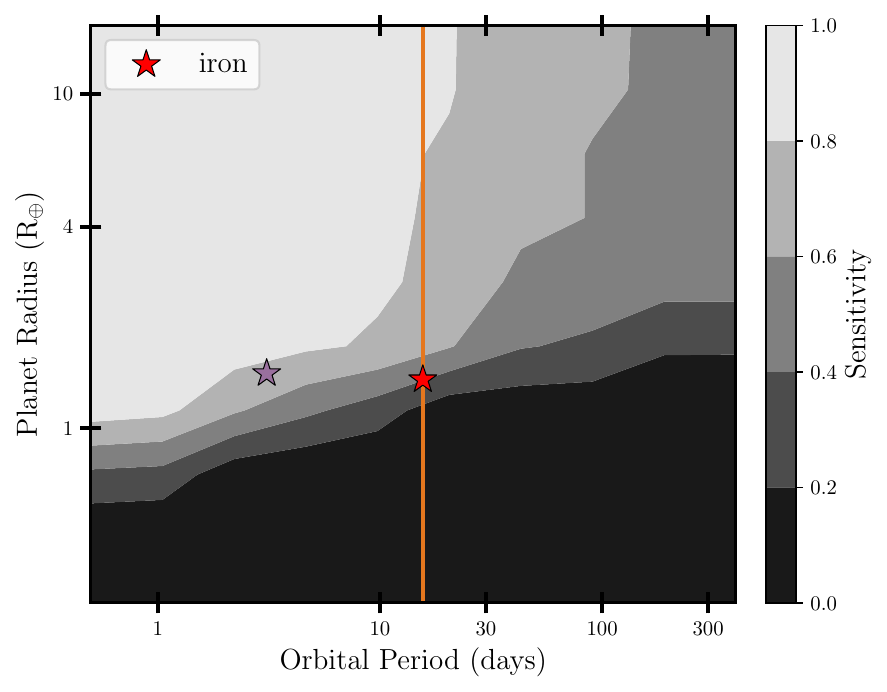}
    \caption{Sensitivity map given the available TESS data. TOI-2211\,b is highlighted by the purple star, while the outer planet candidate is presented by the orange line. The computed radius of this planet candidate for 100\% iron as a lower limit on the radius is shown in red.  }
    \label{fig:sensitivity}
\end{figure}

To investigate whether the 15.6~day planet candidate TOI-2211.02 could be transiting, we performed a sensitivity analysis using the Transit Investigation and Recoverability Application \citep[TIaRA;][]{Rodel+2024,Eschen+2024}. We use the short-cadence TESS data of the four available sectors (as outlined in \autoref{tab:tess_dat}, with \texttt{QUALITY}=0) and the 2-hour CDPP values reported in the header of each file. We generate 1000 planets uniformly distributed in log2 space ranging in period from 0.5 to 400 days and radius from 0.3 to 16 R$_\oplus$. TIaRA computes the signal-to-noise ratio for each of these planets at randomly generated phases and determines the detectability based on the signal-to-noise, number of transits and the corresponding gamma-function. We show the resulting sensitivity map in \autoref{fig:sensitivity}. 

TOI-2211\,b is plotted on top of the computed sensitivities and find that as expected the planet is detectable by TESS. Since we did not detect a transit and hence obtained a radius of the outer planet candidate, we plot a vertical line at its orbital period. As done in \citet{Eschen+2026}, we compute predicted radii for planet compositions of pure iron as a lower limit on the radius using the planet models of \citet{Zeng+2019}. We find that TESS has a sensitivity above 50\% for detecting this planet candidate if it was transiting. However, the sensitivity does not reach 100\% as the TESS data does include gaps. For this reason, we additionally check the available TESS data for the transit windows based on the period and epoch in \autoref{tab:planetary_params}. We find that the TESS data covers the windows in which the transit would fall and hence conclude that this planet candidate does not transit.

\subsubsection{Dynamical stability of the system}

Following \citet{Correia_etal_2005, Correia_etal_2010}, we analysed the dynamical stability of the best-fit two-planet solution in Table~\ref{tab:joint_fit_priors_posteriors}.
We integrated a regular 2D grid of initial conditions around the best fit using the SABAC4 symplectic integrator \citep{Laskar_Robutel_2001}, with a step size of $2.5 \times 10^{-4}$~yr and general relativistic corrections.
Each orbit was evolved for $10^4$~yr, and stability was assessed through a frequency analysis of the mean longitude of TOI-2211.02 \citep{Laskar_1990, Laskar_1993PD}.

Although the period ratio, $P_c/P_b = 5.048$, places the system close to the 5:1 mean motion resonance, the nominal solution lies outside the resonant region and is dynamically stable. The system also remains stable across the current eccentricity uncertainties, and even for configurations with $e_b,e_c \simeq 0.5$.

\begin{figure}
    \centering
	\includegraphics[width=\columnwidth]{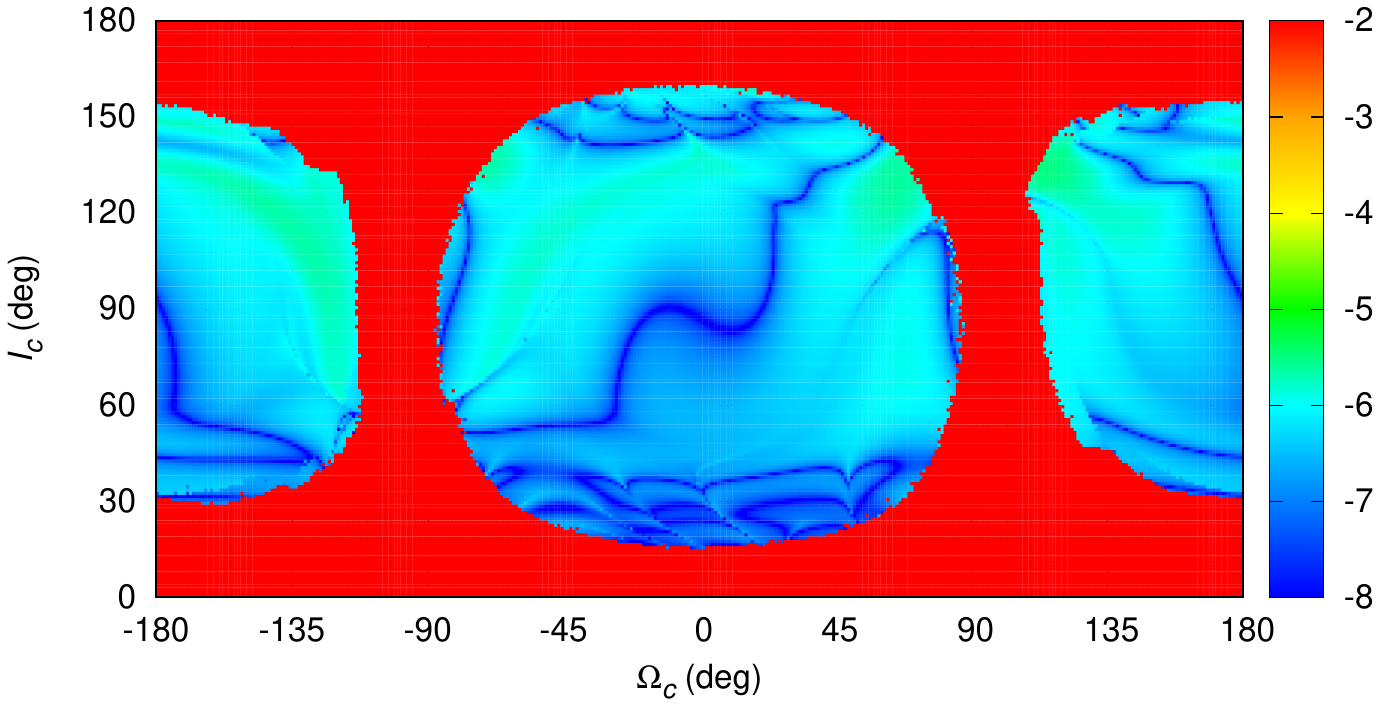}
    \caption{Stability map of the TOI-2211 system in the $(\Omega_c,I_c)$ plane for the outer planet candidate TOI-2211.02. All other orbital parameters were fixed to the nominal solution from Table~\ref{tab:joint_fit_priors_posteriors}. The colour scale shows the adopted stability indicator, with blue regions corresponding to stable configurations and red regions to unstable configurations.}
    \label{fig:stabilityTOI2211}
\end{figure}

Since TOI-2211.02 is not transiting, its true mass depends on the orbital inclination. We explored the stability of the system in the $(\Omega_c,I_c)$ plane (Fig.~\ref{fig:stabilityTOI2211}). Low-inclination solutions are dynamically unstable, while the broad stable prograde region starts at approximately $I_c \simeq 25^\circ$, implying $m_c \lesssim 8.52/\sin 25^\circ \simeq 20\,M_\oplus$ from the measured minimum mass in Table~\ref{tab:joint_fit_priors_posteriors}.

\subsubsection{Inferring the planetary system architecture based on the Bern model of planet formation and evolution} % and possible formation pathways?

Given the detection limits of present day instruments, small planets and planets with long orbital periods remain very challenging to detect. This makes it difficult to gain a full picture of the architecture of a planetary system and constrain the presence, properties, and orbital configuration of potential additional planetary companions. Inferring the broader architecture of the system based on just the observed planets is a degenerate problem and therefore requires a physically motivated prior, which we construct in this section based on synthetic planetary populations.

Our work is based on the Bern model \citep{Alibert+2005, Mordasini+2009, Emsenhuber+2021a, Emsenhuber+2021b}, which is a global end-to-end model of planetary formation and evolution, built on the core-accretion paradigm. It consistently solves the coupled differential equations governing the viscous evolution of the protoplanetary disc, the dynamical state of a planetesimal disc, the internal structure of the forming protoplanets, and the mutual gravitational interactions of concurrently growing embryos through an N-body integrator. The long-term thermodynamic evolution of each planet is computed up to 10 Gyr, including effects such as cooling and contraction, atmospheric escape, radius inflation, and stellar tides.

While the Bern model outputs complete populations of stable planetary systems, the computational cost of these simulations limits the size of the available samples for statistical inference. To overcome this, in a recent work, \cite{Alibert2025} trained a generative model on a set of 25,000 stable planetary systems orbiting solar-type stars produced with the Bern model, enabling the rapid synthesis of additional systems that preserve the statistical properties of the underlying population. It was shown that it effectively generated systems with physical meaning and that it could constrain properly the properties of other planets in the system given at least one detection. In the present work, we adopt the novel framework introduced by \cite{MarquesIEEE2026}, which extends the model of \cite{Alibert2025}, allowing us to infer the properties of yet-to-be detected planets in the same system conditioned on the observed transiting planet(s). This generative model is a character-level transformer that represents each planetary system as a sequence of discrete tokens encoding the planets' properties: radius, orbital period, and mass.

\begin{figure}
        \centering
        \includegraphics[width=0.85\linewidth]{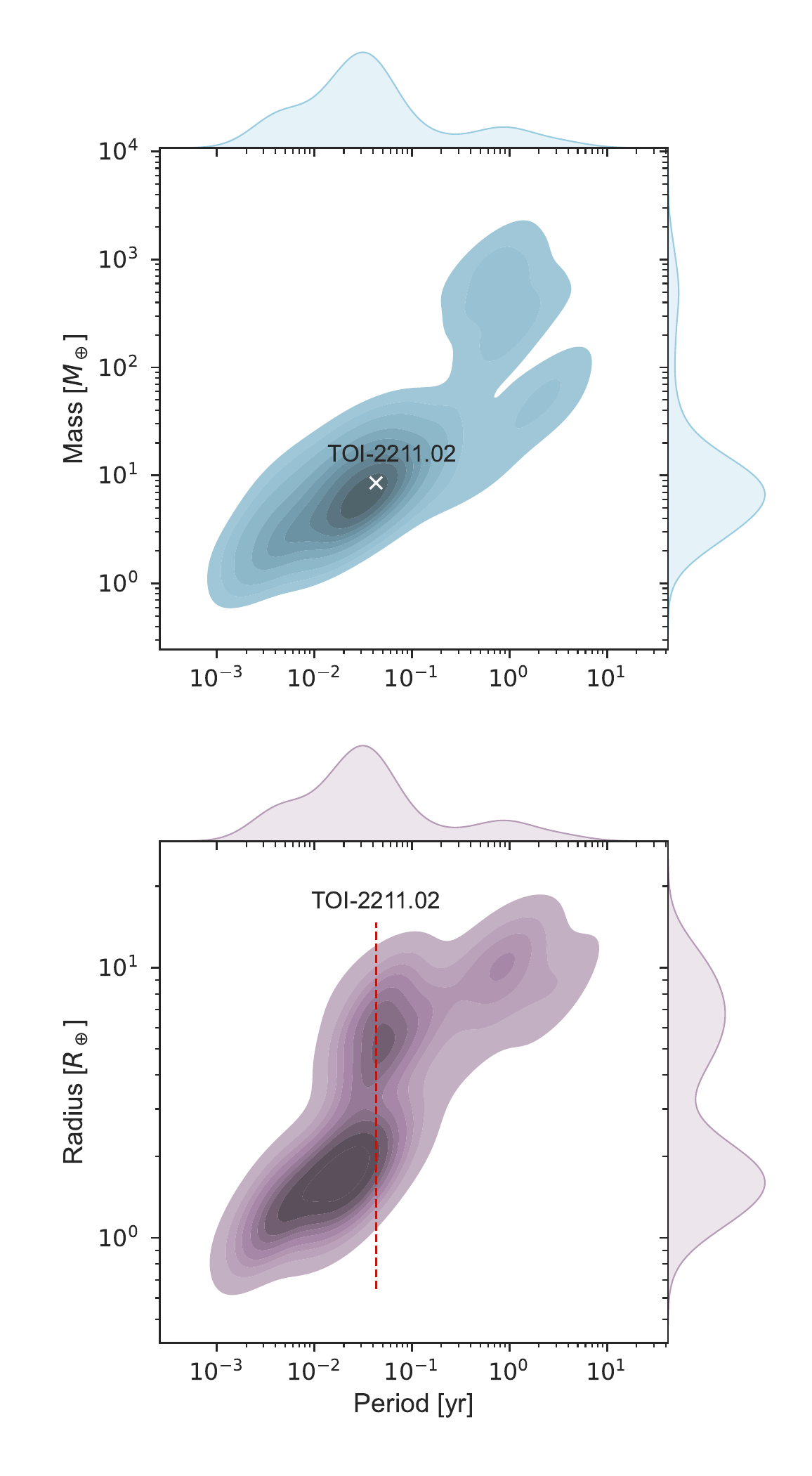}
        \caption{Joint and marginal distributions for the mass, radius and orbital period of the remaining planets in the systems containing a TOI-2211\,b-like planet generated with the \cite{MarquesIEEE2026} framework. An observational filter has been applied to exclude planets with signals with a semi-amplitude less than $K=1$ m/s. The minimal mass and orbital period of the planet candidate TOI-2211.02 were added for reference, following the values in Table~\ref{tab:planetary_params}.}
        \label{fig:analysis_sara}
\end{figure}

To apply this analysis to the TOI-2211 system, a sample of $10^6$ synthetic systems was generated using this new framework. From these, we selected the ones containing a TOI-2211\,b-like planet (allowing for a 1\% uncertainty in orbital period and 10\% uncertainty in radius based on the median values reported in Table \ref{tab:planetary_params}), yielding $\sim$1700 systems. We removed the TOI-2211\,b-like planet from the system to draw the joint distributions over the mass, radius and period for the remaining planets and added an observational filter
at $K=1$ m/s to exclude non-observable planets. The final distributions are shown in Fig.~\ref{fig:analysis_sara}.

These distributions exhibit a clear over-density consistent with the presence of a second planet in the system, whose orbital period and inferred mass are compatible with the ones of the planet candidate identified in the RV analysis in Section~\ref{sec:data+analysis}. In particular, looking at the mass–period distribution, if we apply a Gaussian fit to the predicted logarithmic mass distribution, it yields a normal distribution centred at $0.82\pm0.39$. The minimal mass of candidate TOI-2211.02 lies in the upper portion of this distribution but remains well within $1\sigma$ of the median of the log-normal distribution, 6.7~M$_\oplus$, consistent with the model prediction. 

Furthermore, the framework also provides a way to estimate the radius for the candidate TOI-2211.02 based on a population-level prediction of the adopted synthetic population. The analysis of the radius distribution is a bit more complex than the mass distribution, as it displays a bimodal structure in the logarithmic space with two well-separated peaks. We fit this distribution with two separate Gaussian components in $\log(R_p/R_\oplus)$, which are centred at $0.20\pm0.16$ and $0.84\pm0.20$ in the logarithmic space and correspond to log-normal distributions with median radii of 1.6 and 6.9~R$_\oplus$, respectively. Considering only the marginal radius posterior, the candidate TOI-2211.02 has an approximately equal probability of belonging to either mode. However, conditioning the radius posterior on the measured minimal mass shifts the probability to $\sim$60\% for belonging to the second mode. 
Finally, we note that this overall analysis does not exclude the possibility of additional planets in the system.

%%%%%%%%%%%%%%%%%%%%%%%%%%%%%%%%%%%%%%%%%%%%%%%%%%%%%%%%%%%%%%
\section{Conclusion}
\label{sec:conclusion}
In this paper, we presented a first detailed characterisation of TOI-2211\,b, a recently validated transiting low-density super-Earth orbiting a bright K-type star. This is the second part of a series of papers presenting the results of our CHEOPS GTO programme searching for new hot water world candidates, which we defined in \cite{Egger+2025} as planets that fall into specific triangular regions in the mass-radius space (Hot Water World triangles). In these regions, low-mass primordial H/He envelopes are expected to be lost through atmospheric escape, which means that planets falling into these triangles are expected to host high mean molecular weight atmospheres. 

Here, we combined transit photometry from TESS and CHEOPS with precise radial velocity measurements from HARPS, ESPRESSO, and KPF, arriving at a planetary radius of $1.467 \pm 0.047$~R$_\oplus$ and a mass of $2.69 \pm 0.56$~M$_\oplus$ for the 3.09~d planet TOI-2211\,b. These properties place TOI-2211\,b in a regime where low-mass primordial H/He envelopes are not expected to be stable against atmospheric escape. We verified this using a detailed evaporation analysis, which confirmed that the evaporation timescale of H/He envelopes on this planet does not exceed $\sim$7.65~Myr, much smaller than the age of the system of 9.9$^{+3.9}_{-6.2}$~Gyr and thereby effectively ruling out such an envelope. At the same time, TOI-2211\,b has a lower-than-expected bulk density, with the derived mass and radius values lying above the Earth-like composition line by at least 2$\sigma$ and the median values lying above the composition line for a fully rocky, iron-free core. We therefore concluded based on this observed bulk density and our detailed internal structure and evaporation analyses that TOI-2211\,b is a promising candidate to host a high mean molecular weight volatile envelope.

We also presented an updated version of our method to model the atmospheric lifetimes of water-rich atmospheres, which was first introduced in \cite{Egger+2025}. This updated version is based on the more sophisticated atmospheric escape model of \cite{Kubyshkina+2026}, which makes use of the CHAIN code \citep{kubyshkina2024}. This analysis suggested that water-rich atmospheres could be stable enough to survive on TOI-2211\,b and therefore provide a possible explanation for the observed low mean density of the planet, even if the current data do not uniquely constrain the volatile composition.  We inferred initial atmospheric mass fractions with median values between $\sim$10 and 20\% of the total planetary mass, which is roughly consistent with what is expected from planetary formation models \citep[e.g.][]{Burn+2024}. Overall, this makes TOI-2211\,b an interesting candidate for a hot water world, to be confirmed through future spectroscopic observations. By comparing the results of the new CHAIN-based model to the ones of the energy-limited model used in \cite{Egger+2025}, we confirmed our earlier assumption that choosing a value of $\sim$$5$\% for the heating efficiency coefficient $\eta$ is optimal for modelling the evolution of these hot water worlds using the energy-limited approximation.

In addition, we identified an outer, non-transiting planetary candidate, TOI-2211.02, at an orbital period of 15.6~d with a minimal mass of $8.53 \pm 1.04$~M$_\oplus$ in the radial velocity data. 
Through a dynamical stability analysis, we showed that such a two-planet system would be stable even for configurations with eccentricities up to $\sim$0.5, and concluded that the inclination of TOI-2211.02 needs to be higher than $\sim$25~degrees, leading to an upper mass limit of $\sim$20~M$_\oplus$ for the planet candidate. 
We finally used a generative model based on the Bern model of planet formation and evolution \citep{MarquesIEEE2026, Alibert2025} to infer the full planetary system architecture based on the observed transiting planet TOI-2211\,b. We found that there is a clear over-density in the generated posterior distributions consistent with the presence of a second planet in the system. The properties of this inferred planet are compatible with the orbital period and minimal mass of TOI-2211.02. Additionally, the framework predicts a bimodal radius distribution with two peaks centred at approximately 1.6 and 6.9~R$_\oplus$ for this planet candidate.

%%%%%%%%%%%%%%%%%%%%%%%%%%%%%%%%%%%%%%%%%%%%%%%%%%%%%%%%%%%%%%
\section*{Data availability}
This article uses data from the CHEOPS Guaranteed Time Observation programme CH\_PR140068, and HARPS and ESPRESSO data collected under ESO programmes 113.26G5.001 and 115.2842.001 (PI: Egger). The raw and detrended photometric time-series and radial velocity data are available in electronic form at the CDS via anonymous ftp to cdsarc.ustrasbg.fr (130.79.128.5) or via \url{https://cdsarc.cds.unistra.fr/viz-bin/cat/J/A+A/XXX/YYY}.

\begin{acknowledgements}
    The authors thank François Bouchy and Lucile Mignon for the organisation of the HARPS/NIRPS timeshare, through which the HARPS data presented here were collected.
    CHEOPS is an ESA mission in partnership with Switzerland with important contributions to the payload and the ground segment from Austria, Belgium, France, Germany, Hungary, Italy, Portugal, Spain, Sweden, and the United Kingdom. The CHEOPS Consortium would like to gratefully acknowledge the support received by all the agencies, offices, universities, and industries involved. Their flexibility and willingness to explore new approaches were essential to the success of this mission. CHEOPS data analysed in this article will be made available in the CHEOPS mission archive (\url{https://cheops.unige.ch/archive_browser/}). 
    JAE acknowledges support through the European Space Agency (ESA) Research Fellowship programme in Space Science. 
    This work has been carried out within the framework of the NCCR PlanetS supported by the Swiss National Science Foundation under grants 51NF40\_182901 and 51NF40\_205606. 
    ADe acknowledges financial support from the Swiss National Science Foundation (SNSF) for project 200021\_200726. 
    A. S. acknowledges support from the Swiss Space Office through the ESA PRODEX program. 
    LBo, GBr, VNa, IPa, GPi, RRa, GSc, VSi, and TZi acknowledge support from CHEOPS ASI-INAF agreement n. 2019-29-HH.0. 
    ACMC acknowledges support from the FCT, Portugal, through the CFisUC project UID/04564/2025, with DOI identifier 10.54499/UID/04564/2025. 
    A.C. acknowledge their role as ESA-appointed CHEOPS Science Team Members. 
    YNEE acknowledges support from a Science and Technology Facilities Council (STFC) studentship, grant number ST/Y509693/1. 
    DG gratefully acknowledges financial support from the CRT foundation under Grant No. 2018.2323 ‘Gaseous or rocky? Unveiling the nature of small worlds’. 
    S.G.S. acknowledge support from FCT through FCT contract nr. CEECIND/00826/2018 and POPH/FSE (EC). 
    The Portuguese team thanks the Portuguese Space Agency for the provision of financial support in the framework of the PRODEX Programme of the European Space Agency (ESA) under contract number 4000142255. This work was supported by Fundação para a Ciência e a Tecnologia (FCT) through national funds under the research grant UID/04434/2025 (DOI: 10.54499/UID/04434/2025). 
    TWi acknowledges support from the UKSA and the University of Warwick. 
    YAl acknowledges support from the Swiss National Science Foundation (SNSF) under grant 200020\_192038. 
    TZi acknowledges NVIDIA Academic Hardware Grant Program for the use of the Titan V GPU card and the Italian MUR Departments of Excellence grant 2023-2027 “Quantum Frontiers”. 
    BA acknowledges support of the Swiss National Science Foundation under grant number PCEFP2\_194576. 
    DB, EP, EV, IR and RA acknowledge financial support from the Agencia Estatal de Investigación of the Ministerio de Ciencia e Innovación MCIN/AEI/10.13039/501100011033 and the ERDF “A way of making Europe” through projects PID2021-125627OB-C31, PID2021-125627OB-C32, PID2021-127289NB-I00, PID2023-150468NB-I00 and PID2023-149439NB-C41. 
    from the Centre of Excellence “Severo Ochoa'' award to the Instituto de Astrofísica de Canarias (CEX2019-000920-S), the Centre of Excellence “María de Maeztu” award to the Institut de Ciències de l’Espai (CEX2020-001058-M), and from the Generalitat de Catalunya/CERCA programme. 
    SCCB acknowledges the support from Fundação para a Ciência e Tecnologia (FCT) in the form of work contract through the Scientific Employment Incentive program with reference 2023.06687.CEECIND and DOI 10.54499/2023.06687.CEECIND/CP2839/CT0002. 
    ABr was supported by the SNSA. 
    C.B. acknowledges support from the Swiss Space Office through the ESA PRODEX program. 
    PCZ and ACC acknowledge support from STFC consolidated grant number ST/V000861/1, UKSA grant number ST/X002217/1, and UKRI/ERC Synergy Grant EP/Z000181/1 (REVEAL). 
    P.E.C. is funded by the Austrian Science Fund (FWF) Erwin Schroedinger Fellowship, program J4595-N. 
    This project was supported by the CNES. 
    This work was supported by FCT - Funda\c{c}\~{a}o para a Ci\^{e}ncia e a Tecnologia through national funds and by FEDER through COMPETE2020 through the research grants UIDB/04434/2020, UIDP/04434/2020, 2022.06962.PTDC. 
    O.D.S.D. is supported in the form of work contract (DL 57/2016/CP1364/CT0004) funded by national funds through FCT. 
    B.-O. D. acknowledges support from the Swiss State Secretariat for Education, Research and Innovation (SERI) under contract number MB22.00046. 
    A.D. acknowledge their role as ESA-appointed CHEOPS Science Team Members. 
    B.E. acknowledge their role as ESA-appointed CHEOPS Science Team Members. 
    This project has received funding from the Swiss National Science Foundation for project 200021\_200726. It has also been carried out within the framework of the National Centre of Competence in Research PlanetS supported by the Swiss National Science Foundation under grant 51NF40\_205606. The authors acknowledge the financial support of the SNSF. 
    MF and CMP gratefully acknowledge the support of the Swedish National Space Agency (DNR 65/19, 174/18). 
    K. G. acknowledge their role as ESA-appointed CHEOPS Science Team Members. 
    M.G. is F.R.S.-FNRS Research Director. 
    MNG is the ESA CHEOPS Project Scientist and Mission Representative. BMM is the ESA CHEOPS Project Scientist. KGI was the ESA CHEOPS Project Scientist until the end of December 2022 and Mission Representative until the end of January 2023. All of them are/were responsible for the Guest Observers (GO) Programme. None of them relay/relayed proprietary information between the GO and Guaranteed Time Observation (GTO) Programmes, nor do/did they decide on the definition and target selection of the GTO Programme. 
    J.K. acknowledge their role as ESA-appointed CHEOPS Science Team Members. JK acknowledges support from the Spanish Research Agency of the Ministry of Science, Innovation and Universities (AEI-MICIU) under grant 'Contribution of the IAC to the PLATO Space Mission' with reference PID2023-149439NB-C41. 
    K.W.F.L. was supported by Deutsche Forschungsgemeinschaft grants RA714/14-1 within the DFG Schwerpunkt SPP 1992, Exploring the Diversity of Extrasolar Planets. 
    This work was granted access to the HPC resources of MesoPSL financed by the Region Ile de France and the project Equip@Meso (reference ANR-10-EQPX-29-01) of the programme Investissements d'Avenir supervised by the Agence Nationale pour la Recherche. 
    AL acknowledges support from the Swiss National Science Foundation under grant number TMSGI2\_211697. 
    ML acknowledges support of the Swiss National Science Foundation under grant number PCEFP2\_194576. 
    PM acknowledges support from STFC research grant number ST/R000638/1. 
    This work was also partially supported by a grant from the Simons Foundation (PI Queloz, grant number 327127). 
    NCSa acknowledges funding by the European Union (ERC, FIERCE, 101052347). Views and opinions expressed are however those of the author(s) only and do not necessarily reflect those of the European Union or the European Research Council. Neither the European Union nor the granting authority can be held responsible for them. 
    Gy.M.Sz. thanks the SNN-147362 and the ADVANCED-153410 of the National Research, Development and Innovation Office (NKFIH, Hungary), and the ESA PRODEX Experiment Agreements No. 4000137122 No. 4000149203. 
    V.V.G. is an F.R.S-FNRS Research Associate. 
    JV acknowledges support from the Swiss National Science Foundation (SNSF) under grant PZ00P2\_208945. 
    NAW acknowledges UKSA grant ST/R004838/1. 
\end{acknowledgements}

%%%%%%%%%%%%%%%%%%%%%%%%%%%%%%%%%%%%%%%%%%%%%%%%%%%%%%%%%%%%%%
\bibliographystyle{aa} % style aa.bst
\bibliography{biblio} % your references biblio.bib

@ARTICLE{Laskar_Robutel_2001,
   author = {{Laskar}, J. and {Robutel}, P.},
    title = "{High order symplectic integrators for perturbed Hamiltonian systems}",
  journal = {Celestial Mechanics and Dynamical Astronomy},
   eprint = {arXiv:astro-ph/0005074},
     year = 2001,
    month = jul,
   volume = 80,
    pages = {39-62},
   adsurl = {http://esoads.eso.org/abs/2001CeMDA..80...39L}
}

@ARTICLE{Laskar_1993PD,
   author = {{Laskar}, J.},
    title = "{Frequency analysis for multi-dimensional systems. Global dynamics and diffusion}",
  journal = {Physica D Nonlinear Phenomena},
     year = 1993,
    month = aug,
   volume = 67,
    pages = {257-281},
      doi = {10.1016/0167-2789(93)90210-R},
   adsurl = {http://cdsads.u-strasbg.fr/cgi-bin/nph-bib_query?bibcode=1993PhyD...67..257L&db_key=PHY}
}

@ARTICLE{Laskar_1990,
   author = {{Laskar}, J.},
    title = "{The chaotic motion of the solar system - A numerical estimate of the size of the chaotic zones}",
  journal = {\icarus},
     year = 1990,
    month = dec,
   volume = 88,
    pages = {266-291},
      doi = {10.1016/0019-1035(90)90084-M},
   adsurl = {http://cdsads.u-strasbg.fr/cgi-bin/nph-bib_query?bibcode=1990Icar...88..266L&db_key=AST}
}

@ARTICLE{Correia_etal_2010,
   author = {{Correia}, A.~C.~M. and {Couetdic}, J. and {Laskar}, J. and 
	{Bonfils}, X. and {Mayor}, M. and {Bertaux}, {J.-L.} and {Bouchy}, F. and 
	{Delfosse}, X. and {Forveille}, T. and {Lovis}, C. and {Pepe}, F. and 
	{Perrier}, C. and {Queloz}, D. and {Udry}, S.},
    title = "{The HARPS search for southern extra-solar planets. XIX. Characterization and dynamics of the GJ 876 planetary system}",
  journal = {\aap},
archivePrefix = "arXiv",
   eprint = {1001.4774},
     year = 2010,
    month = feb,
   volume = 511,
    pages = {A21},
      doi = {10.1051/0004-6361/200912700},
   adsurl = {http://esoads.eso.org/abs/2010A%26A...511A..21C}
}

@ARTICLE{Correia_etal_2005,
   author = {{Correia}, A.~C.~M. and {Udry}, S. and {Mayor}, M. and {Laskar}, J. and 
	{Naef}, D. and {Pepe}, F. and {Queloz}, D. and {Santos}, N.~C.},
    title = "{The CORALIE survey for southern extra-solar planets. XIII. A pair of planets around HD202206 or a circumbinary planet?}",
  journal = {\aap},
     year = 2005,
    month = sep,
   volume = 440,
    pages = {751-758},
      doi = {10.1051/0004-6361:20042376},
   adsurl = {http://cdsads.u-strasbg.fr/cgi-bin/nph-bib_query?bibcode=2005A%26A...440..751C&db_key=AST}
}

@ARTICLE{Rajpaul+2015,
       author = {{Rajpaul}, V. and {Aigrain}, S. and {Osborne}, M.~A. and {Reece}, S. and {Roberts}, S.},
        title = "{A Gaussian process framework for modelling stellar activity signals in radial velocity data}",
      journal = {\mnras},
         year = 2015,
        month = sep,
       volume = {452},
       number = {3},
        pages = {2269-2291},
          doi = {10.1093/mnras/stv1428},
archivePrefix = {arXiv},
       eprint = {1506.07304},
 primaryClass = {astro-ph.EP},
       adsurl = {https://ui.adsabs.harvard.edu/abs/2015MNRAS.452.2269R}
}

@ARTICLE{VanEylen+2015,
       author = {{Van Eylen}, Vincent and {Albrecht}, Simon},
        title = "{Eccentricity from Transit Photometry: Small Planets in Kepler Multi-planet Systems Have Low Eccentricities}",
      journal = {The Astrophysical Journal},
         year = 2015,
        month = aug,
       volume = {808},
       number = {2},
          eid = {126},
        pages = {126},
          doi = {10.1088/0004-637X/808/2/126},
archivePrefix = {arXiv},
       eprint = {1505.02814},
 primaryClass = {astro-ph.EP},
       adsurl = {https://ui.adsabs.harvard.edu/abs/2015ApJ...808..126V}
}

@ARTICLE{Barragan+2022,
       author = {{Barrag{\'a}n}, Oscar and {Aigrain}, Suzanne and {Rajpaul}, Vinesh M. and {Zicher}, Norbert},
        title = "{PYANETI - II. A multidimensional Gaussian process approach to analysing spectroscopic time-series}",
      journal = {\mnras},
         year = 2022,
        month = jan,
       volume = {509},
       number = {1},
        pages = {866-883},
          doi = {10.1093/mnras/stab2889},
archivePrefix = {arXiv},
       eprint = {2109.14086},
 primaryClass = {astro-ph.EP},
       adsurl = {https://ui.adsabs.harvard.edu/abs/2022MNRAS.509..866B}
}

@ARTICLE{Noyes+1984,
       author = {{Noyes}, R.~W. and {Hartmann}, L.~W. and {Baliunas}, S.~L. and {Duncan}, D.~K. and {Vaughan}, A.~H.},
        title = "{Rotation, convection, and magnetic activity in lower main-sequence stars.}",
      journal = {\apj},
         year = 1984,
        month = apr,
       volume = {279},
        pages = {763-777},
          doi = {10.1086/161945},
       adsurl = {https://ui.adsabs.harvard.edu/abs/1984ApJ...279..763N}
}

@ARTICLE{Barnes+2007,
       author = {{Barnes}, Sydney A.},
        title = "{Ages for Illustrative Field Stars Using Gyrochronology: Viability, Limitations, and Errors}",
      journal = {\apj},
         year = 2007,
        month = nov,
       volume = {669},
       number = {2},
        pages = {1167-1189},
          doi = {10.1086/519295},
archivePrefix = {arXiv},
       eprint = {0704.3068},
 primaryClass = {astro-ph},
       adsurl = {https://ui.adsabs.harvard.edu/abs/2007ApJ...669.1167B}
}

@article{kima,
      title = {kima: Exoplanet detection in radial velocities},
      author = {Faria, J. P. and Santos, N. C. and Figueira, P. and Brewer, B. J.},
      journal = {Journal of Open Source Software},
      issn = {2475-9066},
      number = {26},
      volume = {3},
      pages = {487},
      date = {2018-06-19},
      year = {2018},
      month = {6},
      day = {19},
      publisher = {The Open Journal},
      doi = {10.21105/joss.00487},
      url = {http://dx.doi.org/10.21105/joss.00487},
}

@ARTICLE{GLS,
       author = {{Zechmeister}, M. and {K{\"u}rster}, M.},
        title = "{The generalised Lomb-Scargle periodogram. A new formalism for the floating-mean and Keplerian periodograms}",
      journal = {\aap},
         year = 2009,
        month = mar,
       volume = {496},
       number = {2},
        pages = {577-584},
          doi = {10.1051/0004-6361:200811296},
archivePrefix = {arXiv},
       eprint = {0901.2573},
 primaryClass = {astro-ph.IM},
       adsurl = {https://ui.adsabs.harvard.edu/abs/2009A&A...496..577Z}
}

@ARTICLE{Lopez+Fortney2014,
       author = {{Lopez}, Eric D. and {Fortney}, Jonathan J.},
        title = "{Understanding the Mass-Radius Relation for Sub-neptunes: Radius as a Proxy for Composition}",
      journal = {\apj},
         year = 2014,
        month = sep,
       volume = {792},
       number = {1},
          eid = {1},
        pages = {1},
          doi = {10.1088/0004-637X/792/1/1},
archivePrefix = {arXiv},
       eprint = {1311.0329},
 primaryClass = {astro-ph.EP},
       adsurl = {https://ui.adsabs.harvard.edu/abs/2014ApJ...792....1L}
}

@ARTICLE{GaiaCollab2021,
       author = {{Gaia Collaboration} and {Brown}, A.~G.~A. and {Vallenari}, A. and {Prusti}, T. and {de Bruijne}, J.~H.~J. and {Babusiaux}, C. and {Biermann}, M. and {Creevey}, O.~L. and {Evans}, D.~W. and {Eyer}, L. and {Hutton}, A. and {Jansen}, F. and {Jordi}, C. and {Klioner}, S.~A. and {Lammers}, U. and {Lindegren}, L. and {Luri}, X. and {Mignard}, F. and {Panem}, C. and {Pourbaix}, D. and {Randich}, S. and {Sartoretti}, P. and {Soubiran}, C. and {Walton}, N.~A. and {Arenou}, F. and {Bailer-Jones}, C.~A.~L. and {Bastian}, U. and {Cropper}, M. and {Drimmel}, R. and {Katz}, D. and {Lattanzi}, M.~G. and {van Leeuwen}, F. and {Bakker}, J. and {Cacciari}, C. and {Casta{\~n}eda}, J. and {De Angeli}, F. and {Ducourant}, C. and {Fabricius}, C. and {Fouesneau}, M. and {Fr{\'e}mat}, Y. and {Guerra}, R. and {Guerrier}, A. and {Guiraud}, J. and {Jean-Antoine Piccolo}, A. and {Masana}, E. and {Messineo}, R. and {Mowlavi}, N. and {Nicolas}, C. and {Nienartowicz}, K. and {Pailler}, F. and {Panuzzo}, P. and {Riclet}, F. and {Roux}, W. and {Seabroke}, G.~M. and {Sordo}, R. and {Tanga}, P. and {Th{\'e}venin}, F. and {Gracia-Abril}, G. and {Portell}, J. and {Teyssier}, D. and {Altmann}, M. and {Andrae}, R. and {Bellas-Velidis}, I. and {Benson}, K. and {Berthier}, J. and {Blomme}, R. and {Brugaletta}, E. and {Burgess}, P.~W. and {Busso}, G. and {Carry}, B. and {Cellino}, A. and {Cheek}, N. and {Clementini}, G. and {Damerdji}, Y. and {Davidson}, M. and {Delchambre}, L. and {Dell'Oro}, A. and {Fern{\'a}ndez-Hern{\'a}ndez}, J. and {Galluccio}, L. and {Garc{\'\i}a-Lario}, P. and {Garcia-Reinaldos}, M. and {Gonz{\'a}lez-N{\'u}{\~n}ez}, J. and {Gosset}, E. and {Haigron}, R. and {Halbwachs}, J. -L. and {Hambly}, N.~C. and {Harrison}, D.~L. and {Hatzidimitriou}, D. and {Heiter}, U. and {Hern{\'a}ndez}, J. and {Hestroffer}, D. and {Hodgkin}, S.~T. and {Holl}, B. and {Jan{\ss}en}, K. and {Jevardat de Fombelle}, G. and {Jordan}, S. and {Krone-Martins}, A. and {Lanzafame}, A.~C. and {L{\"o}ffler}, W. and {Lorca}, A. and {Manteiga}, M. and {Marchal}, O. and {Marrese}, P.~M. and {Moitinho}, A. and {Mora}, A. and {Muinonen}, K. and {Osborne}, P. and {Pancino}, E. and {Pauwels}, T. and {Petit}, J. -M. and {Recio-Blanco}, A. and {Richards}, P.~J. and {Riello}, M. and {Rimoldini}, L. and {Robin}, A.~C. and {Roegiers}, T. and {Rybizki}, J. and {Sarro}, L.~M. and {Siopis}, C. and {Smith}, M. and {Sozzetti}, A. and {Ulla}, A. and {Utrilla}, E. and {van Leeuwen}, M. and {van Reeven}, W. and {Abbas}, U. and {Abreu Aramburu}, A. and {Accart}, S. and {Aerts}, C. and {Aguado}, J.~J. and {Ajaj}, M. and {Altavilla}, G. and {{\'A}lvarez}, M.~A. and {{\'A}lvarez Cid-Fuentes}, J. and {Alves}, J. and {Anderson}, R.~I. and {Anglada Varela}, E. and {Antoja}, T. and {Audard}, M. and {Baines}, D. and {Baker}, S.~G. and {Balaguer-N{\'u}{\~n}ez}, L. and {Balbinot}, E. and {Balog}, Z. and {Barache}, C. and {Barbato}, D. and {Barros}, M. and {Barstow}, M.~A. and {Bartolom{\'e}}, S. and {Bassilana}, J. -L. and {Bauchet}, N. and {Baudesson-Stella}, A. and {Becciani}, U. and {Bellazzini}, M. and {Bernet}, M. and {Bertone}, S. and {Bianchi}, L. and {Blanco-Cuaresma}, S. and {Boch}, T. and {Bombrun}, A. and {Bossini}, D. and {Bouquillon}, S. and {Bragaglia}, A. and {Bramante}, L. and {Breedt}, E. and {Bressan}, A. and {Brouillet}, N. and {Bucciarelli}, B. and {Burlacu}, A. and {Busonero}, D. and {Butkevich}, A.~G. and {Buzzi}, R. and {Caffau}, E. and {Cancelliere}, R. and {C{\'a}novas}, H. and {Cantat-Gaudin}, T. and {Carballo}, R. and {Carlucci}, T. and {Carnerero}, M.~I. and {Carrasco}, J.~M. and {Casamiquela}, L. and {Castellani}, M. and {Castro-Ginard}, A. and {Castro Sampol}, P. and {Chaoul}, L. and {Charlot}, P. and {Chemin}, L. and {Chiavassa}, A. and {Cioni}, M. -R.~L. and {Comoretto}, G. and {Cooper}, W.~J. and {Cornez}, T. and {Cowell}, S. and {Crifo}, F. and {Crosta}, M. and {Crowley}, C. and {Dafonte}, C. and {Dapergolas}, A. and {David}, M. and {David}, P. and {de Laverny}, P. and {De Luise}, F. and {De March}, R. and {De Ridder}, J. and {de Souza}, R. and {de Teodoro}, P. and {de Torres}, A. and {del Peloso}, E.~F. and {del Pozo}, E. and {Delbo}, M. and {Delgado}, A. and {Delgado}, H.~E. and {Delisle}, J. -B. and {Di Matteo}, P. and {Diakite}, S. and {Diener}, C. and {Distefano}, E. and {Dolding}, C. and {Eappachen}, D. and {Edvardsson}, B. and {Enke}, H. and {Esquej}, P. and {Fabre}, C. and {Fabrizio}, M. and {Faigler}, S. and {Fedorets}, G. and {Fernique}, P. and {Fienga}, A. and {Figueras}, F. and {Fouron}, C. and {Fragkoudi}, F. and {Fraile}, E. and {Franke}, F. and {Gai}, M. and {Garabato}, D. and {Garcia-Gutierrez}, A. and {Garc{\'\i}a-Torres}, M. and {Garofalo}, A. and {Gavras}, P. and {Gerlach}, E. and {Geyer}, R. and {Giacobbe}, P. and {Gilmore}, G. and {Girona}, S. and {Giuffrida}, G. and {Gomel}, R. and {Gomez}, A. and {Gonzalez-Santamaria}, I. and {Gonz{\'a}lez-Vidal}, J.~J. and {Granvik}, M. and {Guti{\'e}rrez-S{\'a}nchez}, R. and {Guy}, L.~P. and {Hauser}, M. and {Haywood}, M. and {Helmi}, A. and {Hidalgo}, S.~L. and {Hilger}, T. and {H{\l}adczuk}, N. and {Hobbs}, D. and {Holland}, G. and {Huckle}, H.~E. and {Jasniewicz}, G. and {Jonker}, P.~G. and {Juaristi Campillo}, J. and {Julbe}, F. and {Karbevska}, L. and {Kervella}, P. and {Khanna}, S. and {Kochoska}, A. and {Kontizas}, M. and {Kordopatis}, G. and {Korn}, A.~J. and {Kostrzewa-Rutkowska}, Z. and {Kruszy{\'n}ska}, K. and {Lambert}, S. and {Lanza}, A.~F. and {Lasne}, Y. and {Le Campion}, J. -F. and {Le Fustec}, Y. and {Lebreton}, Y. and {Lebzelter}, T. and {Leccia}, S. and {Leclerc}, N. and {Lecoeur-Taibi}, I. and {Liao}, S. and {Licata}, E. and {Lindstr{\o}m}, E.~P. and {Lister}, T.~A. and {Livanou}, E. and {Lobel}, A. and {Madrero Pardo}, P. and {Managau}, S. and {Mann}, R.~G. and {Marchant}, J.~M. and {Marconi}, M. and {Marcos Santos}, M.~M.~S. and {Marinoni}, S. and {Marocco}, F. and {Marshall}, D.~J. and {Martin Polo}, L. and {Mart{\'\i}n-Fleitas}, J.~M. and {Masip}, A. and {Massari}, D. and {Mastrobuono-Battisti}, A. and {Mazeh}, T. and {McMillan}, P.~J. and {Messina}, S. and {Michalik}, D. and {Millar}, N.~R. and {Mints}, A. and {Molina}, D. and {Molinaro}, R. and {Moln{\'a}r}, L. and {Montegriffo}, P. and {Mor}, R. and {Morbidelli}, R. and {Morel}, T. and {Morris}, D. and {Mulone}, A.~F. and {Munoz}, D. and {Muraveva}, T. and {Murphy}, C.~P. and {Musella}, I. and {Noval}, L. and {Ord{\'e}novic}, C. and {Orr{\`u}}, G. and {Osinde}, J. and {Pagani}, C. and {Pagano}, I. and {Palaversa}, L. and {Palicio}, P.~A. and {Panahi}, A. and {Pawlak}, M. and {Pe{\~n}alosa Esteller}, X. and {Penttil{\"a}}, A. and {Piersimoni}, A.~M. and {Pineau}, F. -X. and {Plachy}, E. and {Plum}, G. and {Poggio}, E. and {Poretti}, E. and {Poujoulet}, E. and {Pr{\v{s}}a}, A. and {Pulone}, L. and {Racero}, E. and {Ragaini}, S. and {Rainer}, M. and {Raiteri}, C.~M. and {Rambaux}, N. and {Ramos}, P. and {Ramos-Lerate}, M. and {Re Fiorentin}, P. and {Regibo}, S. and {Reyl{\'e}}, C. and {Ripepi}, V. and {Riva}, A. and {Rixon}, G. and {Robichon}, N. and {Robin}, C. and {Roelens}, M. and {Rohrbasser}, L. and {Romero-G{\'o}mez}, M. and {Rowell}, N. and {Royer}, F. and {Rybicki}, K.~A. and {Sadowski}, G. and {Sagrist{\`a} Sell{\'e}s}, A. and {Sahlmann}, J. and {Salgado}, J. and {Salguero}, E. and {Samaras}, N. and {Sanchez Gimenez}, V. and {Sanna}, N. and {Santove{\~n}a}, R. and {Sarasso}, M. and {Schultheis}, M. and {Sciacca}, E. and {Segol}, M. and {Segovia}, J.~C. and {S{\'e}gransan}, D. and {Semeux}, D. and {Shahaf}, S. and {Siddiqui}, H.~I. and {Siebert}, A. and {Siltala}, L. and {Slezak}, E. and {Smart}, R.~L. and {Solano}, E. and {Solitro}, F. and {Souami}, D. and {Souchay}, J. and {Spagna}, A. and {Spoto}, F. and {Steele}, I.~A. and {Steidelm{\"u}ller}, H. and {Stephenson}, C.~A. and {S{\"u}veges}, M. and {Szabados}, L. and {Szegedi-Elek}, E. and {Taris}, F. and {Tauran}, G. and {Taylor}, M.~B. and {Teixeira}, R. and {Thuillot}, W. and {Tonello}, N. and {Torra}, F. and {Torra}, J. and {Turon}, C. and {Unger}, N. and {Vaillant}, M. and {van Dillen}, E. and {Vanel}, O. and {Vecchiato}, A. and {Viala}, Y. and {Vicente}, D. and {Voutsinas}, S. and {Weiler}, M. and {Wevers}, T. and {Wyrzykowski}, {\L}. and {Yoldas}, A. and {Yvard}, P. and {Zhao}, H. and {Zorec}, J. and {Zucker}, S. and {Zurbach}, C. and {Zwitter}, T.},
        title = "{Gaia Early Data Release 3. Summary of the contents and survey properties}",
      journal = {\aap},
         year = 2021,
        month = may,
       volume = {649},
          eid = {A1},
        pages = {A1},
          doi = {10.1051/0004-6361/202039657},
archivePrefix = {arXiv},
       eprint = {2012.01533},
 primaryClass = {astro-ph.GA},
       adsurl = {https://ui.adsabs.harvard.edu/abs/2021A&A...649A...1G}
}

@ARTICLE{Luque+2023,
       author = {{Luque}, R. and {Osborn}, H.~P. and {Leleu}, A. and {Pall{\'e}}, E. and {Bonfanti}, A. and {Barrag{\'a}n}, O. and {Wilson}, T.~G. and {Broeg}, C. and {Cameron}, A. Collier and {Lendl}, M. and {Maxted}, P.~F.~L. and {Alibert}, Y. and {Gandolfi}, D. and {Delisle}, J. -B. and {Hooton}, M.~J. and {Egger}, J.~A. and {Nowak}, G. and {Lafarga}, M. and {Rapetti}, D. and {Twicken}, J.~D. and {Morales}, J.~C. and {Carleo}, I. and {Orell-Miquel}, J. and {Adibekyan}, V. and {Alonso}, R. and {Alqasim}, A. and {Amado}, P.~J. and {Anderson}, D.~R. and {Anglada-Escud{\'e}}, G. and {Bandy}, T. and {B{\'a}rczy}, T. and {Barrado Navascues}, D. and {Barros}, S.~C.~C. and {Baumjohann}, W. and {Bayliss}, D. and {Bean}, J.~L. and {Beck}, M. and {Beck}, T. and {Benz}, W. and {Billot}, N. and {Bonfils}, X. and {Borsato}, L. and {Boyle}, A.~W. and {Brandeker}, A. and {Bryant}, E.~M. and {Cabrera}, J. and {Carrazco-Gaxiola}, S. and {Charbonneau}, D. and {Charnoz}, S. and {Ciardi}, D.~R. and {Cochran}, W.~D. and {Collins}, K.~A. and {Crossfield}, I.~J.~M. and {Csizmadia}, Sz. and {Cubillos}, P.~E. and {Dai}, F. and {Davies}, M.~B. and {Deeg}, H.~J. and {Deleuil}, M. and {Deline}, A. and {Delrez}, L. and {Demangeon}, O.~D.~S. and {Demory}, B. -O. and {Ehrenreich}, D. and {Erikson}, A. and {Esparza-Borges}, E. and {Falk}, B. and {Fortier}, A. and {Fossati}, L. and {Fridlund}, M. and {Fukui}, A. and {Garcia-Mejia}, J. and {Gill}, S. and {Gillon}, M. and {Goffo}, E. and {G{\'o}mez Maqueo Chew}, Y. and {G{\"u}del}, M. and {Guenther}, E.~W. and {G{\"u}nther}, M.~N. and {Hatzes}, A.~P. and {Helling}, Ch. and {Hesse}, K.~M. and {Howell}, S.~B. and {Hoyer}, S. and {Ikuta}, K. and {Isaak}, K.~G. and {Jenkins}, J.~M. and {Kagetani}, T. and {Kiss}, L.~L. and {Kodama}, T. and {Korth}, J. and {Lam}, K.~W.~F. and {Laskar}, J. and {Latham}, D.~W. and {Lecavelier des Etangs}, A. and {Leon}, J.~P.~D. and {Livingston}, J.~H. and {Magrin}, D. and {Matson}, R.~A. and {Matthews}, E.~C. and {Mordasini}, C. and {Mori}, M. and {Moyano}, M. and {Munari}, M. and {Murgas}, F. and {Narita}, N. and {Nascimbeni}, V. and {Olofsson}, G. and {Osborne}, H.~L.~M. and {Ottensamer}, R. and {Pagano}, I. and {Parviainen}, H. and {Peter}, G. and {Piotto}, G. and {Pollacco}, D. and {Queloz}, D. and {Quinn}, S.~N. and {Quirrenbach}, A. and {Ragazzoni}, R. and {Rando}, N. and {Ratti}, F. and {Rauer}, H. and {Redfield}, S. and {Ribas}, I. and {Ricker}, G.~R. and {Rudat}, A. and {Sabin}, L. and {Salmon}, S. and {Santos}, N.~C. and {Scandariato}, G. and {Schanche}, N. and {Schlieder}, J.~E. and {Seager}, S. and {S{\'e}gransan}, D. and {Shporer}, A. and {Simon}, A.~E. and {Smith}, A.~M.~S. and {Sousa}, S.~G. and {Stalport}, M. and {Szab{\'o}}, Gy. M. and {Thomas}, N. and {Tuson}, A. and {Udry}, S. and {Vanderburg}, A.~M. and {Van Eylen}, V. and {Van Grootel}, V. and {Venturini}, J. and {Walter}, I. and {Walton}, N.~A. and {Watanabe}, N. and {Winn}, J.~N. and {Zingales}, T.},
        title = "{A resonant sextuplet of sub-Neptunes transiting the bright star HD 110067}",
      journal = {\nat},
         year = 2023,
        month = nov,
       volume = {623},
       number = {7989},
        pages = {932-937},
          doi = {10.1038/s41586-023-06692-3},
archivePrefix = {arXiv},
       eprint = {2311.17775},
 primaryClass = {astro-ph.EP},
       adsurl = {https://ui.adsabs.harvard.edu/abs/2023Natur.623..932L}
}

@ARTICLE{Emsenhuber+2021a,
       author = {{Emsenhuber}, Alexandre and {Mordasini}, Christoph and {Burn}, Remo and {Alibert}, Yann and {Benz}, Willy and {Asphaug}, Erik},
        title = "{The New Generation Planetary Population Synthesis (NGPPS). I. Bern global model of planet formation and evolution, model tests, and emerging planetary systems}",
      journal = {\aap},
         year = 2021,
        month = dec,
       volume = {656},
          eid = {A69},
        pages = {A69},
          doi = {10.1051/0004-6361/202038553},
archivePrefix = {arXiv},
       eprint = {2007.05561},
 primaryClass = {astro-ph.EP},
       adsurl = {https://ui.adsabs.harvard.edu/abs/2021A&A...656A..69E}
}

@ARTICLE{Emsenhuber+2021b,
       author = {{Emsenhuber}, Alexandre and {Mordasini}, Christoph and {Burn}, Remo and {Alibert}, Yann and {Benz}, Willy and {Asphaug}, Erik},
        title = "{The New Generation Planetary Population Synthesis (NGPPS). II. Planetary population of solar-like stars and overview of statistical results}",
      journal = {\aap},
         year = 2021,
        month = dec,
       volume = {656},
          eid = {A70},
        pages = {A70},
          doi = {10.1051/0004-6361/202038863},
archivePrefix = {arXiv},
       eprint = {2007.05562},
 primaryClass = {astro-ph.EP},
       adsurl = {https://ui.adsabs.harvard.edu/abs/2021A&A...656A..70E}
}

@ARTICLE{Burn+2024,
       author = {{Burn}, Remo and {Mordasini}, Christoph and {Mishra}, Lokesh and {Haldemann}, Jonas and {Venturini}, Julia and {Emsenhuber}, Alexandre and {Henning}, Thomas},
        title = "{A radius valley between migrated steam worlds and evaporated rocky cores}",
      journal = {Nat. Astron.},
         year = 2024,
        month = apr,
       volume = {8},
        pages = {463-471},
          doi = {10.1038/s41550-023-02183-7},
archivePrefix = {arXiv},
       eprint = {2401.04380},
 primaryClass = {astro-ph.EP},
       adsurl = {https://ui.adsabs.harvard.edu/abs/2024NatAs...8..463B}
}

@ARTICLE{Emsenhuber+2023,
       author = {{Emsenhuber}, Alexandre and {Mordasini}, Christoph and {Burn}, Remo},
        title = "{Planetary population synthesis and the emergence of four classes of planetary system architectures}",
      journal = {European Physical Journal Plus},
         year = 2023,
        month = feb,
       volume = {138},
       number = {2},
          eid = {181},
        pages = {181},
          doi = {10.1140/epjp/s13360-023-03784-x},
archivePrefix = {arXiv},
       eprint = {2303.00012},
 primaryClass = {astro-ph.EP},
       adsurl = {https://ui.adsabs.harvard.edu/abs/2023EPJP..138..181E}
}

@ARTICLE{Mordasini+2009,
       author = {{Mordasini}, C. and {Alibert}, Y. and {Benz}, W.},
        title = "{Extrasolar planet population synthesis. I. Method, formation tracks, and mass-distance distribution}",
      journal = {\aap},
         year = 2009,
        month = jul,
       volume = {501},
       number = {3},
        pages = {1139-1160},
          doi = {10.1051/0004-6361/200810301},
archivePrefix = {arXiv},
       eprint = {0904.2524},
 primaryClass = {astro-ph.EP},
       adsurl = {https://ui.adsabs.harvard.edu/abs/2009A&A...501.1139M}
}

@ARTICLE{Alibert+2005,
       author = {{Alibert}, Y. and {Mordasini}, C. and {Benz}, W. and {Winisdoerffer}, C.},
        title = "{Models of giant planet formation with migration and disc evolution}",
      journal = {\aap},
         year = 2005,
        month = apr,
       volume = {434},
       number = {1},
        pages = {343-353},
          doi = {10.1051/0004-6361:20042032},
archivePrefix = {arXiv},
       eprint = {astro-ph/0412444},
 primaryClass = {astro-ph},
       adsurl = {https://ui.adsabs.harvard.edu/abs/2005A&A...434..343A}
}

@ARTICLE{Mordasini2020,
       author = {{Mordasini}, C.},
        title = "{Planetary evolution with atmospheric photoevaporation. I. Analytical derivation and numerical study of the evaporation valley and transition from super-Earths to sub-Neptunes}",
      journal = {\aap},
         year = 2020,
        month = jun,
       volume = {638},
          eid = {A52},
        pages = {A52},
          doi = {10.1051/0004-6361/201935541},
archivePrefix = {arXiv},
       eprint = {2002.02455},
 primaryClass = {astro-ph.EP},
       adsurl = {https://ui.adsabs.harvard.edu/abs/2020A&A...638A..52M}
}

@ARTICLE{Haldemann+2024,
       author = {{Haldemann}, Jonas and {Dorn}, Caroline and {Venturini}, Julia and {Alibert}, Yann and {Benz}, Willy},
        title = "{BICEPS: An improved characterization model for low- and intermediate-mass exoplanets}",
      journal = {\aap},
         year = 2024,
        month = jan,
       volume = {681},
          eid = {A96},
        pages = {A96},
          doi = {10.1051/0004-6361/202346965}
}

@ARTICLE{Adibekyan+2021,
       author = {{Adibekyan}, Vardan and {Dorn}, Caroline and {Sousa}, S{\'e}rgio G. and {Santos}, Nuno C. and {Bitsch}, Bertram and {Israelian}, Garik and {Mordasini}, Christoph and {Barros}, Susana C.~C. and {Delgado Mena}, Elisa and {Demangeon}, Olivier D.~S. and {Faria}, Jo{\~a}o P. and {Figueira}, Pedro and {Hakobyan}, Artur A. and {Oshagh}, Mahmoudreza and {Soares}, B{\'a}rbara M.~T.~B. and {Kunitomo}, Masanobu and {Takeda}, Yoichi and {Jofr{\'e}}, Emiliano and {Petrucci}, Romina and {Martioli}, Eder},
        title = "{A compositional link between rocky exoplanets and their host stars}",
      journal = {Science},
         year = 2021,
        month = oct,
       volume = {374},
       number = {6565},
        pages = {330-332},
          doi = {10.1126/science.abg8794},
archivePrefix = {arXiv},
       eprint = {2102.12444},
 primaryClass = {astro-ph.EP},
       adsurl = {https://ui.adsabs.harvard.edu/abs/2021Sci...374..330A}
}

@ARTICLE{marigo2017,
   author = {{Marigo}, P. and {Girardi}, L. and {Bressan}, A. and {Rosenfield}, P. and 
	{Aringer}, B. and {Chen}, Y. and {Dussin}, M. and {Nanni}, A. and 
	{Pastorelli}, G. and {Rodrigues}, T.~S. and {Trabucchi}, M. and 
	{Bladh}, S. and {Dalcanton}, J. and {Groenewegen}, M.~A.~T. and 
	{Montalb{\'a}n}, J. and {Wood}, P.~R.},
    title = "{A New Generation of PARSEC-COLIBRI Stellar Isochrones Including the TP-AGB Phase}",
  journal = {\apj},
archivePrefix = "arXiv",
   eprint = {1701.08510},
 primaryClass = "astro-ph.SR",
     year = 2017,
    month = jan,
   volume = 835,
      eid = {77},
    pages = {77},
      doi = {10.3847/1538-4357/835/1/77},
   adsurl = {http://adsabs.harvard.edu/abs/2017ApJ...835...77M}
}

@ARTICLE{bonfanti2015,
   author = {{Bonfanti}, A. and {Ortolani}, S. and {Piotto}, G. and {Nascimbeni}, V.
	},
    title = "{Revising the ages of planet-hosting stars}",
  journal = {\aap},
archivePrefix = "arXiv",
   eprint = {1411.4302},
 primaryClass = "astro-ph.SR",
     year = 2015,
    month = mar,
   volume = 575,
      eid = {A18},
    pages = {A18},
      doi = {10.1051/0004-6361/201424951},
   adsurl = {http://adsabs.harvard.edu/abs/2015A%26A...575A..18B}
}

@ARTICLE{bonfanti2016,
   author = {{Bonfanti}, A. and {Ortolani}, S. and {Nascimbeni}, V.},
    title = "{Age consistency between exoplanet hosts and field stars}",
  journal = {\aap},
archivePrefix = "arXiv",
   eprint = {1511.01744},
 primaryClass = "astro-ph.SR",
     year = 2016,
    month = jan,
   volume = 585,
      eid = {A5},
    pages = {A5},
      doi = {10.1051/0004-6361/201527297},
   adsurl = {http://adsabs.harvard.edu/abs/2016A%26A...585A...5B}
}

@ARTICLE{bonfanti2021,
       author = {{Bonfanti}, A. and {Delrez}, L. and {Hooton}, M.~J. and {Wilson}, T.~G. and {Fossati}, L. and {Alibert}, Y. and {Hoyer}, S. and {Mustill}, A.~J. and {Osborn}, H.~P. and {Adibekyan}, V. and {Gandolfi}, D. and {Salmon}, S. and {Sousa}, S.~G. and {Tuson}, A. and {Van Grootel}, V. and {Cabrera}, J. and {Nascimbeni}, V. and {Maxted}, P.~F.~L. and {Barros}, S.~C.~C. and {Billot}, N. and {Bonfils}, X. and {Borsato}, L. and {Broeg}, C. and {Davies}, M.~B. and {Deleuil}, M. and {Demangeon}, O.~D.~S. and {Fridlund}, M. and {Lacedelli}, G. and {Lendl}, M. and {Persson}, C. and {Santos}, N.~C. and {Scandariato}, G. and {Szab{\'o}}, Gy. M. and {Collier Cameron}, A. and {Udry}, S. and {Benz}, W. and {Beck}, M. and {Ehrenreich}, D. and {Fortier}, A. and {Isaak}, K.~G. and {Queloz}, D. and {Alonso}, R. and {Asquier}, J. and {Bandy}, T. and {B{\'a}rczy}, T. and {Barrado}, D. and {Barrag{\'a}n}, O. and {Baumjohann}, W. and {Beck}, T. and {Bekkelien}, A. and {Bergomi}, M. and {Brandeker}, A. and {Busch}, M. -D. and {Cessa}, V. and {Charnoz}, S. and {Chazelas}, B. and {Corral Van Damme}, C. and {Demory}, B. -O. and {Erikson}, A. and {Farinato}, J. and {Futyan}, D. and {Garcia Mu{\~n}oz}, A. and {Gillon}, M. and {Guedel}, M. and {Guterman}, P. and {Hasiba}, J. and {Heng}, K. and {Hernandez}, E. and {Kiss}, L. and {Kuntzer}, T. and {Laskar}, J. and {Lecavelier des Etangs}, A. and {Lovis}, C. and {Magrin}, D. and {Malvasio}, L. and {Marafatto}, L. and {Michaelis}, H. and {Munari}, M. and {Olofsson}, G. and {Ottacher}, H. and {Ottensamer}, R. and {Pagano}, I. and {Pall{\'e}}, E. and {Peter}, G. and {Piazza}, D. and {Piotto}, G. and {Pollacco}, D. and {Ragazzoni}, R. and {Rando}, N. and {Ratti}, F. and {Rauer}, H. and {Ribas}, I. and {Rieder}, M. and {Rohlfs}, R. and {Safa}, F. and {Salatti}, M. and {S{\'e}gransan}, D. and {Simon}, A.~E. and {Smith}, A.~M.~S. and {Sordet}, M. and {Steller}, M. and {Thomas}, N. and {Tschentscher}, M. and {Van Eylen}, V. and {Viotto}, V. and {Walter}, I. and {Walton}, N.~A. and {Wildi}, F. and {Wolter}, D.},
        title = "{CHEOPS observations of the HD 108236 planetary system: a fifth planet, improved ephemerides, and planetary radii}",
      journal = {\aap},
         year = 2021,
        month = feb,
       volume = {646},
          eid = {A157},
        pages = {A157},
          doi = {10.1051/0004-6361/202039608},
archivePrefix = {arXiv},
       eprint = {2101.00663},
 primaryClass = {astro-ph.EP},
       adsurl = {https://ui.adsabs.harvard.edu/abs/2021A&A...646A.157B}
}

@ARTICLE{Otegi+2020a,
       author = {{Otegi}, J.~F. and {Bouchy}, F. and {Helled}, R.},
        title = "{Revisited mass-radius relations for exoplanets below 120 M$_{{\ensuremath{\oplus}}}$}",
      journal = {\aap},
         year = 2020,
        month = feb,
       volume = {634},
          eid = {A43},
        pages = {A43},
          doi = {10.1051/0004-6361/201936482},
archivePrefix = {arXiv},
       eprint = {1911.04745},
 primaryClass = {astro-ph.EP},
       adsurl = {https://ui.adsabs.harvard.edu/abs/2020A&A...634A..43O}
}

@ARTICLE{Fulton+2017,
       author = {{Fulton}, Benjamin J. and {Petigura}, Erik A. and {Howard}, Andrew W. and {Isaacson}, Howard and {Marcy}, Geoffrey W. and {Cargile}, Phillip A. and {Hebb}, Leslie and {Weiss}, Lauren M. and {Johnson}, John Asher and {Morton}, Timothy D. and {Sinukoff}, Evan and {Crossfield}, Ian J.~M. and {Hirsch}, Lea A.},
        title = "{The California-Kepler Survey. III. A Gap in the Radius Distribution of Small Planets}",
      journal = {\aj},
         year = 2017,
        month = sep,
       volume = {154},
       number = {3},
          eid = {109},
        pages = {109},
          doi = {10.3847/1538-3881/aa80eb},
archivePrefix = {arXiv},
       eprint = {1703.10375},
 primaryClass = {astro-ph.EP},
       adsurl = {https://ui.adsabs.harvard.edu/abs/2017AJ....154..109F}
}

@ARTICLE{Owen+Wu2013,
       author = {{Owen}, James E. and {Wu}, Yanqin},
        title = "{Kepler Planets: A Tale of Evaporation}",
      journal = {\apj},
         year = 2013,
        month = oct,
       volume = {775},
       number = {2},
          eid = {105},
        pages = {105},
          doi = {10.1088/0004-637X/775/2/105},
archivePrefix = {arXiv},
       eprint = {1303.3899},
 primaryClass = {astro-ph.EP},
       adsurl = {https://ui.adsabs.harvard.edu/abs/2013ApJ...775..105O}
}

@ARTICLE{Jin+2014,
       author = {{Jin}, Sheng and {Mordasini}, Christoph and {Parmentier}, Vivien and {van Boekel}, Roy and {Henning}, Thomas and {Ji}, Jianghui},
        title = "{Planetary Population Synthesis Coupled with Atmospheric Escape: A Statistical View of Evaporation}",
      journal = {\apj},
         year = 2014,
        month = nov,
       volume = {795},
       number = {1},
          eid = {65},
        pages = {65},
          doi = {10.1088/0004-637X/795/1/65},
archivePrefix = {arXiv},
       eprint = {1409.2879},
 primaryClass = {astro-ph.EP},
       adsurl = {https://ui.adsabs.harvard.edu/abs/2014ApJ...795...65J}
}

@ARTICLE{Owen+Wu2017,
       author = {{Owen}, James E. and {Wu}, Yanqin},
        title = "{The Evaporation Valley in the Kepler Planets}",
      journal = {\apj},
         year = 2017,
        month = sep,
       volume = {847},
       number = {1},
          eid = {29},
        pages = {29},
          doi = {10.3847/1538-4357/aa890a},
archivePrefix = {arXiv},
       eprint = {1705.10810},
 primaryClass = {astro-ph.EP},
       adsurl = {https://ui.adsabs.harvard.edu/abs/2017ApJ...847...29O}
}

@ARTICLE{Jin+Mordasini2018,
       author = {{Jin}, Sheng and {Mordasini}, Christoph},
        title = "{Compositional Imprints in Density-Distance-Time: A Rocky Composition for Close-in Low-mass Exoplanets from the Location of the Valley of Evaporation}",
      journal = {\apj},
         year = 2018,
        month = feb,
       volume = {853},
       number = {2},
          eid = {163},
        pages = {163},
          doi = {10.3847/1538-4357/aa9f1e},
archivePrefix = {arXiv},
       eprint = {1706.00251},
 primaryClass = {astro-ph.EP},
       adsurl = {https://ui.adsabs.harvard.edu/abs/2018ApJ...853..163J}
}

@ARTICLE{Venturini+2020b,
       author = {{Venturini}, Julia and {Guilera}, Octavio M. and {Haldemann}, Jonas and {Ronco}, Mar{\'\i}a P. and {Mordasini}, Christoph},
        title = "{The nature of the radius valley. Hints from formation and evolution models}",
      journal = {\aap},
         year = 2020,
        month = nov,
       volume = {643},
          eid = {L1},
        pages = {L1},
          doi = {10.1051/0004-6361/202039141},
archivePrefix = {arXiv},
       eprint = {2008.05513},
 primaryClass = {astro-ph.EP},
       adsurl = {https://ui.adsabs.harvard.edu/abs/2020A&A...643L...1V}
}

@ARTICLE{Zeng+2019,
       author = {{Zeng}, Li and {Jacobsen}, Stein B. and {Sasselov}, Dimitar D. and {Petaev}, Michail I. and {Vanderburg}, Andrew and {Lopez-Morales}, Mercedes and {Perez-Mercader}, Juan and {Mattsson}, Thomas R. and {Li}, Gongjie and {Heising}, Matthew Z. and {Bonomo}, Aldo S. and {Damasso}, Mario and {Berger}, Travis A. and {Cao}, Hao and {Levi}, Amit and {Wordsworth}, Robin D.},
        title = "{Growth model interpretation of planet size distribution}",
      journal = {Proceedings of the National Academy of Science},
         year = 2019,
        month = may,
       volume = {116},
       number = {20},
        pages = {9723-9728},
          doi = {10.1073/pnas.1812905116},
archivePrefix = {arXiv},
       eprint = {1906.04253},
 primaryClass = {astro-ph.EP},
       adsurl = {https://ui.adsabs.harvard.edu/abs/2019PNAS..116.9723Z}
}

@ARTICLE{Izidoro+2022,
       author = {{Izidoro}, Andr{\'e} and {Schlichting}, Hilke E. and {Isella}, Andrea and {Dasgupta}, Rajdeep and {Zimmermann}, Christian and {Bitsch}, Bertram},
        title = "{The Exoplanet Radius Valley from Gas-driven Planet Migration and Breaking of Resonant Chains}",
      journal = {\apjl},
         year = 2022,
        month = nov,
       volume = {939},
       number = {2},
          eid = {L19},
        pages = {L19},
          doi = {10.3847/2041-8213/ac990d},
archivePrefix = {arXiv},
       eprint = {2210.05595},
 primaryClass = {astro-ph.EP},
       adsurl = {https://ui.adsabs.harvard.edu/abs/2022ApJ...939L..19I}
}

@ARTICLE{Tanaka+2002,
       author = {{Tanaka}, Hidekazu and {Takeuchi}, Taku and {Ward}, William R.},
        title = "{Three-Dimensional Interaction between a Planet and an Isothermal Gaseous Disk. I. Corotation and Lindblad Torques and Planet Migration}",
      journal = {\apj},
         year = 2002,
        month = feb,
       volume = {565},
       number = {2},
        pages = {1257-1274},
          doi = {10.1086/324713},
       adsurl = {https://ui.adsabs.harvard.edu/abs/2002ApJ...565.1257T}
}

@ARTICLE{Giacalone+2021,
       author = {{Giacalone}, Steven and {Dressing}, Courtney D. and {Jensen}, Eric L.~N. and {Collins}, Karen A. and {Ricker}, George R. and {Vanderspek}, Roland and {Seager}, S. and {Winn}, Joshua N. and {Jenkins}, Jon M. and {Barclay}, Thomas and {Barkaoui}, Khalid and {Cadieux}, Charles and {Charbonneau}, David and {Collins}, Kevin I. and {Conti}, Dennis M. and {Doyon}, Ren{\'e} and {Evans}, Phil and {Ghachoui}, Mourad and {Gillon}, Micha{\"e}l and {Guerrero}, Natalia M. and {Hart}, Rhodes and {Jehin}, Emmanu{\"e}l and {Kielkopf}, John F. and {McLean}, Brian and {Murgas}, Felipe and {Palle}, Enric and {Parviainen}, Hannu and {Pozuelos}, Francisco J. and {Relles}, Howard M. and {Shporer}, Avi and {Socia}, Quentin and {Stockdale}, Chris and {Tan}, Thiam-Guan and {Torres}, Guillermo and {Twicken}, Joseph D. and {Waalkes}, William C. and {Waite}, Ian A.},
        title = "{Vetting of 384 TESS Objects of Interest with TRICERATOPS and Statistical Validation of 12 Planet Candidates}",
      journal = {\aj},
         year = 2021,
        month = jan,
       volume = {161},
       number = {1},
          eid = {24},
        pages = {24},
          doi = {10.3847/1538-3881/abc6af},
archivePrefix = {arXiv},
       eprint = {2002.00691},
 primaryClass = {astro-ph.EP},
       adsurl = {https://ui.adsabs.harvard.edu/abs/2021AJ....161...24G}
}

@ARTICLE{Benz+2021,
       author = {{Benz}, W. and {Broeg}, C. and {Fortier}, A. and {Rando}, N. and {Beck}, T. and {Beck}, M. and {Queloz}, D. and {Ehrenreich}, D. and {Maxted}, P.~F.~L. and {Isaak}, K.~G. and {Billot}, N. and {Alibert}, Y. and {Alonso}, R. and {Ant{\'o}nio}, C. and {Asquier}, J. and {Bandy}, T. and {B{\'a}rczy}, T. and {Barrado}, D. and {Barros}, S.~C.~C. and {Baumjohann}, W. and {Bekkelien}, A. and {Bergomi}, M. and {Biondi}, F. and {Bonfils}, X. and {Borsato}, L. and {Brandeker}, A. and {Busch}, M. -D. and {Cabrera}, J. and {Cessa}, V. and {Charnoz}, S. and {Chazelas}, B. and {Collier Cameron}, A. and {Corral Van Damme}, C. and {Cortes}, D. and {Davies}, M.~B. and {Deleuil}, M. and {Deline}, A. and {Delrez}, L. and {Demangeon}, O. and {Demory}, B.~O. and {Erikson}, A. and {Farinato}, J. and {Fossati}, L. and {Fridlund}, M. and {Futyan}, D. and {Gandolfi}, D. and {Garcia Munoz}, A. and {Gillon}, M. and {Guterman}, P. and {Gutierrez}, A. and {Hasiba}, J. and {Heng}, K. and {Hernandez}, E. and {Hoyer}, S. and {Kiss}, L.~L. and {Kovacs}, Z. and {Kuntzer}, T. and {Laskar}, J. and {Lecavelier des Etangs}, A. and {Lendl}, M. and {L{\'o}pez}, A. and {Lora}, I. and {Lovis}, C. and {L{\"u}ftinger}, T. and {Magrin}, D. and {Malvasio}, L. and {Marafatto}, L. and {Michaelis}, H. and {de Miguel}, D. and {Modrego}, D. and {Munari}, M. and {Nascimbeni}, V. and {Olofsson}, G. and {Ottacher}, H. and {Ottensamer}, R. and {Pagano}, I. and {Palacios}, R. and {Pall{\'e}}, E. and {Peter}, G. and {Piazza}, D. and {Piotto}, G. and {Pizarro}, A. and {Pollaco}, D. and {Ragazzoni}, R. and {Ratti}, F. and {Rauer}, H. and {Ribas}, I. and {Rieder}, M. and {Rohlfs}, R. and {Safa}, F. and {Salatti}, M. and {Santos}, N.~C. and {Scandariato}, G. and {S{\'e}gransan}, D. and {Simon}, A.~E. and {Smith}, A.~M.~S. and {Sordet}, M. and {Sousa}, S.~G. and {Steller}, M. and {Szab{\'o}}, G.~M. and {Szoke}, J. and {Thomas}, N. and {Tschentscher}, M. and {Udry}, S. and {Van Grootel}, V. and {Viotto}, V. and {Walter}, I. and {Walton}, N.~A. and {Wildi}, F. and {Wolter}, D.},
        title = "{The CHEOPS mission}",
      journal = {Exp. Astron.},
         year = 2021,
        month = feb,
       volume = {51},
       number = {1},
        pages = {109-151},
          doi = {10.1007/s10686-020-09679-4},
archivePrefix = {arXiv},
       eprint = {2009.11633},
 primaryClass = {astro-ph.IM},
       adsurl = {https://ui.adsabs.harvard.edu/abs/2021ExA....51..109B}
}

@ARTICLE{Ricker+2015,
       author = {{Ricker}, George R. and {Winn}, Joshua N. and {Vanderspek}, Roland and {Latham}, David W. and {Bakos}, G{\'a}sp{\'a}r {\'A}. and {Bean}, Jacob L. and {Berta-Thompson}, Zachory K. and {Brown}, Timothy M. and {Buchhave}, Lars and {Butler}, Nathaniel R. and {Butler}, R. Paul and {Chaplin}, William J. and {Charbonneau}, David and {Christensen-Dalsgaard}, J{\o}rgen and {Clampin}, Mark and {Deming}, Drake and {Doty}, John and {De Lee}, Nathan and {Dressing}, Courtney and {Dunham}, Edward W. and {Endl}, Michael and {Fressin}, Francois and {Ge}, Jian and {Henning}, Thomas and {Holman}, Matthew J. and {Howard}, Andrew W. and {Ida}, Shigeru and {Jenkins}, Jon M. and {Jernigan}, Garrett and {Johnson}, John Asher and {Kaltenegger}, Lisa and {Kawai}, Nobuyuki and {Kjeldsen}, Hans and {Laughlin}, Gregory and {Levine}, Alan M. and {Lin}, Douglas and {Lissauer}, Jack J. and {MacQueen}, Phillip and {Marcy}, Geoffrey and {McCullough}, Peter R. and {Morton}, Timothy D. and {Narita}, Norio and {Paegert}, Martin and {Palle}, Enric and {Pepe}, Francesco and {Pepper}, Joshua and {Quirrenbach}, Andreas and {Rinehart}, Stephen A. and {Sasselov}, Dimitar and {Sato}, Bun'ei and {Seager}, Sara and {Sozzetti}, Alessandro and {Stassun}, Keivan G. and {Sullivan}, Peter and {Szentgyorgyi}, Andrew and {Torres}, Guillermo and {Udry}, Stephane and {Villasenor}, Joel},
        title = "{Transiting Exoplanet Survey Satellite (TESS)}",
      journal = {JATIS},
         year = 2015,
        month = jan,
       volume = {1},
          eid = {014003},
        pages = {014003},
          doi = {10.1117/1.JATIS.1.1.014003},
       adsurl = {https://ui.adsabs.harvard.edu/abs/2015JATIS...1a4003R}
}

@ARTICLE{Pepe+2021,
       author = {{Pepe}, F. and {Cristiani}, S. and {Rebolo}, R. and {Santos}, N.~C. and {Dekker}, H. and {Cabral}, A. and {Di Marcantonio}, P. and {Figueira}, P. and {Lo Curto}, G. and {Lovis}, C. and {Mayor}, M. and {M{\'e}gevand}, D. and {Molaro}, P. and {Riva}, M. and {Zapatero Osorio}, M.~R. and {Amate}, M. and {Manescau}, A. and {Pasquini}, L. and {Zerbi}, F.~M. and {Adibekyan}, V. and {Abreu}, M. and {Affolter}, M. and {Alibert}, Y. and {Aliverti}, M. and {Allart}, R. and {Allende Prieto}, C. and {{\'A}lvarez}, D. and {Alves}, D. and {Avila}, G. and {Baldini}, V. and {Bandy}, T. and {Barros}, S.~C.~C. and {Benz}, W. and {Bianco}, A. and {Borsa}, F. and {Bourrier}, V. and {Bouchy}, F. and {Broeg}, C. and {Calderone}, G. and {Cirami}, R. and {Coelho}, J. and {Conconi}, P. and {Coretti}, I. and {Cumani}, C. and {Cupani}, G. and {D'Odorico}, V. and {Damasso}, M. and {Deiries}, S. and {Delabre}, B. and {Demangeon}, O.~D.~S. and {Dumusque}, X. and {Ehrenreich}, D. and {Faria}, J.~P. and {Fragoso}, A. and {Genolet}, L. and {Genoni}, M. and {G{\'e}nova Santos}, R. and {Gonz{\'a}lez Hern{\'a}ndez}, J.~I. and {Hughes}, I. and {Iwert}, O. and {Kerber}, F. and {Knudstrup}, J. and {Landoni}, M. and {Lavie}, B. and {Lillo-Box}, J. and {Lizon}, J. -L. and {Maire}, C. and {Martins}, C.~J.~A.~P. and {Mehner}, A. and {Micela}, G. and {Modigliani}, A. and {Monteiro}, M.~A. and {Monteiro}, M.~J.~P.~F.~G. and {Moschetti}, M. and {Murphy}, M.~T. and {Nunes}, N. and {Oggioni}, L. and {Oliveira}, A. and {Oshagh}, M. and {Pall{\'e}}, E. and {Pariani}, G. and {Poretti}, E. and {Rasilla}, J.~L. and {Rebord{\~a}o}, J. and {Redaelli}, E.~M. and {Santana Tschudi}, S. and {Santin}, P. and {Santos}, P. and {S{\'e}gransan}, D. and {Schmidt}, T.~M. and {Segovia}, A. and {Sosnowska}, D. and {Sozzetti}, A. and {Sousa}, S.~G. and {Span{\`o}}, P. and {Su{\'a}rez Mascare{\~n}o}, A. and {Tabernero}, H. and {Tenegi}, F. and {Udry}, S. and {Zanutta}, A.},
        title = "{ESPRESSO at VLT. On-sky performance and first results}",
      journal = {\aap},
         year = 2021,
        month = jan,
       volume = {645},
          eid = {A96},
        pages = {A96},
          doi = {10.1051/0004-6361/202038306},
archivePrefix = {arXiv},
       eprint = {2010.00316},
 primaryClass = {astro-ph.IM},
       adsurl = {https://ui.adsabs.harvard.edu/abs/2021A&A...645A..96P}
}

@ARTICLE{Seager+2007,
       author = {{Seager}, S. and {Kuchner}, M. and {Hier-Majumder}, C.~A. and {Militzer}, B.},
        title = "{Mass-Radius Relationships for Solid Exoplanets}",
      journal = {\apj},
         year = 2007,
        month = nov,
       volume = {669},
       number = {2},
        pages = {1279-1297},
          doi = {10.1086/521346},
archivePrefix = {arXiv},
       eprint = {0707.2895},
 primaryClass = {astro-ph},
       adsurl = {https://ui.adsabs.harvard.edu/abs/2007ApJ...669.1279S}
}

@ARTICLE{Rogers+Seager2010a,
       author = {{Rogers}, L.~A. and {Seager}, S.},
        title = "{A Framework for Quantifying the Degeneracies of Exoplanet Interior Compositions}",
      journal = {\apj},
         year = 2010,
        month = apr,
       volume = {712},
       number = {2},
        pages = {974-991},
          doi = {10.1088/0004-637X/712/2/974},
archivePrefix = {arXiv},
       eprint = {0912.3288},
 primaryClass = {astro-ph.EP},
       adsurl = {https://ui.adsabs.harvard.edu/abs/2010ApJ...712..974R}
}

@ARTICLE{Dorn+Lichtenberg2021,
       author = {{Dorn}, Caroline and {Lichtenberg}, Tim},
        title = "{Hidden Water in Magma Ocean Exoplanets}",
      journal = {\apjl},
         year = 2021,
        month = nov,
       volume = {922},
       number = {1},
          eid = {L4},
        pages = {L4},
          doi = {10.3847/2041-8213/ac33af},
archivePrefix = {arXiv},
       eprint = {2110.15069},
 primaryClass = {astro-ph.EP},
       adsurl = {https://ui.adsabs.harvard.edu/abs/2021ApJ...922L...4D}
}

@ARTICLE{johnstone2021,
       author = {{Johnstone}, C.~P. and {Bartel}, M. and {G{\"u}del}, M.},
        title = "{The active lives of stars: A complete description of the rotation and XUV evolution of F, G, K, and M dwarfs}",
      journal = {\aap},
         year = 2021,
        month = may,
       volume = {649},
          eid = {A96},
        pages = {A96},
          doi = {10.1051/0004-6361/202038407},
archivePrefix = {arXiv},
       eprint = {2009.07695},
 primaryClass = {astro-ph.SR},
       adsurl = {https://ui.adsabs.harvard.edu/abs/2021A&A...649A..96J}
}

@ARTICLE{kubyshkina2024,
       author = {{Kubyshkina}, Daria and {Fossati}, Luca and {Erkaev}, Nikolai V.},
        title = "{Precise photoionisation treatment and hydrodynamic effects in atmospheric modelling of warm and hot Neptunes}",
      journal = {\aap},
         year = 2024,
        month = apr,
       volume = {684},
          eid = {A26},
        pages = {A26},
          doi = {10.1051/0004-6361/202347837},
archivePrefix = {arXiv},
       eprint = {2312.07236},
 primaryClass = {astro-ph.EP},
       adsurl = {https://ui.adsabs.harvard.edu/abs/2024A&A...684A..26K}
}

@ARTICLE{spada2013,
       author = {{Spada}, F. and {Demarque}, P. and {Kim}, Y. -C. and {Sills}, A.},
        title = "{The Radius Discrepancy in Low-mass Stars: Single versus Binaries}",
      journal = {\apj},
         year = 2013,
        month = oct,
       volume = {776},
       number = {2},
          eid = {87},
        pages = {87},
          doi = {10.1088/0004-637X/776/2/87},
archivePrefix = {arXiv},
       eprint = {1308.5558},
 primaryClass = {astro-ph.SR},
       adsurl = {https://ui.adsabs.harvard.edu/abs/2013ApJ...776...87S}
}

@ARTICLE{Kite+2020,
       author = {{Kite}, Edwin S. and {Fegley}, Bruce, Jr. and {Schaefer}, Laura and {Ford}, Eric B.},
        title = "{Atmosphere Origins for Exoplanet Sub-Neptunes}",
      journal = {\apj},
         year = 2020,
        month = mar,
       volume = {891},
       number = {2},
          eid = {111},
        pages = {111},
          doi = {10.3847/1538-4357/ab6ffb},
archivePrefix = {arXiv},
       eprint = {2001.09269},
 primaryClass = {astro-ph.EP},
       adsurl = {https://ui.adsabs.harvard.edu/abs/2020ApJ...891..111K}
}

@ARTICLE{Kubyshkina+2022,
       author = {{Kubyshkina}, D. and {Fossati}, L.},
        title = "{The mass-radius relation of intermediate-mass planets outlined by hydrodynamic escape and thermal evolution}",
      journal = {\aap},
         year = 2022,
        month = dec,
       volume = {668},
          eid = {A178},
        pages = {A178},
          doi = {10.1051/0004-6361/202244916},
archivePrefix = {arXiv},
       eprint = {2211.10166},
 primaryClass = {astro-ph.EP},
       adsurl = {https://ui.adsabs.harvard.edu/abs/2022A&A...668A.178K}
}

@ARTICLE{Lendl+2020,
       author = {{Lendl}, M. and {Csizmadia}, Sz. and {Deline}, A. and {Fossati}, L. and {Kitzmann}, D. and {Heng}, K. and {Hoyer}, S. and {Salmon}, S. and {Benz}, W. and {Broeg}, C. and {Ehrenreich}, D. and {Fortier}, A. and {Queloz}, D. and {Bonfanti}, A. and {Brandeker}, A. and {Collier Cameron}, A. and {Delrez}, L. and {Garcia Mu{\~n}oz}, A. and {Hooton}, M.~J. and {Maxted}, P.~F.~L. and {Morris}, B.~M. and {Van Grootel}, V. and {Wilson}, T.~G. and {Alibert}, Y. and {Alonso}, R. and {Asquier}, J. and {Bandy}, T. and {B{\'a}rczy}, T. and {Barrado}, D. and {Barros}, S.~C.~C. and {Baumjohann}, W. and {Beck}, M. and {Beck}, T. and {Bekkelien}, A. and {Bergomi}, M. and {Billot}, N. and {Biondi}, F. and {Bonfils}, X. and {Bourrier}, V. and {Busch}, M. -D. and {Cabrera}, J. and {Cessa}, V. and {Charnoz}, S. and {Chazelas}, B. and {Corral Van Damme}, C. and {Davies}, M.~B. and {Deleuil}, M. and {Demangeon}, O.~D.~S. and {Demory}, B. -O. and {Erikson}, A. and {Farinato}, J. and {Fridlund}, M. and {Futyan}, D. and {Gandolfi}, D. and {Gillon}, M. and {Guterman}, P. and {Hasiba}, J. and {Hernandez}, E. and {Isaak}, K.~G. and {Kiss}, L. and {Kuntzer}, T. and {Lecavelier des Etangs}, A. and {L{\"u}ftinger}, T. and {Laskar}, J. and {Lovis}, C. and {Magrin}, D. and {Malvasio}, L. and {Marafatto}, L. and {Michaelis}, H. and {Munari}, M. and {Nascimbeni}, V. and {Olofsson}, G. and {Ottacher}, H. and {Ottensamer}, R. and {Pagano}, I. and {Pall{\'e}}, E. and {Peter}, G. and {Piazza}, D. and {Piotto}, G. and {Pollacco}, D. and {Ratti}, F. and {Rauer}, H. and {Ragazzoni}, R. and {Rando}, N. and {Ribas}, I. and {Rieder}, M. and {Rohlfs}, R. and {Safa}, F. and {Santos}, N.~C. and {Scandariato}, G. and {S{\'e}gransan}, D. and {Simon}, A.~E. and {Singh}, V. and {Smith}, A.~M.~S. and {Sordet}, M. and {Sousa}, S.~G. and {Steller}, M. and {Szab{\'o}}, Gy. M. and {Thomas}, N. and {Tschentscher}, M. and {Udry}, S. and {Viotto}, V. and {Walter}, I. and {Walton}, N.~A. and {Wildi}, F. and {Wolter}, D.},
        title = "{The hot dayside and asymmetric transit of WASP-189 b seen by CHEOPS}",
      journal = {\aap},
         year = 2020,
        month = nov,
       volume = {643},
          eid = {A94},
        pages = {A94},
          doi = {10.1051/0004-6361/202038677},
archivePrefix = {arXiv},
       eprint = {2009.13403},
 primaryClass = {astro-ph.EP},
       adsurl = {https://ui.adsabs.harvard.edu/abs/2020A&A...643A..94L}
}

@ARTICLE{Venturini+2020a,
       author = {{Venturini}, Julia and {Guilera}, Octavio Miguel and {Ronco}, Mar{\'\i}a Paula and {Mordasini}, Christoph},
        title = "{Most super-Earths formed by dry pebble accretion are less massive than 5 Earth masses}",
      journal = {\aap},
         year = 2020,
        month = dec,
       volume = {644},
          eid = {A174},
        pages = {A174},
          doi = {10.1051/0004-6361/202039140},
archivePrefix = {arXiv},
       eprint = {2008.05497},
 primaryClass = {astro-ph.EP},
       adsurl = {https://ui.adsabs.harvard.edu/abs/2020A&A...644A.174V}
}

@ARTICLE{Venturini+2024,
       author = {{Venturini}, J. and {Ronco}, M.~P. and {Guilera}, O.~M. and {Haldemann}, J. and {Mordasini}, C. and {Miller Bertolami}, M.},
        title = "{A fading radius valley towards M dwarfs, a persistent density valley across stellar types}",
      journal = {\aap},
         year = 2024,
        month = jun,
       volume = {686},
          eid = {L9},
        pages = {L9},
          doi = {10.1051/0004-6361/202349088},
archivePrefix = {arXiv},
       eprint = {2404.01967},
 primaryClass = {astro-ph.EP},
       adsurl = {https://ui.adsabs.harvard.edu/abs/2024A&A...686L...9V}
}

@ARTICLE{Bonfanti+2024,
       author = {{Bonfanti}, A. and {Brady}, M. and {Wilson}, T.~G. and {Venturini}, J. and {Egger}, J.~A. and {Brandeker}, A. and {Sousa}, S.~G. and {Lendl}, M. and {Simon}, A.~E. and {Queloz}, D. and {Olofsson}, G. and {Adibekyan}, V. and {Alibert}, Y. and {Fossati}, L. and {Hooton}, M.~J. and {Kubyshkina}, D. and {Luque}, R. and {Murgas}, F. and {Mustill}, A.~J. and {Santos}, N.~C. and {Van Grootel}, V. and {Alonso}, R. and {Asquier}, J. and {Bandy}, T. and {B{\'a}rczy}, T. and {Barrado Navascues}, D. and {Barros}, S.~C.~C. and {Baumjohann}, W. and {Bean}, J. and {Beck}, M. and {Beck}, T. and {Benz}, W. and {Bergomi}, M. and {Billot}, N. and {Borsato}, L. and {Broeg}, C. and {Collier Cameron}, A. and {Csizmadia}, Sz. and {Cubillos}, P.~E. and {Davies}, M.~B. and {Deleuil}, M. and {Deline}, A. and {Delrez}, L. and {Demangeon}, O.~D.~S. and {Demory}, B. -O. and {Ehrenreich}, D. and {Erikson}, A. and {Fortier}, A. and {Fridlund}, M. and {Gandolfi}, D. and {Gillon}, M. and {G{\"u}del}, M. and {G{\"u}nther}, M.~N. and {Heitzmann}, A. and {Helling}, Ch. and {Hoyer}, S. and {Isaak}, K.~G. and {Kasper}, D. and {Kiss}, L.~L. and {Lam}, K.~W.~F. and {Laskar}, J. and {Lecavelier des Etangs}, A. and {Magrin}, D. and {Maxted}, P.~F.~L. and {Mordasini}, C. and {Nascimbeni}, V. and {Ottensamer}, R. and {Pagano}, I. and {Pall{\'e}}, E. and {Peter}, G. and {Piotto}, G. and {Pollacco}, D. and {Ragazzoni}, R. and {Rando}, N. and {Rauer}, H. and {Ribas}, I. and {Scandariato}, G. and {S{\'e}gransan}, D. and {Seifahrt}, A. and {Smith}, A.~M.~S. and {Stalport}, M. and {Stef{\'a}nsson}, G. and {Steinberger}, M. and {St{\"u}rmer}, J. and {Szab{\'o}}, Gy. M. and {Thomas}, N. and {Udry}, S. and {Villaver}, E. and {Walton}, N.~A. and {Westerdorff}, K. and {Zingales}, T.},
        title = "{Characterising TOI-732 b and c: New insights into the M-dwarf radius and density valley}",
      journal = {\aap},
         year = 2024,
        month = feb,
       volume = {682},
          eid = {A66},
        pages = {A66},
          doi = {10.1051/0004-6361/202348180},
archivePrefix = {arXiv},
       eprint = {2311.12577},
 primaryClass = {astro-ph.EP},
       adsurl = {https://ui.adsabs.harvard.edu/abs/2024A&A...682A..66B}
}

@ARTICLE{Benneke+2024,
       author = {{Benneke}, Bj{\"o}rn and {Roy}, Pierre-Alexis and {Coulombe}, Louis-Philippe and {Radica}, Michael and {Piaulet}, Caroline and {Ahrer}, Eva-Maria and {Pierrehumbert}, Raymond and {Krissansen-Totton}, Joshua and {Schlichting}, Hilke E. and {Hu}, Renyu and {Yang}, Jeehyun and {Christie}, Duncan and {Thorngren}, Daniel and {Young}, Edward D. and {Pelletier}, Stefan and {Knutson}, Heather A. and {Miguel}, Yamila and {Evans-Soma}, Thomas M. and {Dorn}, Caroline and {Gagnebin}, Anna and {Fortney}, Jonathan J. and {Komacek}, Thaddeus and {MacDonald}, Ryan and {Raul}, Eshan and {Cloutier}, Ryan and {Acuna}, Lorena and {Lafreni{\`e}re}, David and {Cadieux}, Charles and {Doyon}, Ren{\'e} and {Welbanks}, Luis and {Allart}, Romain},
        title = "{JWST Reveals CH$_4$, CO$_2$, and H$_2$O in a Metal-rich Miscible Atmosphere on a Two-Earth-Radius Exoplanet}",
      journal = {arXiv e-prints},
         year = 2024,
        month = mar,
          eid = {arXiv:2403.03325},
        pages = {arXiv:2403.03325},
          doi = {10.48550/arXiv.2403.03325},
archivePrefix = {arXiv},
       eprint = {2403.03325},
 primaryClass = {astro-ph.EP},
       adsurl = {https://ui.adsabs.harvard.edu/abs/2024arXiv240303325B}
}

@ARTICLE{Holmberg+Madhusudhan2024,
       author = {{Holmberg}, M{\r{a}}ns and {Madhusudhan}, Nikku},
        title = "{Possible Hycean conditions in the sub-Neptune TOI-270 d}",
      journal = {\aap},
         year = 2024,
        month = mar,
       volume = {683},
          eid = {L2},
        pages = {L2},
          doi = {10.1051/0004-6361/202348238},
archivePrefix = {arXiv},
       eprint = {2403.03244},
 primaryClass = {astro-ph.EP},
       adsurl = {https://ui.adsabs.harvard.edu/abs/2024A&A...683L...2H}
}

@ARTICLE{Madhusudhan+2023,
       author = {{Madhusudhan}, Nikku and {Sarkar}, Subhajit and {Constantinou}, Savvas and {Holmberg}, M{\r{a}}ns and {Piette}, Anjali A.~A. and {Moses}, Julianne I.},
        title = "{Carbon-bearing Molecules in a Possible Hycean Atmosphere}",
      journal = {\apjl},
         year = 2023,
        month = oct,
       volume = {956},
       number = {1},
          eid = {L13},
        pages = {L13},
          doi = {10.3847/2041-8213/acf577},
archivePrefix = {arXiv},
       eprint = {2309.05566},
 primaryClass = {astro-ph.EP},
       adsurl = {https://ui.adsabs.harvard.edu/abs/2023ApJ...956L..13M}
}

@ARTICLE{Fortier+2024,
       author = {{Fortier}, A. and {Simon}, A.~E. and {Broeg}, C. and {Olofsson}, G. and {Deline}, A. and {Wilson}, T.~G. and {Maxted}, P.~F.~L. and {Brandeker}, A. and {Collier Cameron}, A. and {Beck}, M. and {Bekkelien}, A. and {Billot}, N. and {Bonfanti}, A. and {Bruno}, G. and {Cabrera}, J. and {Delrez}, L. and {Demory}, B. -O. and {Futyan}, D. and {Flor{\'e}n}, H. -G. and {G{\"u}nther}, M.~N. and {Heitzmann}, A. and {Hoyer}, S. and {Isaak}, K.~G. and {Sousa}, S.~G. and {Stalport}, M. and {Turin}, A. and {Verhoeve}, P. and {Akinsanmi}, B. and {Alibert}, Y. and {Alonso}, R. and {B{\'a}nhidi}, D. and {B{\'a}rczy}, T. and {Barrado}, D. and {Barros}, S.~C.~C. and {Baumjohann}, W. and {Baycroft}, T. and {Beck}, T. and {Benz}, W. and {B{\'\i}r{\'o}}, B.~I. and {B{\'o}di}, A. and {Bonfils}, X. and {Borsato}, L. and {Charnoz}, S. and {Cseh}, B. and {Csizmadia}, Sz. and {Cs{\'a}nyi}, I. and {Cubillos}, P.~E. and {Davies}, M.~B. and {Davis}, Y.~T. and {Deleuil}, M. and {Demangeon}, O.~D.~S. and {Derekas}, A. and {Dransfield}, G. and {Ducrot}, E. and {Ehrenreich}, D. and {Erikson}, A. and {Fari{\~n}a}, C. and {Fossati}, L. and {Fridlund}, M. and {Gandolfi}, D. and {Garai}, Z. and {Garcia}, L. and {Gillon}, M. and {G{\'o}mez Maqueo Chew}, Y. and {G{\'o}mez-Mu{\~n}oz}, M.~A. and {Granata}, V. and {G{\"u}del}, M. and {Guterman}, P. and {Heged{\"u}s}, T. and {Helling}, Ch. and {Jehin}, E. and {Kalup}, Cs. and {Kilkenny}, D. and {Kiss}, L.~L. and {Kriskovics}, L. and {Lam}, K.~W.~F. and {Laskar}, J. and {Lecavelier des Etangs}, A. and {Lendl}, M. and {Lopez Pina}, A. and {Luntzer}, A. and {Magrin}, D. and {Miller}, N.~J. and {Modrego Contreras}, D. and {Mordasini}, C. and {Munari}, M. and {Murray}, C.~A. and {Nascimbeni}, V. and {Ottacher}, H. and {Ottensamer}, R. and {Pagano}, I. and {P{\'a}l}, A. and {Pall{\'e}}, E. and {Pasetti}, A. and {Pedersen}, P.~P. and {Peter}, G. and {Petrucci}, R. and {Piotto}, G. and {Pizarro-Rubio}, A. and {Pollacco}, D. and {Pribulla}, T. and {Queloz}, D. and {Ragazzoni}, R. and {Rando}, N. and {Rauer}, H. and {Ribas}, I. and {Sabin}, L. and {Santos}, N.~C. and {Scandariato}, G. and {Schanche}, N. and {Schroffenegger}, U. and {Scutt}, O.~J. and {Sebastian}, D. and {S{\'e}gransan}, D. and {Seli}, B. and {Smith}, A.~M.~S. and {Southworth}, R. and {Standing}, M.~R. and {Szab{\'o}}, Gy. M. and {Szak{\'a}ts}, R. and {Thomas}, N. and {Timmermans}, M. and {Triaud}, A.~H.~M.~J. and {Udry}, S. and {Van Grootel}, V. and {Venturini}, J. and {Villaver}, E. and {Vink{\'o}}, J. and {Walton}, N.~A. and {Wells}, R. and {Wolter}, D.},
        title = "{CHEOPS in-flight performance. A comprehensive look at the first 3.5 yr of operations}",
      journal = {\aap},
         year = 2024,
        month = jul,
       volume = {687},
          eid = {A302},
        pages = {A302},
          doi = {10.1051/0004-6361/202348576},
archivePrefix = {arXiv},
       eprint = {2406.01716},
 primaryClass = {astro-ph.IM},
       adsurl = {https://ui.adsabs.harvard.edu/abs/2024A&A...687A.302F}
}

@ARTICLE{Egger+2024,
       author = {{Egger}, J.~A. and {Osborn}, H.~P. and {Kubyshkina}, D. and {Mordasini}, C. and {Alibert}, Y. and {G{\"u}nther}, M.~N. and {Lendl}, M. and {Brandeker}, A. and {Heitzmann}, A. and {Leleu}, A. and {Damasso}, M. and {Bonfanti}, A. and {Wilson}, T.~G. and {Sousa}, S.~G. and {Haldemann}, J. and {Delrez}, L. and {Hooton}, M.~J. and {Zingales}, T. and {Luque}, R. and {Alonso}, R. and {Asquier}, J. and {B{\'a}rczy}, T. and {Navascues}, D. Barrado and {Barros}, S.~C.~C. and {Baumjohann}, W. and {Benz}, W. and {Billot}, N. and {Borsato}, L. and {Broeg}, C. and {Buder}, M. and {Castro-Gonz{\'a}lez}, A. and {Cameron}, A. Collier and {Correia}, A.~C.~M. and {Cortes}, D. and {Csizmadia}, Sz. and {Cubillos}, P.~E. and {Davies}, M.~B. and {Deleuil}, M. and {Deline}, A. and {Demangeon}, O.~D.~S. and {Demory}, B. -O. and {Derekas}, A. and {Edwards}, B. and {Ehrenreich}, D. and {Erikson}, A. and {Fortier}, A. and {Fossati}, L. and {Fridlund}, M. and {Gandolfi}, D. and {Gazeas}, K. and {Gillon}, M. and {G{\"u}del}, M. and {Helling}, Ch. and {Isaak}, K.~G. and {Kiss}, L.~L. and {Korth}, J. and {Lam}, K.~W.~F. and {Laskar}, J. and {Lavie}, B. and {des Etangs}, A. Lecavelier and {Lovis}, C. and {Luntzer}, A. and {Magrin}, D. and {Maxted}, P.~F.~L. and {Mer{\'\i}n}, B. and {Munari}, M. and {Nascimbeni}, V. and {Olofsson}, G. and {Ottensamer}, R. and {Pagano}, I. and {Pall{\'e}}, E. and {Peter}, G. and {Piazza}, D. and {Piotto}, G. and {Pollacco}, D. and {Queloz}, D. and {Ragazzoni}, R. and {Rando}, N. and {Rauer}, H. and {Ribas}, I. and {Rodrigues}, J. and {Santos}, N.~C. and {Scandariato}, G. and {S{\'e}gransan}, D. and {Simon}, A.~E. and {Smith}, A.~M.~S. and {Stalport}, M. and {Sulis}, S. and {Szab{\'o}}, Gy. M. and {Udry}, S. and {Van Grootel}, V. and {Venturini}, J. and {Villaver}, E. and {Walton}, N.~A.},
        title = "{Unveiling the internal structure and formation history of the three planets transiting HIP 29442 (TOI-469) with CHEOPS}",
      journal = {\aap},
         year = 2024,
        month = aug,
       volume = {688},
          eid = {A223},
        pages = {A223},
          doi = {10.1051/0004-6361/202450472},
archivePrefix = {arXiv},
       eprint = {2406.18653},
 primaryClass = {astro-ph.EP},
       adsurl = {https://ui.adsabs.harvard.edu/abs/2024A&A...688A.223E}
}

@ARTICLE{Egger+2025,
       author = {{Egger}, J.~A. and {Kubyshkina}, D. and {Alibert}, Y. and {Osborn}, H.~P. and {Bonfanti}, A. and {Wilson}, T.~G. and {Brandeker}, A. and {G{\"u}nther}, M.~N. and {Lendl}, M. and {Kitzmann}, D. and {Fossati}, L. and {Mordasini}, C. and {Sousa}, S.~G. and {Adibekyan}, V. and {Fridlund}, M. and {Pezzotti}, C. and {Gandolfi}, D. and {Ulmer-Moll}, S. and {Alonso}, R. and {B{\'a}rczy}, T. and {Barrado Navascues}, D. and {Barros}, S.~C.~C. and {Baumjohann}, W. and {Benz}, W. and {Billot}, N. and {Borsato}, L. and {Broeg}, C. and {Collier Cameron}, A. and {Correia}, A.~C.~M. and {Csizmadia}, Sz. and {Cubillos}, P.~E. and {Davies}, M.~B. and {Deleuil}, M. and {Deline}, A. and {Demangeon}, O.~D.~S. and {Demory}, B. -O. and {Derekas}, A. and {Edwards}, B. and {Ehrenreich}, D. and {Erikson}, A. and {Fortier}, A. and {Gazeas}, K. and {Gillon}, M. and {G{\"u}del}, M. and {Heitzmann}, A. and {Helling}, Ch. and {Isaak}, K.~G. and {Kiss}, L.~L. and {Korth}, J. and {Lam}, K.~W.~F. and {Laskar}, J. and {Lecavelier des Etangs}, A. and {Luntzer}, A. and {Luque}, R. and {Magrin}, D. and {Maxted}, P.~F.~L. and {Mer{\'\i}n}, B. and {Munari}, M. and {Nascimbeni}, V. and {Olofsson}, G. and {Ottensamer}, R. and {Pagano}, I. and {Pall{\'e}}, E. and {Peter}, G. and {Piazza}, D. and {Piotto}, G. and {Pollacco}, D. and {Queloz}, D. and {Ragazzoni}, R. and {Rando}, N. and {Rauer}, H. and {Ribas}, I. and {Santos}, N.~C. and {Scandariato}, G. and {S{\'e}gransan}, D. and {Simon}, A.~E. and {Smith}, A.~M.~S. and {Southworth}, R. and {Stalport}, M. and {Sulis}, S. and {Szab{\'o}}, Gy. M. and {Udry}, S. and {Van Grootel}, V. and {Venturini}, J. and {Villaver}, E. and {Walton}, N.~A. and {Wolf}, S. and {Wolter}, D.},
        title = "{Searching for hot water world candidates with CHEOPS: Refining the radii and analysing the internal structures and atmospheric lifetimes of TOI-238 b and TOI-1685 b}",
      journal = {\aap},
         year = 2025,
        month = apr,
       volume = {696},
          eid = {A28},
        pages = {A28},
          doi = {10.1051/0004-6361/202453325},
       adsurl = {https://ui.adsabs.harvard.edu/abs/2025A&A...696A..28E}
}

@ARTICLE{Zingales+2025,
       author = {{Zingales}, T. and {Malavolta}, L. and {Borsato}, L. and {Turrini}, D. and {Bonfanti}, A. and {Polychroni}, D. and {Mantovan}, G. and {Nardiello}, D. and {Nascimbeni}, V. and {Lanza}, A.~F. and {Bekkelien}, A. and {Sozzetti}, A. and {Broeg}, C. and {Naponiello}, L. and {Lendl}, M. and {Bonomo}, A.~S. and {Simon}, A.~E. and {Desidera}, S. and {Piotto}, G. and {Mancini}, L. and {Hooton}, M.~J. and {Bignamini}, A. and {Egger}, J.~A. and {Maggio}, A. and {Alibert}, Y. and {Locci}, D. and {Delrez}, L. and {Biassoni}, F. and {Fossati}, L. and {Cabona}, L. and {Lacedelli}, G. and {Carleo}, I. and {Leonardi}, P. and {Andreuzzi}, G. and {Brandeker}, A. and {Cosentino}, R. and {Correia}, A.~C.~M. and {Claudi}, R. and {Alonso}, R. and {Damasso}, M. and {Wilson}, T.~G. and {B{\'a}rczy}, T. and {Pinamonti}, M. and {Baker}, D. and {Barkaoui}, K. and {Barrado Navascues}, D. and {Barros}, S.~C.~C. and {Baumjohann}, W. and {Beck}, T. and {Beichman}, C. and {Benz}, W. and {Bieryla}, A. and {Billot}, N. and {Bosch-Cabot}, P. and {Bouma}, L.~G. and {Ciardi}, D.~R. and {Collier Cameron}, A. and {Collins}, K.~A. and {Crossfield}, Ian J.~M. and {Csizmadia}, Sz. and {Cubillos}, P.~E. and {Davies}, M.~B. and {Deleuil}, M. and {Deline}, A. and {Demangeon}, O.~D.~S. and {Demory}, B. -O. and {Derekas}, A. and {Dragomir}, D. and {Edwards}, B. and {Ehrenreich}, D. and {Erikson}, A. and {Falk}, B. and {Fortier}, A. and {Fridlund}, M. and {Fukui}, A. and {Gandolfi}, D. and {Gazeas}, K. and {Gillon}, M. and {Gonzales}, E. and {G{\"u}del}, M. and {Guerra}, P. and {G{\"u}nther}, M.~N. and {Heitzmann}, A. and {Helling}, Ch. and {Howell}, S.~B. and {Isaak}, K.~G. and {Jenkins}, J. and {Kiss}, L.~L. and {Korth}, J. and {Lam}, K.~W.~F. and {Laskar}, J. and {Lecavelier des Etangs}, A. and {Magrin}, D. and {Matson}, R. and {Matthews}, E.~C. and {Maxted}, P.~F.~L. and {McDermott}, S. and {Munari}, M. and {Mordasini}, C. and {Narita}, N. and {Olofsson}, G. and {Ottensamer}, R. and {Pagano}, I. and {Pall{\'e}}, E. and {Peter}, G. and {Pollacco}, D. and {Queloz}, D. and {Ragazzoni}, R. and {Rando}, N. and {Ratti}, F. and {Rauer}, H. and {Ribas}, I. and {Salmon}, S. and {Santos}, N.~C. and {Scandariato}, G. and {Seager}, S. and {S{\'e}gransan}, D. and {Smith}, A.~M.~S. and {Schlieder}, J. and {Schwarz}, R.~P. and {Shporer}, A. and {Sousa}, S.~G. and {Stalport}, M. and {Steinberger}, M. and {Sulis}, S. and {Szab{\'o}}, Gy. M. and {Twicken}, J.~D. and {Udry}, S. and {Van Grootel}, V. and {Venturini}, J. and {Villaver}, E. and {Walton}, N.~A. and {Winn}, J.~N.},
        title = "{A joint effort to discover and characterize two resonant mini-Neptunes around TOI-1803 with TESS, HARPS-N, and CHEOPS}",
      journal = {\aap},
         year = 2025,
        month = mar,
       volume = {695},
          eid = {A273},
        pages = {A273},
          doi = {10.1051/0004-6361/202451180},
       adsurl = {https://ui.adsabs.harvard.edu/abs/2025A&A...695A.273Z}
}

@ARTICLE{Krenn+2024,
       author = {{Krenn}, A.~F. and {Kubyshkina}, D. and {Fossati}, L. and {Egger}, J.~A. and {Bonfanti}, A. and {Deline}, A. and {Ehrenreich}, D. and {Beck}, M. and {Benz}, W. and {Cabrera}, J. and {Wilson}, T.~G. and {Leleu}, A. and {Sousa}, S.~G. and {Adibekyan}, V. and {Correia}, A.~C.~M. and {Alibert}, Y. and {Delrez}, L. and {Lendl}, M. and {Patel}, J.~A. and {Venturini}, J. and {Alonso}, R. and {Anglada}, G. and {Asquier}, J. and {B{\'a}rczy}, T. and {Barrado Navascues}, D. and {Barros}, S.~C.~C. and {Baumjohann}, W. and {Beck}, T. and {Billot}, N. and {Bonfils}, X. and {Borsato}, L. and {Brandeker}, A. and {Broeg}, C. and {Charnoz}, S. and {Collier Cameron}, A. and {Csizmadia}, Sz. and {Cubillos}, P.~E. and {Davies}, M.~B. and {Deleuil}, M. and {Demangeon}, O.~D.~S. and {Demory}, B. -O. and {Erikson}, A. and {Fortier}, A. and {Fridlund}, M. and {Gandolfi}, D. and {Gillon}, M. and {G{\"u}del}, M. and {G{\"u}nther}, M.~N. and {Hasiba}, J. and {Heitzmann}, A. and {Helling}, C. and {Hoyer}, S. and {Isaak}, K.~G. and {Kiss}, L.~L. and {Lam}, K.~W.~F. and {Laskar}, J. and {Lecavelier des Etangs}, A. and {Lovis}, C. and {Magrin}, D. and {Maxted}, P.~F.~L. and {Mordasini}, C. and {Nascimbeni}, V. and {Olofsson}, G. and {Ottensamer}, R. and {Pagano}, I. and {Pall{\'e}}, E. and {Peter}, G. and {Piotto}, G. and {Pollacco}, D. and {Queloz}, D. and {Ragazzoni}, R. and {Rando}, N. and {Rauer}, H. and {Ribas}, I. and {Rieder}, M. and {Santos}, N.~C. and {Scandariato}, G. and {S{\'e}gransan}, D. and {Simon}, A.~E. and {Smith}, A.~M.~S. and {Stalport}, M. and {Steller}, M. and {Szab{\'o}}, Gy. M. and {Thomas}, N. and {Udry}, S. and {Ulmer}, B. and {Van Grootel}, V. and {Villaver}, E. and {Viotto}, V. and {Walton}, N.~A. and {Zingales}, T.},
        title = "{Characterisation of the TOI-421 planetary system using CHEOPS, TESS, and archival radial velocity data}",
      journal = {\aap},
         year = 2024,
        month = jun,
       volume = {686},
          eid = {A301},
        pages = {A301},
          doi = {10.1051/0004-6361/202348584},
archivePrefix = {arXiv},
       eprint = {2404.11074},
 primaryClass = {astro-ph.EP},
       adsurl = {https://ui.adsabs.harvard.edu/abs/2024A&A...686A.301K}
}

@ARTICLE{Rosario+2024,
       author = {{Ros{\'a}rio}, N.~M. and {Demangeon}, O.~D.~S. and {Barros}, S.~C.~C. and {Gandolfi}, D. and {Egger}, J.~A. and {Serrano}, L.~M. and {Osborn}, H.~P. and {Beck}, M. and {Benz}, W. and {Flor{\'e}n}, H. -G. and {Guterman}, P. and {Wilson}, T.~G. and {Alibert}, Y. and {Fossati}, L. and {Hooton}, M.~J. and {Delrez}, L. and {Santos}, N.~C. and {Sousa}, S.~G. and {Bonfanti}, A. and {Salmon}, S. and {Adibekyan}, V. and {Nigioni}, A. and {Venturini}, J. and {Alonso}, R. and {Anglada}, G. and {Asquier}, J. and {B{\'a}rczy}, T. and {Barrado Navascues}, D. and {Barrag{\'a}n}, O. and {Baumjohann}, W. and {Beck}, T. and {Billot}, N. and {Biondi}, F. and {Bonfils}, X. and {Borsato}, L. and {Brandeker}, A. and {Broeg}, C. and {Cessa}, V. and {Charnoz}, S. and {Collier Cameron}, A. and {Csizmadia}, Sz. and {Cubillos}, P.~E. and {Davies}, M.~B. and {Deleuil}, M. and {Deline}, A. and {Demory}, B. -O. and {Ehrenreich}, D. and {Erikson}, A. and {Esposito}, M. and {Fortier}, A. and {Fridlund}, M. and {Gillon}, M. and {G{\"u}del}, M. and {G{\"u}nther}, M.~N. and {Helling}, Ch. and {Hoyer}, S. and {Isaak}, K.~G. and {Kiss}, L.~L. and {Lam}, K.~W.~F. and {Laskar}, J. and {Lecavelier des Etangs}, A. and {Lendl}, M. and {Luntzer}, A. and {Magrin}, D. and {Maxted}, P.~F.~L. and {Mordasini}, C. and {Nascimbeni}, V. and {Olofsson}, G. and {Osborne}, H.~L.~M. and {Ottensamer}, R. and {Pagano}, I. and {Pall{\'e}}, E. and {Peter}, G. and {Piotto}, G. and {Pollacco}, D. and {Queloz}, D. and {Ragazzoni}, R. and {Rando}, N. and {Rauer}, H. and {Ribas}, I. and {Scandariato}, G. and {S{\'e}gransan}, D. and {Simon}, A.~E. and {Smith}, A.~M.~S. and {Stalport}, M. and {Szab{\'o}}, Gy. M. and {Thomas}, N. and {Udry}, S. and {Van Eylen}, V. and {Van Grootel}, V. and {Villaver}, E. and {Walter}, I. and {Walton}, N.~A.},
        title = "{Precise characterisation of HD 15337 with CHEOPS: A laboratory for planet formation and evolution}",
      journal = {\aap},
         year = 2024,
        month = jun,
       volume = {686},
          eid = {A282},
        pages = {A282},
          doi = {10.1051/0004-6361/202347759},
archivePrefix = {arXiv},
       eprint = {2403.16621},
 primaryClass = {astro-ph.EP},
       adsurl = {https://ui.adsabs.harvard.edu/abs/2024A&A...686A.282R}
}

@ARTICLE{Gardner+2006,
       author = {{Gardner}, Jonathan P. and {Mather}, John C. and {Clampin}, Mark and {Doyon}, Rene and {Greenhouse}, Matthew A. and {Hammel}, Heidi B. and {Hutchings}, John B. and {Jakobsen}, Peter and {Lilly}, Simon J. and {Long}, Knox S. and {Lunine}, Jonathan I. and {McCaughrean}, Mark J. and {Mountain}, Matt and {Nella}, John and {Rieke}, George H. and {Rieke}, Marcia J. and {Rix}, Hans-Walter and {Smith}, Eric P. and {Sonneborn}, George and {Stiavelli}, Massimo and {Stockman}, H.~S. and {Windhorst}, Rogier A. and {Wright}, Gillian S.},
        title = "{The James Webb Space Telescope}",
      journal = {\ssr},
         year = 2006,
        month = apr,
       volume = {123},
       number = {4},
        pages = {485-606},
          doi = {10.1007/s11214-006-8315-7},
archivePrefix = {arXiv},
       eprint = {astro-ph/0606175},
 primaryClass = {astro-ph},
       adsurl = {https://ui.adsabs.harvard.edu/abs/2006SSRv..123..485G}
}

@ARTICLE{Mayor+2011,
       author = {{Mayor}, M. and {Marmier}, M. and {Lovis}, C. and {Udry}, S. and {S{\'e}gransan}, D. and {Pepe}, F. and {Benz}, W. and {Bertaux}, J. -L. and {Bouchy}, F. and {Dumusque}, X. and {Lo Curto}, G. and {Mordasini}, C. and {Queloz}, D. and {Santos}, N.~C.},
        title = "{The HARPS search for southern extra-solar planets XXXIV. Occurrence, mass distribution and orbital properties of super-Earths and Neptune-mass planets}",
      journal = {arXiv e-prints},
         year = 2011,
        month = sep,
          eid = {arXiv:1109.2497},
        pages = {arXiv:1109.2497},
          doi = {10.48550/arXiv.1109.2497},
archivePrefix = {arXiv},
       eprint = {1109.2497},
 primaryClass = {astro-ph.EP},
       adsurl = {https://ui.adsabs.harvard.edu/abs/2011arXiv1109.2497M}
}

@ARTICLE{Fressin+2013,
       author = {{Fressin}, Fran{\c{c}}ois and {Torres}, Guillermo and {Charbonneau}, David and {Bryson}, Stephen T. and {Christiansen}, Jessie and {Dressing}, Courtney D. and {Jenkins}, Jon M. and {Walkowicz}, Lucianne M. and {Batalha}, Natalie M.},
        title = "{The False Positive Rate of Kepler and the Occurrence of Planets}",
      journal = {\apj},
         year = 2013,
        month = apr,
       volume = {766},
       number = {2},
          eid = {81},
        pages = {81},
          doi = {10.1088/0004-637X/766/2/81},
archivePrefix = {arXiv},
       eprint = {1301.0842},
 primaryClass = {astro-ph.EP},
       adsurl = {https://ui.adsabs.harvard.edu/abs/2013ApJ...766...81F}
}

@ARTICLE{Kunimoto+2020,
       author = {{Kunimoto}, Michelle and {Matthews}, Jaymie M.},
        title = "{Searching the Entirety of Kepler Data. II. Occurrence Rate Estimates for FGK Stars}",
      journal = {\aj},
         year = 2020,
        month = jun,
       volume = {159},
       number = {6},
          eid = {248},
        pages = {248},
          doi = {10.3847/1538-3881/ab88b0},
archivePrefix = {arXiv},
       eprint = {2004.05296},
 primaryClass = {astro-ph.EP},
       adsurl = {https://ui.adsabs.harvard.edu/abs/2020AJ....159..248K}
}

@ARTICLE{Hsu+2019,
       author = {{Hsu}, Danley C. and {Ford}, Eric B. and {Ragozzine}, Darin and {Ashby}, Keir},
        title = "{Occurrence Rates of Planets Orbiting FGK Stars: Combining Kepler DR25, Gaia DR2, and Bayesian Inference}",
      journal = {\aj},
         year = 2019,
        month = sep,
       volume = {158},
       number = {3},
          eid = {109},
        pages = {109},
          doi = {10.3847/1538-3881/ab31ab},
archivePrefix = {arXiv},
       eprint = {1902.01417},
 primaryClass = {astro-ph.EP},
       adsurl = {https://ui.adsabs.harvard.edu/abs/2019AJ....158..109H}
}

@ARTICLE{Ward1997,
       author = {{Ward}, William R.},
        title = "{Survival of Planetary Systems}",
      journal = {\apjl},
         year = 1997,
        month = jun,
       volume = {482},
       number = {2},
        pages = {L211-L214},
          doi = {10.1086/310701},
       adsurl = {https://ui.adsabs.harvard.edu/abs/1997ApJ...482L.211W}
}

@ARTICLE{VanEylen+2018,
       author = {{Van Eylen}, V. and {Agentoft}, Camilla and {Lundkvist}, M.~S. and {Kjeldsen}, H. and {Owen}, J.~E. and {Fulton}, B.~J. and {Petigura}, E. and {Snellen}, I.},
        title = "{An asteroseismic view of the radius valley: stripped cores, not born rocky}",
      journal = {\mnras},
         year = 2018,
        month = oct,
       volume = {479},
       number = {4},
        pages = {4786-4795},
          doi = {10.1093/mnras/sty1783},
archivePrefix = {arXiv},
       eprint = {1710.05398},
 primaryClass = {astro-ph.EP},
       adsurl = {https://ui.adsabs.harvard.edu/abs/2018MNRAS.479.4786V}
}

@ARTICLE{Fulton+Petigura2018,
       author = {{Fulton}, Benjamin J. and {Petigura}, Erik A.},
        title = "{The California-Kepler Survey. VII. Precise Planet Radii Leveraging Gaia DR2 Reveal the Stellar Mass Dependence of the Planet Radius Gap}",
      journal = {\aj},
         year = 2018,
        month = dec,
       volume = {156},
       number = {6},
          eid = {264},
        pages = {264},
          doi = {10.3847/1538-3881/aae828},
archivePrefix = {arXiv},
       eprint = {1805.01453},
 primaryClass = {astro-ph.EP},
       adsurl = {https://ui.adsabs.harvard.edu/abs/2018AJ....156..264F}
}

@ARTICLE{Ho+VanEylen2023,
       author = {{Ho}, Cynthia S.~K. and {Van Eylen}, Vincent},
        title = "{A deep radius valley revealed by Kepler short cadence observations}",
      journal = {\mnras},
         year = 2023,
        month = mar,
       volume = {519},
       number = {3},
        pages = {4056-4073},
          doi = {10.1093/mnras/stac3802},
archivePrefix = {arXiv},
       eprint = {2301.04062},
 primaryClass = {astro-ph.EP},
       adsurl = {https://ui.adsabs.harvard.edu/abs/2023MNRAS.519.4056H}
}

@ARTICLE{Ho+2024,
       author = {{Ho}, Cynthia S.~K. and {Rogers}, James G. and {Van Eylen}, Vincent and {Owen}, James E. and {Schlichting}, Hilke E.},
        title = "{Shallower radius valley around low-mass hosts: evidence for icy planets, collisions, or high-energy radiation scatter}",
      journal = {\mnras},
         year = 2024,
        month = jul,
       volume = {531},
       number = {3},
        pages = {3698-3714},
          doi = {10.1093/mnras/stae1376},
archivePrefix = {arXiv},
       eprint = {2401.12378},
 primaryClass = {astro-ph.EP},
       adsurl = {https://ui.adsabs.harvard.edu/abs/2024MNRAS.531.3698H}
}

@ARTICLE{Parc+2024,
       author = {{Parc}, L{\'e}na and {Bouchy}, Fran{\c{c}}ois and {Venturini}, Julia and {Dorn}, Caroline and {Helled}, Ravit},
        title = "{From super-Earths to sub-Neptunes: Observational constraints and connections to theoretical models}",
      journal = {\aap},
         year = 2024,
        month = aug,
       volume = {688},
          eid = {A59},
        pages = {A59},
          doi = {10.1051/0004-6361/202449911},
       adsurl = {https://ui.adsabs.harvard.edu/abs/2024A&A...688A..59P}
}

@ARTICLE{Piaulet+2024,
       author = {{Piaulet-Ghorayeb}, Caroline and {Benneke}, Bj{\"o}rn and {Radica}, Michael and {Raul}, Eshan and {Coulombe}, Louis-Philippe and {Ahrer}, Eva-Maria and {Kubyshkina}, Daria and {Howard}, Ward S. and {Krissansen-Totton}, Joshua and {MacDonald}, Ryan J. and {Roy}, Pierre-Alexis and {Louca}, Amy and {Christie}, Duncan and {Fournier-Tondreau}, Marylou and {Allart}, Romain and {Miguel}, Yamila and {Schlichting}, Hilke E. and {Welbanks}, Luis and {Cadieux}, Charles and {Dorn}, Caroline and {Evans-Soma}, Thomas M. and {Fortney}, Jonathan J. and {Pierrehumbert}, Raymond and {Lafreni{\`e}re}, David and {Acu{\~n}a}, Lorena and {Komacek}, Thaddeus and {Innes}, Hamish and {Beatty}, Thomas G. and {Cloutier}, Ryan and {Doyon}, Ren{\'e} and {Gagnebin}, Anna and {Gapp}, Cyril and {Knutson}, Heather A.},
        title = "{JWST/NIRISS Reveals the Water-rich ``Steam World'' Atmosphere of GJ 9827 d}",
      journal = {\apjl},
         year = 2024,
        month = oct,
       volume = {974},
       number = {1},
          eid = {L10},
        pages = {L10},
          doi = {10.3847/2041-8213/ad6f00},
archivePrefix = {arXiv},
       eprint = {2410.03527},
 primaryClass = {astro-ph.EP},
       adsurl = {https://ui.adsabs.harvard.edu/abs/2024ApJ...974L..10P}
}

@ARTICLE{Piaulet+2023,
       author = {{Piaulet}, Caroline and {Benneke}, Bj{\"o}rn and {Almenara}, Jose M. and {Dragomir}, Diana and {Knutson}, Heather A. and {Thorngren}, Daniel and {Peterson}, Merrin S. and {Crossfield}, Ian J.~M. and {Kempton}, Eliza M. -R. and {Kubyshkina}, Daria and {Howard}, Andrew W. and {Angus}, Ruth and {Isaacson}, Howard and {Weiss}, Lauren M. and {Beichman}, Charles A. and {Fortney}, Jonathan J. and {Fossati}, Luca and {Lammer}, Helmut and {McCullough}, P.~R. and {Morley}, Caroline V. and {Wong}, Ian},
        title = "{Evidence for the volatile-rich composition of a 1.5-Earth-radius planet}",
      journal = {Nature Astronomy},
         year = 2023,
        month = feb,
       volume = {7},
        pages = {206-222},
          doi = {10.1038/s41550-022-01835-4},
archivePrefix = {arXiv},
       eprint = {2212.08477},
 primaryClass = {astro-ph.EP},
       adsurl = {https://ui.adsabs.harvard.edu/abs/2023NatAs...7..206P}
}

@ARTICLE{Luo+2024,
       author = {{Luo}, Haiyang and {Dorn}, Caroline and {Deng}, Jie},
        title = "{The interior as the dominant water reservoir in super-Earths and sub-Neptunes}",
      journal = {Nature Astronomy},
         year = 2024,
        month = nov,
       volume = {8},
        pages = {1399-1407},
          doi = {10.1038/s41550-024-02347-z},
archivePrefix = {arXiv},
       eprint = {2401.16394},
 primaryClass = {astro-ph.EP},
       adsurl = {https://ui.adsabs.harvard.edu/abs/2024NatAs...8.1399L}
}

@ARTICLE{Valencia+2007,
       author = {{Valencia}, Diana and {Sasselov}, Dimitar D. and {O'Connell}, Richard J.},
        title = "{Radius and Structure Models of the First Super-Earth Planet}",
      journal = {\apj},
         year = 2007,
        month = feb,
       volume = {656},
       number = {1},
        pages = {545-551},
          doi = {10.1086/509800},
archivePrefix = {arXiv},
       eprint = {astro-ph/0610122},
 primaryClass = {astro-ph},
       adsurl = {https://ui.adsabs.harvard.edu/abs/2007ApJ...656..545V}
}

@INPROCEEDINGS{Pepe+2000,
       author = {{Pepe}, Francesco and {Mayor}, Michel and {Delabre}, Bernard and {Kohler}, Dominique and {Lacroix}, Daniel and {Queloz}, Didier and {Udry}, Stephane and {Benz}, Willy and {Bertaux}, Jean-Loup and {Sivan}, Jean-Pierre},
        title = "{HARPS: a new high-resolution spectrograph for the search of extrasolar planets}",
    booktitle = {Optical and IR Telescope Instrumentation and Detectors},
         year = 2000,
       editor = {{Iye}, Masanori and {Moorwood}, Alan F.},
       series = {Society of Photo-Optical Instrumentation Engineers (SPIE) Conference Series},
       volume = {4008},
        month = aug,
        pages = {582-592},
          doi = {10.1117/12.395516},
       adsurl = {https://ui.adsabs.harvard.edu/abs/2000SPIE.4008..582P}
}

@ARTICLE{Crossfield+2025,
       author = {{Crossfield}, Ian J.~M. and {Polanski}, Alex S. and {Robertson}, Paul and {Murphy}, Joseph Akana and {Turtelboom}, Emma V. and {Luque}, Rafael and {Beatty}, Thomas and {Daylan}, Tansu and {Isaacson}, Howard and {Brande}, Jonathan and {Kreidberg}, Laura and {Batalha}, Natalie M. and {Huber}, Daniel and {Rhem}, Maleah and {Dressing}, Courtney and {Kane}, Stephen R. and {Bossett}, Malik and {Gagnebin}, Anna and {Kroft}, Maxwell A. and {Premnath}, Pranav H. and {Rogers}, Claire J. and {Collins}, Karen A. and {Latham}, David W. and {Watkins}, Cristilyn N. and {Ciardi}, David R. and {Howell}, Steve B. and {Savel}, Arjun B. and {Berlind}, Perry and {Calkins}, Michael L. and {Esquerdo}, Gilbert A. and {Mink}, Jessica and {Clark}, Catherine A. and {Lund}, Michael B. and {Matson}, Rachel A. and {Everett}, Mark E. and {Schlieder}, Joshua E. and {Matthews}, Elisabeth C. and {Giacalone}, Steven and {Barclay}, Thomas and {Zambelli}, Roberto and {Plavchan}, Peter and {Ellingson}, Taylor and {Bowen}, Michael and {Srdoc}, Gregor and {McLeod}, Kim K. and {Schwarz}, Richard P. and {Barkaoui}, Khalid and {Kamler}, Jacob and {Murgas}, Felipe and {Palle}, Enric and {Narita}, Norio and {Fukui}, Akihiko and {Relles}, Howard M. and {Bieryla}, Allyson and {Girardin}, Eric and {Massey}, Bob and {Stockdale}, Chris and {Lewin}, Pablo and {Papini}, Riccardo and {Guerra}, Pere and {Conti}, Dennis M. and {Yal{\c{c}}inkaya}, Sel{\c{c}}uk and {Ba{\textcommabelow s}t{\"u}rk}, {\"O}zg{\"u}r and {Mourad}, Ghachoui},
        title = "{OrCAS: Origins, Compositions, and Atmospheres of Sub-Neptunes. I. Survey Definition}",
      journal = {\aj},
         year = 2025,
        month = feb,
       volume = {169},
       number = {2},
          eid = {89},
        pages = {89},
          doi = {10.3847/1538-3881/ad9aa6},
archivePrefix = {arXiv},
       eprint = {2411.16836},
 primaryClass = {astro-ph.EP},
       adsurl = {https://ui.adsabs.harvard.edu/abs/2025AJ....169...89C}
}

@software{Brandeker+2024,
       author = {{Brandeker}, Alexis and {Patel}, Jayshil A. and {Morris}, Brett M.},
        title = "{PIPE: Extracting PSF photometry from CHEOPS data}",
 howpublished = {Astrophysics Source Code Library, record ascl:2404.002},
         year = 2024,
        month = apr,
          eid = {ascl:2404.002},
       adsurl = {https://ui.adsabs.harvard.edu/abs/2024ascl.soft04002B}
}

@INPROCEEDINGS{Gibson+2016,
       author = {{Gibson}, Steven R. and {Howard}, Andrew W. and {Marcy}, Geoffrey W. and {Edelstein}, Jerry and {Wishnow}, Edward H. and {Poppett}, Claire L.},
        title = "{KPF: Keck Planet Finder}",
    booktitle = {Ground-based and Airborne Instrumentation for Astronomy VI},
         year = 2016,
       editor = {{Evans}, Christopher J. and {Simard}, Luc and {Takami}, Hideki},
       series = {Society of Photo-Optical Instrumentation Engineers (SPIE) Conference Series},
       volume = {9908},
        month = aug,
          eid = {990870},
        pages = {990870},
          doi = {10.1117/12.2233334},
       adsurl = {https://ui.adsabs.harvard.edu/abs/2016SPIE.9908E..70G}
}

@ARTICLE{Roy+2026,
       author = {{Roy}, Pierre-Alexis and {Benneke}, Bj{\"o}rn and {Fournier-Tondreau}, Marylou and {Coulombe}, Louis-Philippe and {Piaulet-Ghorayeb}, Caroline and {Lafreni{\`e}re}, David and {Allart}, Romain and {Cowan}, Nicolas B. and {Dang}, Lisa and {Johnstone}, Doug and {Langeveld}, Adam B. and {Pelletier}, Stefan and {Radica}, Michael and {Taylor}, Jake and {Albert}, Lo{\"\i}c and {Doyon}, Ren{\'e} and {Flagg}, Laura and {Jayawardhana}, Ray and {MacDonald}, Ryan J. and {Turner}, Jake D.},
        title = "{Diversity in the haziness and chemistry of temperate sub-Neptunes}",
      journal = {Nature Astronomy},
         year = 2026,
        month = mar,
       volume = {10},
        pages = {371-384},
          doi = {10.1038/s41550-025-02723-3},
archivePrefix = {arXiv},
       eprint = {2512.10876},
 primaryClass = {astro-ph.EP},
       adsurl = {https://ui.adsabs.harvard.edu/abs/2026NatAs..10..371R}
}

@ARTICLE{Chakrabarty+2026,
       author = {{Chakrabarty}, Aritra and {Mulders}, Gijs D. and {Aguichine}, Artyom and {Batalha}, Natalie},
        title = "{The Radius Cliff is a Waterfall: Explaining Sub-Neptune Exoplanets with Steam Worlds}",
      journal = {\apj},
         year = 2026,
        month = mar,
       volume = {999},
       number = {2},
          eid = {188},
        pages = {188},
          doi = {10.3847/1538-4357/ae3da8},
archivePrefix = {arXiv},
       eprint = {2602.11923},
 primaryClass = {astro-ph.EP},
       adsurl = {https://ui.adsabs.harvard.edu/abs/2026ApJ...999..188C}
}

@ARTICLE{Shibata+Izidoro2026,
       author = {{Shibata}, Sho and {Izidoro}, Andre},
        title = "{Elevated Eccentricities in the Radius Valley Hint at Water-rich Mini-Neptunes}",
      journal = {\apj},
         year = 2026,
        month = apr,
       volume = {1000},
       number = {2},
          eid = {224},
        pages = {224},
          doi = {10.3847/1538-4357/ae4a28},
archivePrefix = {arXiv},
       eprint = {2602.23250},
 primaryClass = {astro-ph.EP},
       adsurl = {https://ui.adsabs.harvard.edu/abs/2026ApJ..1000..224S}
}

@ARTICLE{Steinmeyer+2026,
       author = {{Steinmeyer}, Marie-Luise and {Dorn}, Caroline and {Werlen}, Aaron and {Grimm}, Simon L.},
        title = "{Coupled Thermal─Chemical Evolution Models of Sub-Neptunes Reveal Atmospheric Signatures of Their Formation Location}",
      journal = {\apj},
         year = 2026,
        month = apr,
       volume = {1001},
       number = {1},
          eid = {36},
        pages = {36},
          doi = {10.3847/1538-4357/ae4c47},
archivePrefix = {arXiv},
       eprint = {2601.21377},
 primaryClass = {astro-ph.EP},
       adsurl = {https://ui.adsabs.harvard.edu/abs/2026ApJ..1001...36S}
}

@ARTICLE{Li+2026,
       author = {{Li}, Jie and {Bergin}, Edwin A. and {Hirschmann}, Marc M. and {Blake}, Geoffrey A. and {Ciesla}, Fred J. and {Kempton}, Eliza M.-R.},
        title = "{Soot Planets Instead of Water Worlds}",
      journal = {\apjl},
         year = 2026,
        month = jan,
       volume = {997},
       number = {1},
          eid = {L29},
        pages = {L29},
          doi = {10.3847/2041-8213/ae29a6},
archivePrefix = {arXiv},
       eprint = {2508.16781},
 primaryClass = {astro-ph.EP},
       adsurl = {https://ui.adsabs.harvard.edu/abs/2026ApJ...997L..29L}
}

@ARTICLE{Ito+2025,
       author = {{Ito}, Yuichi and {Kimura}, Tadahiro and {Ohno}, Kazumasa and {Fujii}, Yuka and {Ikoma}, Masahiro},
        title = "{Monosilane Worlds: Sub-Neptunes with Atmospheres Shaped by Reduced Magma Oceans}",
      journal = {\apj},
         year = 2025,
        month = jul,
       volume = {987},
       number = {2},
          eid = {174},
        pages = {174},
          doi = {10.3847/1538-4357/add3fe},
archivePrefix = {arXiv},
       eprint = {2505.03200},
 primaryClass = {astro-ph.EP},
       adsurl = {https://ui.adsabs.harvard.edu/abs/2025ApJ...987..174I}
}

@ARTICLE{Madhusudhan+2025,
       author = {{Madhusudhan}, Nikku and {Holmberg}, M{\^a}ns and {Constantinou}, Savvas and {Cooke}, Gregory J.},
        title = "{Exploring the sub-Neptune frontier with JWST}",
      journal = {Proceedings of the National Academy of Science},
         year = 2025,
        month = sep,
       volume = {122},
       number = {39},
          eid = {e2416194122},
        pages = {e2416194122},
          doi = {10.1073/pnas.2416194122},
       adsurl = {https://ui.adsabs.harvard.edu/abs/2025PNAS..12216194M}
}

@ARTICLE{Werlen+2025,
       author = {{Werlen}, Aaron and {Dorn}, Caroline and {Burn}, Remo and {Schlichting}, Hilke E. and {Grimm}, Simon L. and {Young}, Edward D.},
        title = "{Sub-Neptunes Are Drier than They Seem: Rethinking the Origins of Water-rich Worlds}",
      journal = {\apjl},
         year = 2025,
        month = sep,
       volume = {991},
       number = {1},
          eid = {L16},
        pages = {L16},
          doi = {10.3847/2041-8213/adff73},
archivePrefix = {arXiv},
       eprint = {2507.00765},
 primaryClass = {astro-ph.EP},
       adsurl = {https://ui.adsabs.harvard.edu/abs/2025ApJ...991L..16W}
}

@ARTICLE{Cherubim+2025,
       author = {{Cherubim}, Collin and {Wordsworth}, Robin and {Bower}, Dan J. and {Sossi}, Paolo A. and {Adams}, Danica and {Hu}, Renyu},
        title = "{An Oxidation Gradient Straddling the Small Planet Radius Valley}",
      journal = {\apj},
         year = 2025,
        month = apr,
       volume = {983},
       number = {2},
          eid = {97},
        pages = {97},
          doi = {10.3847/1538-4357/adbca9},
archivePrefix = {arXiv},
       eprint = {2503.05055},
 primaryClass = {astro-ph.EP},
       adsurl = {https://ui.adsabs.harvard.edu/abs/2025ApJ...983...97C}
}

@ARTICLE{Bergin+2023,
       author = {{Bergin}, Edwin A. and {Kempton}, Eliza M.-R. and {Hirschmann}, Marc and {Bastelberger}, Sandra T. and {Teal}, D.~J. and {Blake}, Geoffrey A. and {Ciesla}, Fred J. and {Li}, Jie},
        title = "{Exoplanet Volatile Carbon Content as a Natural Pathway for Haze Formation}",
      journal = {\apjl},
         year = 2023,
        month = may,
       volume = {949},
       number = {1},
          eid = {L17},
        pages = {L17},
          doi = {10.3847/2041-8213/acd377},
archivePrefix = {arXiv},
       eprint = {2305.05056},
 primaryClass = {astro-ph.EP},
       adsurl = {https://ui.adsabs.harvard.edu/abs/2023ApJ...949L..17B}
}

@ARTICLE{Schlichting+Young2022,
       author = {{Schlichting}, Hilke E. and {Young}, Edward D.},
        title = "{Chemical Equilibrium between Cores, Mantles, and Atmospheres of Super-Earths and Sub-Neptunes and Implications for Their Compositions, Interiors, and Evolution}",
      journal = {\psj},
         year = 2022,
        month = may,
       volume = {3},
       number = {5},
          eid = {127},
        pages = {127},
          doi = {10.3847/PSJ/ac68e6},
archivePrefix = {arXiv},
       eprint = {2107.10405},
 primaryClass = {astro-ph.EP},
       adsurl = {https://ui.adsabs.harvard.edu/abs/2022PSJ.....3..127S}
}

@ARTICLE{Seo+2024,
       author = {{Seo}, Chanoul and {Ito}, Yuichi and {Fujii}, Yuka},
        title = "{Role of Magma Oceans in Controlling Carbon and Oxygen of Sub-Neptune Atmospheres}",
      journal = {\apj},
         year = 2024,
        month = nov,
       volume = {975},
       number = {1},
          eid = {14},
        pages = {14},
          doi = {10.3847/1538-4357/ad7461},
archivePrefix = {arXiv},
       eprint = {2408.17056},
 primaryClass = {astro-ph.EP},
       adsurl = {https://ui.adsabs.harvard.edu/abs/2024ApJ...975...14S}
}

@ARTICLE{Rigby+2024,
       author = {{Rigby}, Frances E. and {Pica-Ciamarra}, Lorenzo and {Holmberg}, M{\r{a}}ns and {Madhusudhan}, Nikku and {Constantinou}, Savvas and {Schaefer}, Laura and {Deng}, Jie and {Lee}, Kanani K.~M. and {Moses}, Julianne I.},
        title = "{Toward a Self-consistent Evaluation of Gas Dwarf Scenarios for Temperate Sub-Neptunes}",
      journal = {\apj},
         year = 2024,
        month = nov,
       volume = {975},
       number = {1},
          eid = {101},
        pages = {101},
          doi = {10.3847/1538-4357/ad6c38},
archivePrefix = {arXiv},
       eprint = {2409.03683},
 primaryClass = {astro-ph.EP},
       adsurl = {https://ui.adsabs.harvard.edu/abs/2024ApJ...975..101R}
}

@ARTICLE{Bower+2025,
       author = {{Bower}, Dan J. and {Thompson}, Maggie A. and {Hakim}, Kaustubh and {Tian}, Meng and {Sossi}, Paolo A.},
        title = "{Diversity of Low-mass Planet Atmospheres in the C─H─O─N─S─Cl System with Interior Dissolution, Nonideality, and Condensation: Application to TRAPPIST-1e and Sub-Neptunes}",
      journal = {\apj},
         year = 2025,
        month = dec,
       volume = {995},
       number = {1},
          eid = {59},
        pages = {59},
          doi = {10.3847/1538-4357/ae1479},
archivePrefix = {arXiv},
       eprint = {2507.00499},
 primaryClass = {astro-ph.EP},
       adsurl = {https://ui.adsabs.harvard.edu/abs/2025ApJ...995...59B}
}

@ARTICLE{Heng+2025,
       author = {{Heng}, Kevin and {Owen}, James E. and {Tian}, Meng},
        title = "{The Gradient of Mean Molecular Weight across the Radius Valley}",
      journal = {\apj},
         year = 2025,
        month = nov,
       volume = {994},
       number = {1},
          eid = {28},
        pages = {28},
          doi = {10.3847/1538-4357/ae0acc},
archivePrefix = {arXiv},
       eprint = {2504.02499},
 primaryClass = {astro-ph.EP},
       adsurl = {https://ui.adsabs.harvard.edu/abs/2025ApJ...994...28H}
}

@ARTICLE{Davenport+2025,
       author = {{Davenport}, Brian and {Kempton}, Eliza M.-R. and {Nixon}, Matthew C. and {Ih}, Jegug and {Deming}, Drake and {Fu}, Guangwei and {May}, E.~M. and {Bean}, Jacob L. and {Gao}, Peter and {Rogers}, Leslie and {Malik}, Matej},
        title = "{TOI-421 b: A Hot Sub-Neptune with a Haze-free, Low Mean Molecular Weight Atmosphere}",
      journal = {\apjl},
         year = 2025,
        month = may,
       volume = {984},
       number = {2},
          eid = {L44},
        pages = {L44},
          doi = {10.3847/2041-8213/adcd76},
archivePrefix = {arXiv},
       eprint = {2501.01498},
 primaryClass = {astro-ph.EP},
       adsurl = {https://ui.adsabs.harvard.edu/abs/2025ApJ...984L..44D}
}

@ARTICLE{Wallack+2026,
       author = {{Wallack}, Nicole L. and {Gao}, Peter and {Greklek-McKeon}, Michael and {Meech}, Annabella and {Aguichine}, Artyom and {Alam}, Munazza K. and {Alderson}, Lili and {Batalha}, Natasha E. and {Batalha}, Natalie M. and {Gagnebin}, Anna and {Gordon}, Tyler A. and {Kirk}, James and {L{\'o}pez-Morales}, Mercedes and {Moran}, Sarah E. and {Redai}, Jea Iyanla and {Scarsdale}, Nicholas and {Teske}, Johanna and {Wakeford}, Hannah R. and {Wogan}, Nicholas F. and {Wolfgang}, Angie},
        title = "{JWST COMPASS: NIRSpec/G395H Transmission Observations of the Sub-Neptune HD 15337 c}",
      journal = {\aj},
         year = 2026,
        month = mar,
       volume = {171},
       number = {3},
          eid = {180},
        pages = {180},
          doi = {10.3847/1538-3881/ae2d12},
archivePrefix = {arXiv},
       eprint = {2602.22327},
 primaryClass = {astro-ph.EP},
       adsurl = {https://ui.adsabs.harvard.edu/abs/2026AJ....171..180W}
}

@ARTICLE{Wallack+2024,
       author = {{Wallack}, Nicole L. and {Batalha}, Natasha E. and {Alderson}, Lili and {Scarsdale}, Nicholas and {Adams Redai}, Jea I. and {Aguichine}, Artyom and {Alam}, Munazza K. and {Gao}, Peter and {Wolfgang}, Angie and {Batalha}, Natalie M. and {Kirk}, James and {L{\'o}pez-Morales}, Mercedes and {Moran}, Sarah E. and {Teske}, Johanna and {Wakeford}, Hannah R. and {Wogan}, Nicholas F.},
        title = "{JWST COMPASS: A NIRSpec/G395H Transmission Spectrum of the Sub-Neptune TOI-836c}",
      journal = {\aj},
         year = 2024,
        month = aug,
       volume = {168},
       number = {2},
          eid = {77},
        pages = {77},
          doi = {10.3847/1538-3881/ad3917},
archivePrefix = {arXiv},
       eprint = {2404.01264},
 primaryClass = {astro-ph.EP},
       adsurl = {https://ui.adsabs.harvard.edu/abs/2024AJ....168...77W}
}

@ARTICLE{Lopez2017,
       author = {{Lopez}, Eric D.},
        title = "{Born dry in the photoevaporation desert: Kepler's ultra-short-period planets formed water-poor}",
      journal = {\mnras},
         year = 2017,
        month = nov,
       volume = {472},
       number = {1},
        pages = {245-253},
          doi = {10.1093/mnras/stx1558},
archivePrefix = {arXiv},
       eprint = {1610.01170},
 primaryClass = {astro-ph.EP},
       adsurl = {https://ui.adsabs.harvard.edu/abs/2017MNRAS.472..245L}
}

@ARTICLE{kubyshkina_2021mesa,
       author = {{Kubyshkina}, Daria and {Vidotto}, Aline A.},
        title = "{How does the mass and activity history of the host star affect the population of low-mass planets?}",
      journal = {\mnras},
         year = 2021,
        month = jun,
       volume = {504},
       number = {2},
        pages = {2034-2050},
          doi = {10.1093/mnras/stab897},
archivePrefix = {arXiv},
       eprint = {2103.13117},
 primaryClass = {astro-ph.EP},
       adsurl = {https://ui.adsabs.harvard.edu/abs/2021MNRAS.504.2034K}
}

@ARTICLE{scuflaire2008,
   author = {{Scuflaire}, R. and {Th{\'e}ado}, S. and {Montalb{\'a}n}, J. and 
	{Miglio}, A. and {Bourge}, P.-O. and {Godart}, M. and {Thoul}, A. and 
	{Noels}, A.},
    title = "{CL{\'E}S, Code Li{\'e}geois d'{\'E}volution Stellaire}",
  journal = {\apss},
archivePrefix = "arXiv",
   eprint = {0712.3471},
     year = 2008,
    month = aug,
   volume = 316,
    pages = {83-91},
      doi = {10.1007/s10509-007-9650-1},
   adsurl = {http://adsabs.harvard.edu/abs/2008Ap%26SS.316...83S}
}

@ARTICLE{salmon2021,
       author = {{Salmon}, S.~J.~A.~J. and {Van Grootel}, V. and {Buldgen}, G. and {Dupret}, M. -A. and {Eggenberger}, P.},
        title = "{Reinvestigating {\ensuremath{\alpha}} Centauri AB in light of asteroseismic forward and inverse methods}",
      journal = {\aap},
         year = 2021,
        month = feb,
       volume = {646},
          eid = {A7},
        pages = {A7},
          doi = {10.1051/0004-6361/201937174},
archivePrefix = {arXiv},
       eprint = {2011.14932},
 primaryClass = {astro-ph.SR},
       adsurl = {https://ui.adsabs.harvard.edu/abs/2021A&A...646A...7S}
}

@BOOK{Kurucz-93,
   author = {{Kurucz}, R.~L.},
    title = "{SYNTHE spectrum synthesis programs and line data}",
booktitle = {Kurucz CD-ROM, Cambridge, MA: Smithsonian Astrophysical Observatory, |c1993, December 4, 1993},
publisher = {Astrophysics Source Code Library},
     year = 1993,
   adsurl = {http://adsabs.harvard.edu/abs/1993sssp.book.....K}
}

@PHDTHESIS{Sneden-73,
   author = {{Sneden}, C.~A.},
    title = "{Carbon and Nitrogen Abundances in Metal-Poor Stars.}",
   school = {The University of Texas at Austin},
     year = 1973,
   adsurl = {http://adsabs.harvard.edu/abs/1973PhDT.......180S}
}

@INCOLLECTION{Sousa-14,
       author = {{Sousa}, S{\'e}rgio G.},
        title = "{ARES + MOOG: A Practical Overview of an Equivalent Width (EW) Method to Derive Stellar Parameters}",
    booktitle = {Determination of Atmospheric Parameters of B, A, F and G Type Stars},
         year = 2014,
       editor = {{Niemczura}, Ewa and {Smalley}, Barry and {Pych}, Wojtek},
        pages = {297-310},
          doi = {10.1007/978-3-319-06956-2_26},
       adsurl = {https://ui.adsabs.harvard.edu/abs/2014dapb.book..297S}
}

@ARTICLE{Sousa-07,
   author = {{Sousa}, S.~G. and {Santos}, N.~C. and {Israelian}, G. and {Mayor}, M. and 
	{Monteiro}, M.~J.~P.~F.~G.},
    title = "{A new code for automatic determination of equivalent widths: Automatic Routine for line Equivalent widths in stellar Spectra (ARES)}",
  journal = {A\&A},
   eprint = {arXiv:astro-ph/0703696},
     year = 2007,
    month = jul,
   volume = 469,
    pages = {783-791},
      doi = {10.1051/0004-6361:20077288},
   adsurl = {http://adsabs.harvard.edu/abs/2007A%26A...469..783S}
}

@ARTICLE{Sousa-15,
   author = {{Sousa}, S.~G. and {Santos}, N.~C. and {Adibekyan}, V. and {Delgado-Mena}, E. and 
	{Israelian}, G.},
    title = "{ARES v2: new features and improved performance}",
  journal = {\aap},
archivePrefix = "arXiv",
   eprint = {1504.02725},
 primaryClass = "astro-ph.IM",
     year = 2015,
    month = may,
   volume = 577,
      eid = {A67},
    pages = {A67},
      doi = {10.1051/0004-6361/201425463},
   adsurl = {http://adsabs.harvard.edu/abs/2015A%26A...577A..67S}
}

@ARTICLE{Sousa-21,
       author = {{Sousa}, S.~G. and {Adibekyan}, V. and {Delgado-Mena}, E. and {Santos}, N.~C. and {Rojas-Ayala}, B. and {Soares}, B.~M.~T.~B. and {Legoinha}, H. and {Ulmer-Moll}, S. and {Camacho}, J.~D. and {Barros}, S.~C.~C. and {Demangeon}, O.~D.~S. and {Hoyer}, S. and {Israelian}, G. and {Mortier}, A. and {Tsantaki}, M. and {Monteiro}, M.~A.},
        title = "{SWEET-Cat 2.0: The Cat just got SWEETer. Higher quality spectra and precise parallaxes from Gaia eDR3}",
      journal = {\aap},
         year = 2021,
        month = dec,
       volume = {656},
          eid = {A53},
        pages = {A53},
          doi = {10.1051/0004-6361/202141584},
archivePrefix = {arXiv},
       eprint = {2109.04781},
 primaryClass = {astro-ph.EP},
       adsurl = {https://ui.adsabs.harvard.edu/abs/2021A&A...656A..53S}
}

@ARTICLE{Santos-13,
   author = {{Santos}, N.~C. and {Sousa}, S.~G. and {Mortier}, A. and {Neves}, V. and 
	{Adibekyan}, V. and {Tsantaki}, M. and {Delgado Mena}, E. and 
	{Bonfils}, X. and {Israelian}, G. and {Mayor}, M. and {Udry}, S.
	},
    title = "{SWEET-Cat: A catalogue of parameters for Stars With ExoplanETs. I. New atmospheric parameters and masses for 48 stars with planets}",
  journal = {\aap},
archivePrefix = "arXiv",
   eprint = {1307.0354},
 primaryClass = "astro-ph.SR",
     year = 2013,
    month = aug,
   volume = 556,
      eid = {A150},
    pages = {A150},
      doi = {10.1051/0004-6361/201321286},
   adsurl = {http://adsabs.harvard.edu/abs/2013A%26A...556A.150S}
}

@ARTICLE{Tsantaki-2013,
       author = {{Tsantaki}, M. and {Sousa}, S.~G. and {Adibekyan}, V. Zh. and {Santos}, N.~C. and {Mortier}, A. and {Israelian}, G.},
        title = "{Deriving precise parameters for cool solar-type stars. Optimizing the iron line list}",
      journal = {\aap},
         year = 2013,
        month = jul,
       volume = {555},
          eid = {A150},
        pages = {A150},
          doi = {10.1051/0004-6361/201321103},
archivePrefix = {arXiv},
       eprint = {1304.6639},
 primaryClass = {astro-ph.SR},
       adsurl = {https://ui.adsabs.harvard.edu/abs/2013A&A...555A.150T}
}

@ARTICLE{Adibekyan-15,
       author = {{Adibekyan}, V. and {Figueira}, P. and {Santos}, N.~C. and
         {Sousa}, S.~G. and {Faria}, J.~P. and {Delgado-Mena}, E. and
         {Oshagh}, M. and {Tsantaki}, M. and {Hakobyan}, A.~A. and
         {Gonz{\'a}lez Hern{\'a}ndez}, J.~I. and {Su{\'a}rez-Andr{\'e}s}, L. and
         {Israelian}, G.},
        title = "{Identifying the best iron-peak and {\ensuremath{\alpha}}-capture elements for chemical tagging: The impact of the number of lines on measured scatter}",
      journal = {\aap},
         year = "2015",
        month = "Nov",
       volume = {583},
          eid = {A94},
        pages = {A94},
          doi = {10.1051/0004-6361/201527120},
archivePrefix = {arXiv},
       eprint = {1509.02419},
 primaryClass = {astro-ph.SR},
       adsurl = {https://ui.adsabs.harvard.edu/abs/2015A&A...583A..94A}
}

@ARTICLE{Adibekyan-12,
   author = {{Adibekyan}, V.~Z. and {Sousa}, S.~G. and {Santos}, N.~C. and 
	{Delgado Mena}, E. and {Gonz{\'a}lez Hern{\'a}ndez}, J.~I. and 
	{Israelian}, G. and {Mayor}, M. and {Khachatryan}, G.},
    title = "{Chemical abundances of 1111 FGK stars from the HARPS GTO planet search program. Galactic stellar populations and planets}",
  journal = {\aap},
archivePrefix = "arXiv",
   eprint = {1207.2388},
 primaryClass = "astro-ph.EP",
     year = 2012,
    month = sep,
   volume = 545,
      eid = {A32},
    pages = {A32},
      doi = {10.1051/0004-6361/201219401},
   adsurl = {http://adsabs.harvard.edu/abs/2012A%26A...545A..32A}
}

@article{blackwell_1977,
	adsurl = {https://ui.adsabs.harvard.edu/abs/1977MNRAS.180..177B},
	author = {{Blackwell}, D.~E. and {Shallis}, M.~J.},
	doi = {10.1093/mnras/180.2.177},
	journal = {\mnras},
	month = jul,
	pages = {177-191},
	title = {{Stellar angular diameters from infrared photometry. Application to Arcturus and other stars; with effective temperatures.}},
	volume = {180},
	year = 1977}

@inproceedings{castelli_2003,
	adsurl = {https://ui.adsabs.harvard.edu/abs/2003IAUS..210P.A20C},
	archiveprefix = {arXiv},
	author = {{Castelli}, F. and {Kurucz}, R.~L.},
	booktitle = {Modelling of Stellar Atmospheres},
	editor = {{Piskunov}, N. and {Weiss}, W.~W. and {Gray}, D.~F.},
	eprint = {astro-ph/0405087},
	month = jan,
	pages = {A20},
	primaryclass = {astro-ph},
	series = {IAU Symposium},
	title = {{New Grids of ATLAS9 Model Atmospheres}},
	volume = {210},
	year = 2003}

@ARTICLE{gaiacollaboration_2022,
       author = {{Gaia Collaboration} and {Vallenari}, A. and {Brown}, A.~G.~A. and {Prusti}, T. and {de Bruijne}, J.~H.~J. and {Arenou}, F. and {Babusiaux}, C. and {Biermann}, M. and {Creevey}, O.~L. and {Ducourant}, C. and {Evans}, D.~W. and {Eyer}, L. and {Guerra}, R. and {Hutton}, A. and {Jordi}, C. and {Klioner}, S.~A. and {Lammers}, U.~L. and {Lindegren}, L. and {Luri}, X. and {Mignard}, F. and {Panem}, C. and {Pourbaix}, D. and {Randich}, S. and {Sartoretti}, P. and {Soubiran}, C. and {Tanga}, P. and {Walton}, N.~A. and {Bailer-Jones}, C.~A.~L. and {Bastian}, U. and {Drimmel}, R. and {Jansen}, F. and {Katz}, D. and {Lattanzi}, M.~G. and {van Leeuwen}, F. and {Bakker}, J. and {Cacciari}, C. and {Casta{\~n}eda}, J. and {De Angeli}, F. and {Fabricius}, C. and {Fouesneau}, M. and {Fr{\'e}mat}, Y. and {Galluccio}, L. and {Guerrier}, A. and {Heiter}, U. and {Masana}, E. and {Messineo}, R. and {Mowlavi}, N. and {Nicolas}, C. and {Nienartowicz}, K. and {Pailler}, F. and {Panuzzo}, P. and {Riclet}, F. and {Roux}, W. and {Seabroke}, G.~M. and {Sordo}, R. and {Th{\'e}venin}, F. and {Gracia-Abril}, G. and {Portell}, J. and {Teyssier}, D. and {Altmann}, M. and {Andrae}, R. and {Audard}, M. and {Bellas-Velidis}, I. and {Benson}, K. and {Berthier}, J. and {Blomme}, R. and {Burgess}, P.~W. and {Busonero}, D. and {Busso}, G. and {C{\'a}novas}, H. and {Carry}, B. and {Cellino}, A. and {Cheek}, N. and {Clementini}, G. and {Damerdji}, Y. and {Davidson}, M. and {de Teodoro}, P. and {Nu{\~n}ez Campos}, M. and {Delchambre}, L. and {Dell'Oro}, A. and {Esquej}, P. and {Fern{\'a}ndez-Hern{\'a}ndez}, J. and {Fraile}, E. and {Garabato}, D. and {Garc{\'\i}a-Lario}, P. and {Gosset}, E. and {Haigron}, R. and {Halbwachs}, J. -L. and {Hambly}, N.~C. and {Harrison}, D.~L. and {Hern{\'a}ndez}, J. and {Hestroffer}, D. and {Hodgkin}, S.~T. and {Holl}, B. and {Jan{\ss}en}, K. and {Jevardat de Fombelle}, G. and {Jordan}, S. and {Krone-Martins}, A. and {Lanzafame}, A.~C. and {L{\"o}ffler}, W. and {Marchal}, O. and {Marrese}, P.~M. and {Moitinho}, A. and {Muinonen}, K. and {Osborne}, P. and {Pancino}, E. and {Pauwels}, T. and {Recio-Blanco}, A. and {Reyl{\'e}}, C. and {Riello}, M. and {Rimoldini}, L. and {Roegiers}, T. and {Rybizki}, J. and {Sarro}, L.~M. and {Siopis}, C. and {Smith}, M. and {Sozzetti}, A. and {Utrilla}, E. and {van Leeuwen}, M. and {Abbas}, U. and {{\'A}brah{\'a}m}, P. and {Abreu Aramburu}, A. and {Aerts}, C. and {Aguado}, J.~J. and {Ajaj}, M. and {Aldea-Montero}, F. and {Altavilla}, G. and {{\'A}lvarez}, M.~A. and {Alves}, J. and {Anders}, F. and {Anderson}, R.~I. and {Anglada Varela}, E. and {Antoja}, T. and {Baines}, D. and {Baker}, S.~G. and {Balaguer-N{\'u}{\~n}ez}, L. and {Balbinot}, E. and {Balog}, Z. and {Barache}, C. and {Barbato}, D. and {Barros}, M. and {Barstow}, M.~A. and {Bartolom{\'e}}, S. and {Bassilana}, J. -L. and {Bauchet}, N. and {Becciani}, U. and {Bellazzini}, M. and {Berihuete}, A. and {Bernet}, M. and {Bertone}, S. and {Bianchi}, L. and {Binnenfeld}, A. and {Blanco-Cuaresma}, S. and {Blazere}, A. and {Boch}, T. and {Bombrun}, A. and {Bossini}, D. and {Bouquillon}, S. and {Bragaglia}, A. and {Bramante}, L. and {Breedt}, E. and {Bressan}, A. and {Brouillet}, N. and {Brugaletta}, E. and {Bucciarelli}, B. and {Burlacu}, A. and {Butkevich}, A.~G. and {Buzzi}, R. and {Caffau}, E. and {Cancelliere}, R. and {Cantat-Gaudin}, T. and {Carballo}, R. and {Carlucci}, T. and {Carnerero}, M.~I. and {Carrasco}, J.~M. and {Casamiquela}, L. and {Castellani}, M. and {Castro-Ginard}, A. and {Chaoul}, L. and {Charlot}, P. and {Chemin}, L. and {Chiaramida}, V. and {Chiavassa}, A. and {Chornay}, N. and {Comoretto}, G. and {Contursi}, G. and {Cooper}, W.~J. and {Cornez}, T. and {Cowell}, S. and {Crifo}, F. and {Cropper}, M. and {Crosta}, M. and {Crowley}, C. and {Dafonte}, C. and {Dapergolas}, A. and {David}, M. and {David}, P. and {de Laverny}, P. and {De Luise}, F. and {De March}, R.},
        title = "{Gaia Data Release 3. Summary of the content and survey properties}",
      journal = {\aap},
         year = 2023,
        month = jun,
       volume = {674},
          eid = {A1},
        pages = {A1},
          doi = {10.1051/0004-6361/202243940},
archivePrefix = {arXiv},
       eprint = {2208.00211},
 primaryClass = {astro-ph.GA},
       adsurl = {https://ui.adsabs.harvard.edu/abs/2023A&A...674A...1G}
}

@article{lindegren_2021,
	adsurl = {https://ui.adsabs.harvard.edu/abs/2021A&A...649A...4L},
	archiveprefix = {arXiv},
	author = {{Lindegren}, L. and {Bastian}, U. and {Biermann}, M. and {Bombrun}, A. and {de Torres}, A. and {Gerlach}, E. and {Geyer}, R. and {Hern{\'a}ndez}, J. and {Hilger}, T. and {Hobbs}, D. and {Klioner}, S.~A. and {Lammers}, U. and {McMillan}, P.~J. and {Ramos-Lerate}, M. and {Steidelm{\"u}ller}, H. and {Stephenson}, C.~A. and {van Leeuwen}, F.},
	doi = {10.1051/0004-6361/202039653},
	eid = {A4},
	eprint = {2012.01742},
	journal = {\aap},
	month = may,
	pages = {A4},
	primaryclass = {astro-ph.IM},
	title = {{Gaia Early Data Release 3. Parallax bias versus magnitude, colour, and position}},
	volume = {649},
	year = 2021}

@article{schanche_2020,
	adsurl = {https://ui.adsabs.harvard.edu/abs/2020MNRAS.499..428S},
	archiveprefix = {arXiv},
	author = {{Schanche}, N. and {H{\'e}brard}, G. and {Collier Cameron}, A. and {Dalal}, S. and {Smalley}, B. and {Wilson}, T.~G. and {Boisse}, I. and {Bouchy}, F. and {Brown}, D.~J.~A. and {Demangeon}, O. and {Haswell}, C.~A. and {Hellier}, C. and {Kolb}, U.~C. and {Lopez}, T. and {Maxted}, P.~F.~L. and {Pollacco}, D.~L. and {West}, R.~G. and {Wheatley}, P.~J.},
	doi = {10.1093/mnras/staa2848},
	eprint = {2010.15205},
	journal = {\mnras},
	month = nov,
	number = {1},
	pages = {428-440},
	primaryclass = {astro-ph.EP},
	title = {{WASP-186 and WASP-187: two hot Jupiters discovered by SuperWASP and SOPHIE with additional observations by TESS}},
	volume = {499},
	year = 2020}

@article{Alibert2025,
	author = {Alibert, Yann and Davoult, Jeanne and Marques, Sara},
	title = {A transformer-based generative model for planetary systems},
	DOI= "10.1051/0004-6361/202452297",
	url= "https://doi.org/10.1051/0004-6361/202452297",
	journal = {A\&A},
	year = 2025,
	volume = 701,
	pages = "A149",
}

%%%%%%%%%%%%%%%%%%%%%%%%%%%%%%%%%%%%%%%%%%%%%%%%%%%%%%%%%%%%%%
\begin{appendix}
\onecolumn
\section{Observations and data analysis}
\begin{table}[H]
\renewcommand{\arraystretch}{1.4}
\caption{List of TESS observations of TOI-2211.}
\centering
\begin{tabular}{ccccccc}
\hline \hline
Sector & Camera & CCD & Start date [UTC] & End date [UTC] & Cadence [s] & \# Transits of TOI-2211\,b \\
\hline
27 & 1 & 3 & 2020-07-05 & 2020-07-30 & 120 & 8 \\
67 & 1 & 3 & 2023-07-01 & 2023-07-29 & 20 & 7 \\
92 & 2 & 1 & 2025-05-07 & 2025-06-03 & 20 & 7 \\
95 & 1 & 4 & 2025-07-25 & 2020-08-20 & 20 & 8 \\
\hline
\end{tabular}
\label{tab:tess_dat}
\end{table}
\renewcommand{\arraystretch}{1}

\begin{table}[H]
\renewcommand{\arraystretch}{1.4}
\caption{List of CHEOPS observations of TOI-2211.}
\centering
\begin{tabular}{cccccccc}
\hline \hline
Index & Start date [UTC] & Start date [BJD] & Dur [orbits] & File key & Av. eff. [\%] & RMS [ppm] \\
\hline
LC1 & 2024-06-17T12:00:36.753 & $ 2460479.000425386 $ & $ 7.40 $ & PR140068\_TG004401\_V0300 & $ 57 $ & $ 360 $ \\
LC2 & 2024-08-06T02:01:09.155 & $ 2460528.584133741 $ & $ 5.30 $ & PR140068\_TG005101\_V0300 & $ 80 $ & $ 330 $ \\
LC3 & 2024-08-21T13:19:48.799 & $ 2460544.0554259107 $ & $ 5.09 $ & PR140068\_TG005102\_V0300 & $ 76 $ & $ 343 $ \\
LC4 & 2024-09-18T10:20:01.845 & $ 2460571.9305769103 $ & $ 4.53 $ & PR140068\_TG005601\_V0300 & $ 61 $ & $ 371 $ \\
LC5 & 2025-07-18T12:33:15.184 & $ 2460875.023092412 $ & $ 4.89 $ & PR140068\_TG006901\_V0300 & $ 74 $ & $ 522 $ \\
LC6 & 2025-08-02T23:09:54.345 & $ 2460890.4652123223 $ & $ 4.73 $ & PR140068\_TG007301\_V0300 & $ 82 $ & $ 376 $ \\
LC7 & 2025-08-18T10:12:29.200 & $ 2460905.9253379656 $ & $ 4.80 $ & PR140068\_TG007302\_V0300 & $ 82 $ & $ 495 $ \\
LC8 & 2025-08-24T15:02:59.224 & $ 2460912.1270743515 $ & $ 4.89 $ & PR140068\_TG007303\_V0300 & $ 76 $ & $ 704 $ \\
\hline
\end{tabular}
\label{tab:cheops_dat}
\tablefoot{One CHEOPS orbit is 98.77~minutes. The average efficiency is the time fraction spent on the target during the visit and quantifies interruptions due to Earth occultations and crossings over the South-Atlantic anomaly.}
\end{table}
\renewcommand{\arraystretch}{1}

\begin{table}[H]
\centering
\footnotesize
 \caption[]{Priors of the RV model.}\label{tab:RVprior}
\begin{tabular}{c| c}
\hline \hline
Parameter & Prior\\
\hline
Semi-amplitude $K$ [m/s] & $\rm \mathcal{U}(0,10)$ \\
Orbital period $P$ [days] & $\rm \mathcal{LU}(1,2*time\,span)$ \\
Eccentricity $e$   & $\rm \mathcal{K}(0.867, 3.03)$\\
$\omega$ argument of periastron & $\rm \mathcal{U}(0,180)$\\
$\phi$ mean anomaly & $\rm \mathcal{U}(0,180)$ \\
\hline
\hline

\end{tabular}
\tablefoot{$ \rm \mathcal{U}(a,b)$ describes a uniform prior, with $a$ and $b$ being the minimum and maximum limits, respectively. $ \rm \mathcal{LU}(a,b)$ describes a log uniform prior, with $a$ and $b$ being the minimum and maximum limits, respectively. $ \rm \mathcal{K}(a,b)$ describes a Kumaraswamy distribution with shape parameters $a$ and $b$.}
\end{table}

\newpage
\begin{table}[H]
\renewcommand{\arraystretch}{1.4}
\caption{List of priors and posteriors for the joint photometry and RV fit with chexoplanet.}
\centering
\begin{tabular}{llr}
\hline
Parameter & Prior & Posterior\\
\hline
Stellar Radius, $R_s$ [R$_\odot$] & $\mathcal{N}(\mu=0.839,\sigma=0.007) $ &  $ 0.8386 \pm 0.007 $  \\
Stellar Mass, $M_s$ [M$_\odot$] & $\mathcal{N}(\mu=0.81,\sigma=0.05) $ &  $ 0.812 \pm 0.049 $  \\
Quadratic LD param $u_{\rm tess,0}$ & $\mathcal{N}_{\mathcal{U}}(\mu=0.4,\sigma=0.1,a=0.0000,b=1.0000)$ &  $ 0.33^{+0.08}_{-0.2} $  \\
Quadratic LD param $u_{\rm tess,0}$ & $\mathcal{N}_{\mathcal{U}}(\mu=0.21,\sigma=0.1,a=0.0000,b=1.0000)$ &  $ 0.32^{+0.1}_{-0.2} $  \\
Quadratic LD param $u_{\rm cheops,0}$ & $\mathcal{N}_{\mathcal{U}}(\mu=0.53,\sigma=0.1,a=0.0000,b=1.0000)$ &  $ 0.37 \pm 0.21 $  \\
Quadratic LD param $u_{\rm cheops,0}$ & $\mathcal{N}_{\mathcal{U}}(\mu=0.17,\sigma=0.1,a=0.0000,b=1.0000)$ &  $ 0.39^{+0.19}_{-0.3} $  \\
\hline
Epoch, $t_{{0,{{\rm b}}}}$ [BJD] & $\mathcal{N}(\mu=2459038.104,\sigma=0.009) $ &  $ 2459038.1045 \pm 0.002 $  \\
Period, $P_{{\rm b}}$ [d] & $\mathcal{N}_{\mathcal{U}}(\mu=3.09268,\sigma=7e-05,a=3.0926,b=3.0928)$ &  $ 3.0926838 \pm 0.0000040 $  \\
RV semi-amplitude, $K_{b}$ [ms$^{-1}$] & $\mathcal{N}(\mu=0,\sigma=10) $ &  $ 1.36 \pm 0.27 $  \\
Eccentricity, $e_{pl,b}$& $\mathcal{R}(\sigma=0.0490)$ &  $ 0.056^{+0.035}_{-0.028} $  \\
Argument of Periastron, $\omega_{b}$  [rad] & $ \mathcal{U}(a=0,b=6.3) $ &  $ 0.2 \pm 1.7 $  \\
log Radius Ratio, $\log{R_{\rm b}/R_s}$ & $ \mathcal{U}(a=-6.9,b=-2.3) $ &  $ -4.14 \pm 0.031 $  \\
Impact Param, $b_{\rm b}$ & $ \mathcal{{U}}(a=0.0,b=1+R_{p,b}/R_s)$ &  $ 0.52 \pm 0.09 $  \\
\hline
Epoch, $t_{{0,{{\rm c}}}}$ [BJD] & $\mathcal{N}(\mu=2460821.1,\sigma=0.6) $ &  $ 2460820.91 \pm 0.33 $  \\
Period, $P_{{\rm c}}$ [d] & $\mathcal{N}(\mu=15.6,\sigma=0.1) $ &  $ 15.611^{+0.050}_{-0.056} $  \\
RV semi-amplitude, $K_{c}$ [ms$^{-1}$] & $\mathcal{N}(\mu=0,\sigma=5) $ &  $ 2.52 \pm 0.28 $  \\
Eccentricity, $e_{pl,c}$ & $\mathcal{R}(\sigma=0.0490)$ &  $ 0.058^{+0.035}_{-0.028} $  \\
Argument of Periastron, $\omega_{c}$ [rad] & $ \mathcal{U}(a=0,b=6.3) $ &  $ -0.2^{+1.5}_{-1.2} $  \\
\hline
ESPRESSO RV Offset, $\mu_{\rm RV,ESPRESSO}$ [ms$^{-1}$] & $\mathcal{N}(\mu=-46748.0,\sigma=2.0) $ &  $ -46747.65 \pm 0.21 $  \\
ESPRESSO log RV jitter, $\log{\sigma_{RV,ESPRESSO}}$  & $\mathcal{N}_{\mathcal{U}}(\mu=-1.7,\sigma=1,a=-10.,b=5)$ &  $ -0.1 \pm 0.21 $  \\
HARPS RV Offset, $\mu_{\rm RV,HARPS}$ [ms$^{-1}$] & $\mathcal{N}(\mu=-46749.0,\sigma=3.0) $ &  $ -46748.46 \pm 0.79 $  \\
HARPS log RV jitter, $\log{\sigma_{RV,HARPS}}$ & $\mathcal{N}_{\mathcal{U}}(\mu=-0.3,\sigma=1,a=-10,b=5)$ &  $ 0.81 \pm 0.31 $  \\
KPF RV Offset, $\mu_{\rm RV,KPF}$ [ms$^{-1}$] & $\mathcal{N}(\mu=-46721.0,\sigma=2.0) $ &  $ -46721.09 \pm 0.66 $  \\
KPF log RV jitter, $\log{\sigma_{RV,KPF}}$ & $\mathcal{N}_{\mathcal{U}}(\mu=-1.0,\sigma=1,a=-10.,b=5)$ &  $ 0.89 \pm 0.19 $  \\
TESS jitter $\log{\sigma_{\rm TESS}}$ & $\mathcal{N}(\mu=-0.3,\sigma=1) $ &  $ -1.59^{+0.07}_{-0.08} $  \\
CHEOPS jitter $\log{\sigma_{\rm CHEOPS}}$ & $\mathcal{N}(\mu=-1.0,\sigma=3) $ &  $ -1.09^{+0.025}_{-0.025} $  \\
CHEOPS LC1 mean flux [ppt] & $\mathcal{N}(\mu=-0.0,\sigma=0.4) $ &  $ -0.028 \pm 0.059 $  \\
CHEOPS LC2 mean flux [ppt] & $\mathcal{N}(\mu=0.1,\sigma=0.4) $ &  $ 0.266 \pm 0.064 $  \\
CHEOPS LC3 mean flux [ppt] & $\mathcal{N}(\mu=0.0,\sigma=0.4) $ &  $ 0.098 \pm 0.058 $  \\
CHEOPS LC4 mean flux [ppt] & $\mathcal{N}(\mu=0.0,\sigma=0.4) $ &  $ 0.127 \pm 0.060 $  \\
CHEOPS LC5 mean flux [ppt] & $\mathcal{N}(\mu=0.0,\sigma=0.6) $ &  $ -0.04 \pm 0.13 $  \\
CHEOPS LC6 mean flux [ppt] & $\mathcal{N}(\mu=0.1,\sigma=0.5) $ &  $ -0.30 \pm 0.13 $  \\
CHEOPS LC7 mean flux [ppt] & $\mathcal{N}(\mu=0.1,\sigma=0.6) $ &  $ -1.20 \pm 0.15 $  \\
CHEOPS LC8 mean flux [ppt] & $\mathcal{N}(\mu=0.0,\sigma=1.0) $ &  $ -0.449 \pm 0.082 $  \\
\hline
\end{tabular}
\label{tab:joint_fit_priors_posteriors}
\tablefoot{$\mathcal{U}$ denotes a uniform distribution, $\mathcal{N}$ a normal distribution, $\mathcal{N}_{\mathcal{U}}$ a truncated normal distribution, and $\mathcal{R}$ a Rayleigh distribution. Not listed are the linear and quadratic decorrelation terms for the CHEOPS lightcurves as well as the parameters of the cubic spline used to model the CHEOPS roll angle modulations.}
\end{table}
\renewcommand{\arraystretch}{1}

% \newpage
% \begin{figure}[H]
%     \centering
%     \includegraphics[width=0.84\linewidth]{figures/TOI2211_RV_posteriors.pdf}
%     \caption{Corner plot of the parameters derived with \texttt{pyaneti}}
%     \label{fig:rv_posteriors}
% \end{figure}

\newpage
\section{Internal structure modelling}
\begin{table}[H]
\renewcommand{\arraystretch}{1.5}
\caption{Medians and one-sigma errors for the \texttt{plaNETic} internal structure modelling posteriors for TOI-2211 b.}
\centering
\begin{tabular}{r|ccc|ccc}
\hline \hline
Water prior &              \multicolumn{3}{c|}{Water-rich} & \multicolumn{3}{c}{Water-poor} \\
Si/Mg/Fe prior &           Stellar (A1) &       Iron-enriched (A2) &      Free (A3) &
                           Stellar (B1) &       Iron-enriched (B2) &      Free (B3) \\
\hline
w$_\textrm{core}$ [\%] &        $14_{-10}^{+9}$ &    $21_{-15}^{+17}$ &    $23_{-16}^{+21}$ &
                           $16_{-11}^{+9}$ &    $28_{-19}^{+18}$ &    $33_{-22}^{+22}$ \\
w$_\textrm{mantle}$ [\%] &      $80_{-10}^{+11}$ &    $71_{-18}^{+16}$ &    $69_{-22}^{+17}$ &
                           $83_{-9}^{+11}$ &    $72_{-18}^{+19}$ &    $67_{-22}^{+22}$ \\
w$_\textrm{envelope}$ [\%] &    $4.5_{-3.1}^{+5.8}$ &    $6.4_{-4.1}^{+6.6}$ &    $7.1_{-4.6}^{+6.9}$ &
$\left(1.2_{-0.2}^{+0.4}\right)$ $10^{-4}$ &    $\left(1.7_{-0.5}^{+1.4}\right)$ $10^{-4}$ &    $\left(2.1_{-0.9}^{+2.6}\right)$ $10^{-4}$ \\
\hline
%w$_\textrm{water}$ [\%] &       $4.4742699252_{-3.1117831108}^{+5.8016872131}$ &    $6.3296422294_{-4.1325983242}^{+6.5817859470}$ &    $7.0509794713_{-4.6190692256}^{+6.9472319884}$ &
%                           $0.0000006523_{-0.0000003122}^{+0.0000004206}$ &    $0.0000008659_{-0.0000004369}^{+0.0000008521}$ &    $0.0000010582_{-0.0000005670}^{+0.0000014800}$ \\
%w$_\textrm{H/He}$ [\%] &        $0.0031245290_{-0.0028242644}^{+0.0327786537}$ &    $0.0043006302_{-0.0039664315}^{+0.0529586081}$ &    $0.0045557763_{-0.0042158215}^{+0.0597117979}$ &
%                           $0.0001216801_{-0.0000166747}^{+0.0000425337}$ &    $0.0001689932_{-0.0000530304}^{+0.0001375988}$ &    $0.0002132513_{-0.0000885650}^{+0.0002581016}$ \\
Z$_\textrm{envelope}$ [\%] &        $99.9_{-1.0}^{+0.1}$ &    $99.9_{-1.2}^{+0.1}$ &    $99.9_{-1.2}^{+0.1}$ &
                           $0.5_{-0.2}^{+0.2}$ &    $0.5_{-0.2}^{+0.2}$ &    $0.5_{-0.2}^{+0.2}$ \\
\hline
x$_\textrm{Fe,core}$ [\%] &     $90.3_{-6.4}^{+6.6}$ &    $90.3_{-6.4}^{+6.6}$ &    $90.4_{-6.4}^{+6.5}$ &
                           $90.0_{-6.3}^{+6.8}$ &    $90.3_{-6.4}^{+6.6}$ &    $90.4_{-6.4}^{+6.5}$ \\
x$_\textrm{S,core}$ [\%] &      $9.7_{-6.6}^{+6.4}$ &    $9.7_{-6.6}^{+6.4}$ &    $9.6_{-6.5}^{+6.4}$ &
                           $10.0_{-6.8}^{+6.3}$ &    $9.7_{-6.6}^{+6.4}$ &    $9.6_{-6.5}^{+6.4}$ \\
\hline
x$_\textrm{Si,mantle}$ [\%] &   $37_{-5}^{+6}$ &    $32_{-8}^{+8}$ &    $29_{-20}^{+29}$ &
                           $37_{-5}^{+6}$ &    $30_{-7}^{+9}$ &    $24_{-17}^{+27}$ \\
x$_\textrm{Mg,mantle}$ [\%] &   $49_{-6}^{+7}$ &    $42_{-10}^{+10}$ &    $34_{-22}^{+27}$ &
                           $49_{-6}^{+7}$ &    $40_{-10}^{+11}$ &    $34_{-21}^{+26}$ \\
x$_\textrm{Fe,mantle}$ [\%] &   $14_{-9}^{+8}$ &    $25_{-16}^{+17}$ &    $29_{-20}^{+23}$ &
                           $14_{-9}^{+9}$ &    $30_{-18}^{+16}$ &    $36_{-22}^{+21}$ \\
\hline
\end{tabular}
\label{tab:internal_structure_results_b}
\end{table}
\renewcommand{\arraystretch}{1.0}
\end{appendix}

\end{document}